\documentclass[aps,groupedaddress,floatfix,nofootinbib,10pt,twocolumn]{revtex4}

\usepackage{graphicx}
\usepackage{amsmath}
\usepackage{amssymb}
\usepackage{mathrsfs}

\usepackage{dcolumn}

\usepackage{color}

\newcommand{\beq}{\begin{equation}}
\newcommand{\eeq}{\end{equation}}
\newcommand{\ba}{\begin{array}}
\newcommand{\ea}{\end{array}}
\newcommand{\beqa}{\begin{eqnarray}}
\newcommand{\eeqa}{\end{eqnarray}}
\newcommand{\bal}{\begin{align}}

\newcommand{\trib}{|\!|\!|}
\newcommand{\bes}{\begin{split}}
\newcommand{\ees}{\end{split}}

\def\Hb{\mathbb{H}}   
\def\Rb{\mathbb{R}}
\def\Lb{\mathbb{L}}
\def\Qb{\mathbb{Q}}
\def\Cb{\mathbb{C}}

\DeclareRobustCommand{\scrD}{\mathscr{D}}
\DeclareRobustCommand{\iQ}{\rm{Q}}
\newcommand{\mc}[1]{\mathcal{#1}}

\def\t{\textstyle}

\begin{document}

\title{The many-nucleon theory of nuclear collective structure and its
  macroscopic limits: an algebraic perspective\footnote{This is an
    author-created, un-copyedited version of an article published in
    Physica Scripta [Phys.~Scr.~\textbf{91}, 033003 (2016)]. IOP
    Publishing Ltd.\ is not responsible for any errors or omissions in
    this version of the manuscript or any version derived from it. The
    Version of Record is available online at
    doi:10.1088/0031-8949/91/3/033003.}}

 \author{ D J Rowe}
\address{Department of Physics,
  University of Toronto, Toronto, Ontario M5S\,1A7, Canada}
\author{A E McCoy}
\author{M A Caprio}
\address{Department of Physics, University of Notre Dame,
  Notre Dame, Indiana 46556-5670, USA}



\date{\today}

\begin{abstract}
The nuclear collective models introduced by Bohr, Mottelson and Rainwater, together with the Mayer-Jensen shell model, have provided the central framework for the development of nuclear physics.
This paper reviews the microscopic evolution of the collective  models and their underlying foundations.
In particular, it is shown that the Bohr-Mottelson  models have expressions as macroscopic limits of  microscopic  models that have precisely-defined expressions in many-nucleon quantum mechanics.
Understanding  collective models in this way is especially useful because it enables the analysis of nuclear properties in terms of them to be revisited and reassessed in the light of their  microscopic foundations.
\end{abstract}

\maketitle
  

\section{Introduction}
During the time that the independent-particle shell model of the nucleus was being developed, the quadrupole moments of many odd nuclei were measured and determined to have small values for nuclei close to  magic numbers.  
However,  nuclei with intermediate nucleon numbers 
\cite{Schmidt40, TownesFL49} were often observed to have quadrupole moments much larger than those of a single nucleon.
This was interpreted by Rainwater \cite{Rainwater50} to mean that such nuclei must have spheroidal as opposed to spherical shapes 
and challenged a standard belief of that time that even numbers of neutrons or protons would pair off to give states of zero spin so that the  spin and the quadrupole moment of an odd nucleus would be that of the single unpaired nucleon.
With considerable insight, Rainwater  suggested that the independent-particle states of nucleons in a spheroidal potential should be expected to have density distributions of the same shape as the potential.
Thus, on the basis of shape consistency he concluded that nuclei with spheroidal shapes could be expected.

Following Rainwater's observations, 
Bohr \cite{Bohr52} considered the quantum mechanics for the shape vibrations and deformations of a liquid-drop model of the nucleus.
He also considered the possibility that the interactions between the nucleons and the deformed field that they generated could enhance the deformation of nuclei.
In parallel with developments of the shell model
\cite{Mayer49,HaxelJS49},
Bohr and Mottelson \cite{BohrM53} then extended the Bohr model to a unified model of a nucleus with coupled independent-particle and collective degrees of freedom.

A limitation of the Bohr model is that, as a quantum fluid,  it has irrotational-flow moments of inertia which are much smaller than those needed to describe the low-energy rotational states of deformed nuclei.
In the unified model, it was supposed, again with remarkable insight, that low-energy states would result with the introduction of vorticity associated with the nucleons in  partially-filled shells.
Thus, the unified model was developed 
in a strong-coupling approximation in which the collective core of the nucleus would be polarised  by the added nucleons and assume a spheroidal shape that would  rotate adiabatically relative to the faster intrinsic dynamics of the coupled system.
An important contribution to this development  
was the calculation by Nilsson \cite{Nilsson55} of  single-particle states in a spheroidal shell-model potential.

It would be an understatement to say that the Bohr-Mottelson-Nilsson model  and its many developments (see, for example, 
\cite{DavydovF58,GneussG71,HessSMG80,Stephens75,RoweWC09}
and the many references therein)
have been successful in the interpretation of collective structure in nuclei.
One measure of its success is that it continues to provide the basic concepts and language in terms of which nuclear collective phenomena are described.
Thus, a great deal of effort has been expended by numerous researchers from a variety of geometrical, mean-field and algebraic perspectives,
 to give this phenomenologically-conceived model a microscopic foundation.
The several perspectives ultimately led to a 
formally-precise many-nucleon theory of nuclear monopole  
and quadrupole collective dynamics.

Justifying this statement and explaining  what it means is a primary objective of this paper.
The many-nucleon theory that emerges is the so-called symplectic shell model.
Like the standard  shell model, the symplectic shell model is 
a formal  framework in which simple models of  nuclear properties can  be understood and developed.
The standard shell model expresses the  many-nucleon Hilbert space as an ordered sum of spherical harmonic-oscillator subspaces, whereas the symplectic shell model expresses it as a sum of  spaces
that carry representations of a collective model.
Solvable models then emerge, e.g., by restriction of the shell-model space to a subspace and by considering approximate model interactions.

The  organisation of this paper is first 
to take stock of what has been achieved 
by extracting from the large body of literature 
a short and direct route to the many-nucleon theory of collective dynamics. 
The plan is to first identify the physical content and properties of the theory and then, by  considering its macroscopic limits, explore how it can be used in simple ways to interpret experimental data and  other models with microscopic observables.
It is shown that the macroscopic limits are approached to an extraordinarily high level of accuracy in medium-to-heavy nuclei.
It is also shown that the underlying microscopic structure of the unified models has strong implications for the low-energy beta- and gamma-vibrational states of rotational nuclei.

Models that are  usefully considered from the symplectic-model perspective are those based  on mean-field methods
\cite{BenderHR03,MatsuyanagiMNHS10}.
A popular example is the pairing plus quadrupole model 
\cite{Belyaev59,KisslingerS63,BarangerK68V} 
which explores the competition
\cite{Mottelson62} between 
pair-coupling of nuclei, as in the model of superconductivity
\cite{BohrMP58}, and the $Q.Q$ interaction of aligned coupling,
as in the SU(3) model \cite{Elliott58ab}.
Such models are discussed briefly in Section \ref{sect:MFmodels}.

We would like this presentation to be easy to understand.
At the same time, we wish to avoid suppressing the rigorous foundations on which the models are based.
Thus, where appropriate, we start sections with a brief outline of their  content so that a reader can obtain a quick understanding of its substance and return later to the more challenging details, which can be read as desired.

\section{Early approaches}\label{sect:2}
The search for a microscopic version of the collective model was initiated by Tomonaga
\cite{Tomonaga55}
in two dimensions and extended by Miyazima and Tamura
\cite{MiyazimaT56} to three. 
It successfully identified  irrotational-flow momenta  that  become canonically conjugate to quadrupole moments in a macroscopic limit.
Had this approach been pursued, it could have led more directly to the  expression of the Bohr model as a microscopic irrotational hydrodynamic model, as shown in Section \ref{sect:Bohr}, and even to the model  given in Section \ref{sect:cm3} that included vorticity.

A few years later,  Elliott introduced an SU(3) model as a
coupling scheme for the spherical shell model and showed 
that the states of an irreducible SU(3) representation, 
within the space of valence-shell nucleons in a single spherical harmonic-oscillator  shell, 
has properties that resemble those of a truncated rotor model.
This showed that states with rotation-like properties can emerge in the shell model with effective interactions and other effective observables that conserve the symmetries of the SU(3) model.
Subsequently, and unexpectedly, the SU(3) model re-emerged as a sub-model of the sought-after collective model 
and as the projected image of a rigid-rotor model onto the states of a single harmonic-oscillator shell-model energy.
Thus, although Elliott's model was only intended to apply to light nuclei, it  acquired a relationship to the rotational states of well-deformed nuclei, albeit in spaces very different from those of the original SU(3) model.

The unified model of  Bohr, Mottelson and Nilsson was based on the premiss that a separation could be made between  collective and intrinsic particle degrees of freedom of a nucleus.
Thus, in seeking a derivation of this model, a suitable separation of the nucleon variables was pursued by many authors
\cite{ScheidG68,Cusson68,Rowe70,VillarsC70,%
Zickendraht71,DzyublikOSF72,GulshaniR76,BuckBC79}.
The several results were subsequently obtained
\cite{RoweR79,RoweR80}  as simple expressions of the nuclear kinetic energy as a Laplacian on a classical many-nucleon space of which  collective model spaces were subspaces.
Such methods also showed \cite{Rowe70} that collective and intrinsic dynamics could only be completely decoupled  if the intrinsic structure had an unphysical degree of rigidity. 
Otherwise the collective rotations would necessarily be strongly coupled to intrinsic vorticity. 
This posed a serious challenge because the vortex spin operators did not have simple expressions as one-body operators.
However, the challenge evaporated with the recognition  of an intimate relationship  \cite{RoweR79,RoweR80} between the canonical transformation method
and simple models based on the quantisation of  
Lie algebras of collective observables expressed in microscopic terms.

An algebraic expression of a rigid-rotor was given by Ui \cite{Ui70} and an algebraic expression for a rotor-vibrator model was given by Weaver, Biedenharn and Cusson \cite{Ui70,WeaverBC73,BuckBC79}.
Neither of them quite achieved the desired results;
the first because a completely-rigid rotor
is unphysical and the second because non-zero
 vorticity, while clearly essential, is not conserved in nuclear collective dynamics.
However, they were very influential and were quickly followed by the symplectic model \cite{RosensteelR77a,RosensteelR80,Rowe85}
which  contains these two models as submodels but without their limitations. 
As this review will show, the symplectic model is  a microscopic version of the Bohr-Motteson-Nilsson collective model and defines a practical shell-model coupling scheme.

It should be mentioned that the symplectic model and
 many of the results reviewed in this paper  
can also be obtained from a complementary approach based on the
O$(A-1)$ symmetry group introduced,  by Dzyublik, Filippov, Simonov, and colleagues \cite{DzyublikOSF72,Filippov73,Filippov78,Simonov68},
as the group that leaves collective model observables invariant.
It transpires \cite{MoshinskyQ71}
that O$(A-1)$ is a maximal group of transformations of a many-particle space that commute with the dynamical group of the collective model  that we consider.
This group plays a central role in the formulation by Kretzchmar, Smirnov, Vanagas, and many colleagues of  a translationally-invariant shell model \cite{Kretzchmar60,NeudachinS69,Vanagas80}.
Such dual pairs of commuting symmetry and dynamical groups occur frequently in nuclear physics and give important relationships between  collective models and shell-model coupling schemes 
\cite{RoweCR12}.

\section{Realisations of microscopic collective models as algebraic models}
\label{sect:3}
A quick and simple approach to a microscopic theory of collective  states in nuclei is obtained by identifying the essential observables that characterise their properties and expressing them in terms of a  basic Lie algebra of  observables.%
\footnote{The concept of a Lie algebra of observables was introduced in Dirac's formulation of quantum mechanics as a set of, in principle, measurable properties of a system that  are represented in quantum mechanics as Hermitian operators on the states of a Hilbert space and which close under commutation (or a Poisson bracket in classical mechanics)  to generate a Lie algebra.
For example,  the elementary position and momentum coordinates 
$\{ x_i,p_i\}$ of a particle are represented in quantum mechanics as operators $\{ \hat x_i, \hat p_i\}$ that satisfy the  commutation relations
$[ \hat x_j, \hat p_k] = {\rm i} \hbar\delta_{j,k} \hat I$ of a Heisenberg algebra, where $\hat I$ is the identity operator.
Thus, any set of variables that have representations as Hermitian operators and close under commutation to form a Lie algebra will be described as a Lie algebra of observables.
Note that, if $\hat X$ and $\hat Y$ are  Hermitian operators,  
their commutator will not be Hermitian.
Hence the factor ${\rm i} = \sqrt{-1}$ is required in the commutation relations of a quantum-mechanical Lie algebra of observables.
}
The elements of such a Lie algebra are the infinitesimal generators of a so-called dynamical group  for the model.
Any operator that is a polynomial in the elements of such a Lie algebra then has a well-defined representation  on the states of an irreducible representation of that Lie algebra.
In fact, the whole underlying many-nucleon quantum mechanics of  nuclear structure theory makes use of this approach by recognising that the many-nucleon observables of a given nucleus are all expressible in terms of the finite Lie algebra, 
with simple commutation relations, generated by the position, momentum and intrinsic spins of nucleons.
 The remarkable property of this Lie algebra, without which nuclear physics would be impossibly complex, arises from the observation that nucleons are identical particles and obey the Pauli exclusion principle.
This means that the  many-nucleon quantum mechanics  of a given nucleus has a single unique fully anti-symmetric representation in accordance with  the Stone-von Neumann theorem.%
\footnote{The Heisenberg algebra for a single particle in ordinary 3-dimensional space has a unitary representation on the infinite-dimensional Hilbert space of square-integrable functions of its 
$\{ x_1,x_2,x_3\}$ coordinates; 
this is the Hilbert space of a 3-dimensional harmonic oscillator.
 However, whereas Lie algebras, in general, have infinite numbers of inequivalent unitary representations, the
 Stone-von Neumann theorem \cite{vonNeumann32} shows that the Heisenberg algebra has only one.
This uniqueness is generally taken for granted. 
In fact, it is quite remarkable and results in the simplification of many-particle quantum mechanics by many orders of magnitude.  
Had it not been true, the discovery of quantum mechanics would surely have been delayed by many years.
}

In spite of its underlying simplicity, the representation of many-nucleon quantum mechanics is infinite-dimensional  and has the potential for unlimited complexity.
However, in common with most zero-temperature systems in physics, 
its lowest-energy states tend to be ones that maximise the available symmetries and its lowest-energy dynamics tend to be collective in nature and characterised by corresponding dynamical symmetries.
This section shows how simple collective models emerge in many-nucleon quantum mechanics from this perspective.

The basic observables of all the models discussed in this section are fully symmetric quadratic functions of the nucleon position and momentum observables.
Thus, in all applications, it is a simple matter to subtract the spurious 
centre-of-mass contributions to these observables.  
In effect, this operation reduces the effective number of particles from $A$ to $A-1$ but doesn't change  the commutation relations of their observables. 
Neither does it change the permutation symmetry of the model wave functions because the centre-of-mass wave function is always fully symmetric.  
This means that totally antisymmetric wave functions  can be constructed as linear combinations of  products of spatial
 and spin-isospin wave functions of conjugate permutation symmetry.

\subsection{A microscopic rotor model}\label{sect:Ui_rotor}
A simple  rotor model was formulated by Ui \cite{Ui70} as an algebraic model, with a Lie algebra of basic observables that contains a set of commuting quadrupole operators, expressed as components of an $L=2$ spherical tensor,
$\{ \hat Q_{2\nu} , \nu = 0, \pm1,\pm2\}$,
and angular momentum operators 
$\{ \hat L_0,\hat L_{\pm 1}= \mp \hat L_\pm/ \sqrt{2}\}$ with commutation relations
\begin{subequations}
\bal
& [\hat L_0, \hat L_\pm] = \pm \hat L_\pm, \quad
 [\hat L_+, \hat L_-] = 2\hat L_0 ,  \\
& [\hat L_\pm, \hat Q_{2\nu}] = 
   \sqrt{(2\mp \nu)(2\pm \nu+1)}\, \hat Q_{2,\nu\pm 1} , \\
&   [\hat L_0, \hat Q_{2\nu}] = \nu \hat Q_{2\nu}, \\
& [ \hat Q_{2\mu},\hat Q_{2\nu}]=0. 
\end{align}
\end{subequations}
 The quadrupole observables are elements of an Abelian Lie algebra and the angular momenta span an SO(3) Lie algebra. 
Together they form a 
a so-called ROT(3) Lie algebra whose elements
 have  expressions
\begin{subequations}
 \bal
&\hat L_0 = \hat L_{23}  , \quad 
\hat L_{\pm 1} 
= \mp \tfrac{1}{\sqrt{2}} (\hat L_{31} \pm {\rm i} \hat L_{12}) ,    \\
& \hat Q_{2,0}  
 = {\textstyle \frac{1}{\sqrt{2}}}\, (2\hat Q_{11} - \hat Q_{22} -\hat  Q_{33}), \\
&   \hat Q_{2,\pm 1} = {\textstyle\sqrt{3}}\,
(\hat Q_{12} \pm {\rm i} \hat Q_{13}), \\
&\hat Q_{2,\pm 2} =  
{\textstyle \frac{\sqrt{3}}{2}}\, 
(\hat Q_{22}- \hat Q_{33} \pm 2{\rm i} \hat Q_{23}), 
\end{align}
\end{subequations}
in terms of the microscopic observables, 
\beq
\label{eq:Uiobs}
 \hat Q_{ij} =\sum_{n=1}^A \hat x_{ni} \hat x_{nj} , \quad
\hbar \hat L_{ij} =
 \sum_{n=1}^A \big(\hat x_{ni}\hat p_{nj}  -\hat x_{nj}\hat p_{ni}\big)  .
\eeq 
 
 The construction of the unitary representations of this Lie algebra,
cf.\ Section \ref{sect:rotor_rep},
is straightforward and provides a prototype for the construction of more general rotor models with appropriate intrinsic degrees of freedom such as those of the unified model.
However, as it stands, 
its intrinsic states are eigenstates of the nuclear quadrupole moments with no vibrational fluctuations.
Thus, the model admits no Coriolis or centrifugal stretching associated with the coupling between the intrinsic and rotational dynamics.
In fact, its intrinsic states are effectively those of a rigid body and should have rigid-body moments of inertia
\cite{Rowe70, RoweR79}.
As a consequence, the irreducible representations 
of the Ui rotor model can only be realised on a many-nucleon Hilbert 
space in a  limit in which their intrinsic wave functions approach delta functions in their quadrupole shapes.
Nevertheless, it provided a microscopic version of the 
rigid-rotor model and was an important
step towards the microscopic theory of nuclear rotations that was being sought.

 \subsection{The microscopic Bohr model}  \label{sect:Bohr}
The original Bohr model of a nucleus was a hydrodynamic model  with a sharply-defined surface defined by its radius expressed as a function of  angles in a spherical-harmonic expansion
 \beq R(\theta, \phi) =  R_0\big( 1+ 
  \sum_{LM} \alpha_{LM}^* Y_{LM} (\theta, \phi) + O(\alpha^2)\big).
  \eeq
Collective Hamiltonians were then considered of the form
\beq H = \tfrac12 \sum_{M} B_L |\dot \alpha_{LM}|^2 + V(\alpha) ,
\eeq
where the $B_L$ coefficients are mass parameters.
The  model was quantized by mapping the classical momenta
$\pi_{LM} = B_L \dot \alpha_{LM} $ to operators 
$\hat\pi_{LM} = -{\rm i}\hbar \partial/\partial \alpha^*_{LM}$.
The various multipole moments of a nuclear density with shapes given by these parameters  then defined corresponding collective models of nuclear vibrations and rotations.

 A microscopic  version of the Bohr model for combined monopole and quadrupole collective dynamics
is obtained, as in Ui's model, by  replacing the surface monopole/quadrupole shape coordinates of the Bohr model with many-nucleon Cartesian quadrupole moments 
$\{Q_{ij} = \sum_{n=1}^A x_{ni} x_{nj}\}$.
Time derivatives and corresponding momenta are then given  by
\begin{subequations}
\bal
& \dot Q_{ij} = \frac{dQ_{ij}}{dt} =\sum_n (\dot x_{ni} x_{nj} + x_{ni} \dot x_{nj}) , \\
& P_{ij} = M \dot Q_{ij} =\sum_n ( p_{ni} x_{nj} + x_{ni} p_{nj}), 
\end{align}
\end{subequations}
where $M$ is the nucleon mass.
These  moments and momenta are then  quantised as appropriate
  for  nucleons by replacing the 
$x_{ni}$ and $p_{ni}$ coordinates by operators
$ \hat x_{ni}$ and $\hat p_{ni}$ with  commutation relations
$[\hat x_{ni}, \hat p_{mj}] = {\rm i}\hbar \delta_{i,j} \delta_{m,n}$, to obtain quantum observables
\beq \hat Q_{ij} =\sum_n \hat x_{ni} \hat x_{nj} , \quad
\hat P_{ij} =  \sum_n ( \hat p_{ni}\hat x_{nj} +\hat x_{ni} \hat p_{nj}), \label{eq:7.hatQhatP}
\eeq
that satisfy the commutation relations
\beq [\hat  Q_{ij}, \hat P_{kl}] =  {\rm i}\hbar \big( \delta_{il}\hat Q_{jk} +  \delta_{ik}  \hat Q_{jl} + \delta_{jl} \hat  Q_{ik} + \delta_{jk}  \hat Q_{il}  \big) . \label{eq:QPcr} \eeq

The classical Bohr model, with a Heisenberg algebra of observables,
is now regained by a contraction corresponding to a hydrodynamic limit of these commutation relations.
The essential property of nuclear matter that is presumed in taking its so-called hydrodynamic limit is that it is essentially incompressible, to within quantum mechanical limits, over the range of  states of interest.
For practical purposes, it is a limit in which the volume of a nucleus, as characterised by its monopole moment, takes an essentially constant value. 
The contraction process we consider is  similar to that introduced by  
\.In\"onu and Wigner \cite{InonuW53} to describe the contraction of the Lie algebra of the Lorentz group to that of the Galilean group  when the observables of the Lie algebra are restricted to states of an object of velocities of small magnitude relative to the speed of light.

The contraction is obtained by expressing
the monopole and quadrupole moments and momenta of a nucleus 
in units of a small parameter $\epsilon$ that is assigned a value such that $4\epsilon^2 = 1/\langle \hat Q_0\rangle$, 
where 
$\langle \hat Q_0\rangle$ is the mean value of the close-to-constant monopole moment of the nucleus in its low-energy states.
In these units, the monopole/quadrupole observables are given by
\begin{subequations}
\bal 
&\hat q_0 = \epsilon  \hat Q_{0} 
=\epsilon (\hat Q_{11} + \hat Q_{22} +\hat  Q_{33}) , \\
& \hat p_0 = \epsilon (\hat P_{11} + \hat P_{22} +\hat  P_{33}),
\end{align}
\end{subequations}
 and 
\begin{subequations}
\bal 
& \hat q_{2,0} = \epsilon  \hat Q_{2,0} 
=\tfrac{1}{\sqrt{2}}\,\epsilon (2\hat Q_{11} - \hat Q_{22} -\hat  Q_{33}), \\
&  \hat p^{2,0} =\tfrac{1}{\sqrt{2}}\,\epsilon 
(2\hat P_{11} - \hat P_{22} -\hat  P_{33}), \\
& \hat q_{2,\pm 1} =  \epsilon  \hat Q_{2,\pm 1} 
= {\textstyle\sqrt{3}}\,\epsilon  (\hat Q_{12} \pm {\rm i} \hat Q_{13}),\\
&  \hat p^{2,\pm 1} = {\textstyle\sqrt{3}}\,\epsilon 
(\hat P_{12} \mp {\rm i} \hat P_{13}),\\
&  \hat q_{2,\pm 2}= \epsilon  \hat Q_{2,\pm 2} =  
(\hat Q_{22}- \hat Q_{33} \pm 2{\rm i} \hat Q_{23}),\\
&  \hat p^{2,\pm 2} =  {\textstyle \frac{\sqrt{3}}{2}}\,\epsilon 
(\hat P_{22}- \hat P_{33} \mp 2{\rm i} \hat P_{23}).
\end{align}
\end{subequations}
Thus, with the commutation relations of equation  (\ref{eq:QPcr}),
it is  determined that
\beq [\hat q_0, \hat p_0] = {\rm i} \hbar\, \hat I + O(\epsilon^2), \quad
[\hat q_{2\mu}, \hat p^{2\nu}]  =  {\rm i} \hbar\,  \delta_\mu^\nu + O(\epsilon)
,\eeq
and that the  commutators, $[\hat q_0, \hat p^{2\nu}] $, 
 $[\hat q_{2\nu}, \hat p_0]$, $[\hat q_{2\mu}, \hat q_{2\nu}]$, 
 and $[\hat p^{2\mu}, \hat p^{2\nu}]$ are of order $O(\epsilon)$.

Two significant observations follow for the Bohr model derived in this way.   
One is that its quadrupole moments are the infinitesimal generators of irrotational flows, consistent with Bohr's formulation of the model as a quantum fluid model.  
The second is that the simplest shell-model representation 
of the model in its spherical vibrational limit is given
for a closed harmonic-oscillator shell-model nucleus.
In such a representation, the one-phonon monopole excitation is at a  high excitation energy as expected for almost-incompressible nuclear matter and the quadrupole excitations are high-energy  giant-quadrupole resonance excitations.
However, it is clear that, without the additional  degrees of freedom of valence-shell nucleons, as introduced in the unified model, the microscopic Bohr model, is not able to provide a description of low-energy rotational states.

\subsection{A microscopic Bohr model with vorticity} \label{sect:cm3}
The previous section showed that the operators 
$\{\hat Q_{ij} ,\hat P_{ij}\}$ 
become the elements of a Heisenberg algebra when restricted
 to a subspace of the nuclear Hilbert space involving only giant-resonance degrees of freedom and in which nuclear matter is  incompressible.
 In this respect the Bohr model is seen to be complementary to the SU(3) model which is the restriction of Ui's rigid rotor model to a subspace that excludes the giant-resonance excitations.
 However, as observed by Weaver, Biedenharn and Cusson \cite{WeaverBC73,WeaverCB76}, neither of these restrictions is necessary because  the commutation relations of the 
$\{\hat Q_{ij} ,\hat P_{ij}\}$ operators are all expressible
in terms  of angular-momentum operators $\{ \hat L_{ij}\}$
defined by
\beq  \hbar \hat L_{ij} 
= \sum_{n=1}^A \big(\hat x_{ni}\hat p_{nj}  -\hat x_{nj}\hat p_{ni}\big) .
\eeq
Moreover, the commutation relations between all pairs of operators
in the augmented set $\{\hat Q_{ij} ,\hat P_{ij}, \hat L_{ij}\}$ produce no new linearly-independent operators and are a basis for a Lie algebra.

A model with this Lie algebra of observables is the so-called 
GCM(3) model \cite{WeaverBC73,WeaverCB76} where
GCM(3) is a mnemonic for a generalised collective model in
three-dimensions.%
\footnote{The GCM(3) model is a minor extension of the original CM(3) model of Weaver \emph{et al.}\ \cite{WeaverBC73,WeaverCB76}
to include the giant monopole degrees of freedom and which, as a result, fits more naturally within the chain of models under discussion.}

Construction of the irreps (irreducible representations) of the GCM(3) model \cite{RosensteelR76,WeaverCB76} shows that it
extends the irrotational flows of the Bohr model to include an intrinsic angular momentum corresponding to quantised vorticity.
Thus, it is significant  that, with the admission of nuclear compressibility, however small, and the inclusion of the angular momentum to form a Lie algebra, the model acquires more general representations.

It is instructive to consider the microscopic version of the Bohr model with vorticity in the light of an argument \cite{Rowe70}
that the  states generated by rotation of a given intrinsic state can only have moments of inertia that differ from the rigid-body moments of inertia of the intrinsic density distribution  if the intrinsic state includes cluster-like components of zero angular momentum.
The complementary interpretation is now that the intrinsic states of a quantum fluid can only have moments of inertia different from those of irrotational flow if it has cluster-like components of non-zero vortex angular momentum.
 
\subsection{The symplectic model}
A serious problem with the GCM(3) model is that there is no easy way to construct a Hamiltonian for the model that gives energy spectra comparable to those observed.
To describe rotational states, the Hamiltonian should clearly contain a kinetic-energy term.
One possibility would be a kinetic energy  proportional to the  rotationally-invariant quadratic $\sum_{ij} \hat P_{ij} \hat P_{ji}$.
A more meaningful choice would be the GCM(3)-conserving component  of the many-nucleon kinetic energy.
However, the many expansions 
\cite{Zickendraht71,DzyublikOSF72,GulshaniR76,RoweR79}
of the nuclear kinetic energy in terms of collective and 
complementary intrinsic coordinates, show  this choice to be complicated.
More significantly, the expansions show that the nuclear kinetic energy
contains large components that do not conserve the 
vorticity quantum number of a GCM(3) irrep.
After many deliberations as to how to handle this problem, the answer that emerged was to simply add the many-nucleon kinetic energy to the Lie algebra of the GCM(3) model to form a still larger algebra.
This is possible because the commutator of two bilinear combinations of nucleon position and momentum coordinates is again a bilinear combination of these coordinates.
The algebra that emerges is the Lie algebra of the  Sp$(3,\Rb)$ symplectic group  \cite{RosensteelR77a}.

In retrospect, it can be seen that the Sp$(3,\Rb)$ algebra is the smallest Lie algebra that contains the nuclear quadrupole moments and its kinetic energy, both of which are essential components of a microscopic model of nuclear collective states.
It is the Lie algebra of observables spanned by the operators
\begin{subequations}\label{eq:CartesianSp3Rops}
\bal 
&\hat Q_{ij} = \sum_{n=1}^A \hat x_{ni} \hat x_{nj} , \quad 
\hat K_{ij}= \sum_{n=1}^A  \hat p_{ni}  \hat p_{nj},
\\
&\hbar \hat L_{ij} 
=  \sum_{n=1}^A \big(\hat x_{ni}\hat p_{nj}  -\hat x_{nj}\hat p_{ni}\big), 
\\
& \hat P_{ij} 
=  \sum_{n=1}^A ( \hat x_{ni} \hat p_{nj} + \hat p_{ni} \hat x_{nj}) .
\end{align}
\end{subequations}
Thus, it includes monopole and quadrupole moments,  given by linear combinations of the elements 
$\{ \hat Q_{ij}, i,j = 1,2,3\}$ which define the size and shape of a many-nucleon nucleus.
It includes infinitesimal generators of size and shape change, 
given  by linear combinations of $\{ \hat P_{ij} \}$, 
and infinitesimal generators of rotations given by the angular momenta $\{ \hat L_{ij} \}$  in units of $\hbar$.
The nuclear kinetic energy is proportional to $\sum_{n}  \hat p_{ni}^2$  and the other elements
$\{ \hat K_{ij}\}$ are included to close the Lie algebra.  
It is interesting to note that elements of the symplectic model algebra are expressed in equation (\ref{eq:CartesianSp3Rops}) in terms of elements of the   subalgebras of the following chain, that are added one set at a time, 
\beq 
\begin{array}{ccccccccc}
 {\rm SO}(3)&\!\!\subset \!\! &   {\rm ROT}(3)    &\!\! \subset\!\! 
 &  {\rm GCM(3)}  &\!\! \subset\!\! &      {\rm Sp}(3,\Rb)
  \\
 \hat L_{ij} &&  \hat Q_{ij}&& \hat P_{ij} &&\hat K_{ij} 
\end{array} .
\eeq
The Sp$(3,\Rb)$ Lie group that finally emerges is recognised as the group of all linear canonical transfomations of a
6-dimensional phase space \cite{MoshinskyQ71} 
with 3 position coordinates and 3 canonical momenta.

An alternative route to the symplectic model is  obtained from the developments by Filippov, Vanagas, Smirnov, and many colleagues
\cite{FilippovOS72,Vanagas80}
of a translationally-invariant shell model for which  O$(A-1)$, where $A$ is the nuclear mass number, is a symmetry group and for which 
Sp$(3,\Rb)$ is a complementary dual group as recognised in nuclear physics by Moshinsky and Quesne \cite{MoshinskyQ71}.

\section{The symplectic model as a unified model}
\label{sect:Sp3Rmodel}
The most valuable property of the symplectic model is that it is a bridge between the shell model and the collective model.
Thus, it is a microscopic unified model in every respect.
From a collective model perspective, it is an algebraic  model with  observables defined in terms of nucleon position and momentum coordinates.
However, when its dynamical group  Sp$(3,\Rb)$ is combined with the 
spin and isospin SU(2) groups, it  defines a complete coupling scheme for the many-nucleon shell-model Hilbert space in a spherical harmonic-oscillator basis.
In fact, as this section will show,  the whole many-nucleon Hilbert space is a sum of symplectic-model spaces with subspaces defined by the subgroups in the chain
\beq {\rm Sp}(3,\Rb) \supset {\rm U(3)}
 \supset {\rm SU}(3) \supset {\rm SO}(3)
 \eeq 
that are identical to those of the spherical harmonic-oscillator shell model.

\subsection{\boldmath The Sp$(3,\Rb)$ Lie algebra of U(3) tensor operators}
\label{sect:SU3<Sp3}
As a prelude to constructing symplectic-model spaces in
$ {\rm Sp}(3,\Rb) \supset {\rm U(3)}
 \supset {\rm SU}(3) \supset {\rm SO}(3)$ coupled basis states, 
 it is useful to express the Sp$(3,\Rb)$ Lie algebra  in terms of U(3) tensor operators obtained by expressing the position and momentum coordinates 
of a nucleon in terms of harmonic-oscillator raising and lowering (boson) operators
\beq \hat x_{ni} = \frac{1}{\sqrt{2}\,a} (c^\dagger_{ni}+ c_{ni} ) , \quad
\hat p_{ni} = {\rm i} \hbar\frac{a}{\sqrt{2}} (c^\dagger_{ni} - c_{ni} ) ,
\label{eq:xpccdag} \eeq
where $a  = \sqrt{M\omega/\hbar}$ is the harmonic-oscillator unit of inverse length.
The elements of the Sp$(3,\Rb)$ algebra then have expansions
\begin{subequations}
\bal 
&  \hat Q_{ij} =
 \frac{1}{2a^2}\big(2\hat{\mathcal{Q}}_{ij}  + \hat {\mathcal{A}}_{ij} + \hat {\mathcal{B}}_{ij} \big), \\
& \hat K_{ij} = \tfrac12 a^2 \hbar^2 (2 \hat{\mathcal{Q}}_{ij} 
- \hat {\mathcal{A}}_{ij} - \hat {\mathcal{B}}_{ij} ) , \\
& \hat P_{ij} = {\rm i}\hbar (\hat {\mathcal{A}}_{ij} - \hat {\mathcal{B}}_{ij}) , \quad
\hat L_{ij}  = -{\rm i} ( \hat {\mathcal{C}}_{ij} - \hat {\mathcal{C}}_{ji} ) , 
\end{align}
\end{subequations}
with
\begin{subequations}  \label{eq:8.ABCQops}
\bal 
& \hat {\mathcal{A}}_{ij} = \hat {\mathcal{A}}_{ji} = \sum_n c^\dagger_{ni} c^\dagger_{nj} ,\quad
  \hat {\mathcal{B}}_{ij} = \hat{\mathcal{B}}_{ji} =\sum_n c_{ni} c_{nj} , \\
&\hat {\mathcal{C}}_{ij} =  \sum_n \big( c^\dagger_{ni} c_{nj} 
+ \textstyle\frac12 \delta_{i,j}\big), \quad
\hat{\mathcal{Q}}_{ij} = \textstyle\frac12\big( \hat {\mathcal{C}}_{ij} + 
\hat {\mathcal{C}}_{ji} \big) .
\end{align}
\end{subequations}
The U(3) $\subset$ Sp$(3,\Rb)$ subalgebra is then spanned by 
the angular-momentum operators $\{ \hat L_{ij}\}$  and the U(3) components 
$\{ \hat{\mathcal{Q}}_{ij}\}$ of the quadrupole operators $\{ \hat Q_{ij}\}$.
These U(3) operators are  also more usefully expressed in terms of SO(3) spherical tensors, which include  $L=1$ angular momentum operators
\beq \hat L_0 = \hat L_{23}  ,\quad 
\hat L_{\pm 1} = \mp  \frac{1}{\sqrt{2}}
(\hat L_{31} \pm {\rm i} \hat L_{12})  , 
\label{eq:6.L_k}
\eeq
an $L=0$ monopole operator
\bal 
&\hat{\mc{Q}}_0 =\hat{\mc{Q}}_{11} + \hat{\mc{Q}}_{22} +\hat{\mc{Q}}_{33},
\label{eq:Q0moments} 
\end{align}
and  $L=2$ quadrupole operators
\begin{subequations} \label{eq:Q2moments}
\bal
& \hat{\mc{Q}}_{2,0}  = 
    2  \hat{\mc{Q}}_{11} -  \hat{\mc{Q}}_{22}-  \hat{\mc{Q}}_{33}, \\
&  \hat{\mc{Q}}_{2,\pm 1} = 
  \mp \sqrt{6}\, ( \hat{\mc{Q}}_{12} \pm {\rm i}  \hat{\mc{Q}}_{13}), \\
& \hat{\mc{Q}}_{2,\pm 2} =  \textstyle\sqrt{\frac{3}{2}}\, 
 ( \hat{\mc{Q}}_{22}-  \hat{\mc{Q}}_{33} \pm 2{\rm i}  \hat{\mc{Q}}_{23}). 
\end{align}
\end{subequations}

In addition, the Sp$(3,\Rb)$ Lie algebra contains Hermitian linear combinations of the creation and annihilation operators,
$\hat {\mathcal{A}}_{ij}$ and $\hat {\mathcal{B}}_{ij}$, 
 of two harmonic-oscillator quanta.
They are, respectively,  the $2\hbar\omega$  raising and lowering operators of giant-resonance excitations.
Monopole and quadrupole giant-resonance raising and lowering operators are  expressed in spherical-tensor notation by
\begin{subequations}\bal
&\hat {\mathcal{A}}_{0} = \sqrt{\tfrac16}\sum_i \hat {\mathcal{A}}_{ii} , \\
& \hat {\mathcal{A}}_{2,0} 
= \sqrt{\tfrac{1}{12}}\, (2\hat {\mathcal{A}}_{11}-\hat {\mathcal{A}}_{22}-\hat {\mathcal{A}}_{33}),\\
&\hat {\mathcal{A}}_{2,\pm 1}
 =\mp \sqrt{\tfrac{1}{2}}\, (\hat{\mathcal{A}}_{12} \pm {\rm i}\hat {\mathcal{A}}_{13}), \\
&  \hat {\mathcal{A}}_{2,\pm 2} 
      = \sqrt{\tfrac{1}{8}}\, (\hat {\mathcal{A}}_{22}- \hat {\mathcal{A}}_{33} 
          \pm 2{\rm i}\hat {\mathcal{A}}_{23}) ,    
\end{align}
\end{subequations}
and
\begin{subequations}\bal
&\hat {\mathcal{B}}_{0} = \sqrt{\tfrac16}\sum_i \hat {\mathcal{B}}_{ii}  \\    
& \hat {\mathcal{B}}_{2,0} 
= \sqrt{\tfrac{1}{12}}\, (2\hat {\mathcal{B}}_{11}-\hat {\mathcal{B}}_{22}-\hat {\mathcal{B}}_{33}), \\
& \hat {\mathcal{B}}_{2,\pm 1}
 =\mp \sqrt{\tfrac{1}{2}}\, (\hat{\mathcal{B}}_{12} \mp {\rm i}\hat {\mathcal{B}}_{13}), \\
& \hat {\mathcal{B}}_{2,\pm 2} 
 = \sqrt{\tfrac{1}{8}}\, (\hat {\mathcal{B}}_{22}- \hat {\mathcal{B}}_{33} 
          \mp 2{\rm i}\hat {\mathcal{B}}_{23}) . 
\end{align}
\end{subequations}
(The normalisation of these tensors is chosen for convenience and historical reasons.)

The above expressions show that the Sp$(3,\Rb)$ Lie algebra is the union of two 
subsets of operators: those of a U(3) subalgebra, and  giant resonance raising and lowering operators.
The SU(3) subalgebra is  spanned by the subset of operators 
$\{ \hat L_{k}, \hat{\mathcal{Q}}_{2,\nu}\}$ and the total  monopole and quadrupole operators are the combinations given in harmonic-oscillator  ($a=1$) units by
\begin{subequations}\label{eq:SpQops}
\bal 
&\hat Q_0 = 
\hat{\mc{Q}}_{0} + \sqrt{3} \, (\hat{\mc{A}}_0 + \hat{\mc{B}}_0 ) ,\\
&\hat Q_{2} = 
\hat{\mc{Q}}_2+ \sqrt{3} \, (\hat{\mc{A}}_2 + \hat{\mc{B}}_2) . 
\end{align}
\end{subequations}
It is also seen that, whereas the operators 
$\hat{\mc{Q}}_{0}$, $\hat L_k$ and $\hat{\mc{Q}}_{2\nu}$ are infinitesimal generators of a U(3) Lie algebra, the giant-resonance raising operators $\hat {\mathcal{A}}_{0}$ and  $\hat {\mathcal{A}}_{2\nu}$ 
are the $L=0$ and 2 components of a U(3) $\{ 2,0,0\}$ tensor, and the lowering operators 
$\hat {\mathcal{B}}_{0}$ and  $\hat {\mathcal{B}}_{2\nu}$
 are their Hermitian adjoints.

\subsection{\boldmath The properties of an Sp$(3,\Rb)$ representation}
\label{sect:4.2}
The many-nucleon representations of the symplectic model are simply defined.
They comprise sets of states based on those of a so-called lowest-grade U(3) irrep, which are states that are annihilated by the giant-resonance lowering operators, and form an infinite tower of
$2\hbar\omega$, $4\hbar\omega$, $6\hbar\omega$, $\cdots$
giant-resonance U(3) representations created by the repeated action
 of the U(3)-coupled giant-resonance excitation operators on the 
 lowest-grade U(3) states.

These many states have properties that relate naturally to those of the Bohr-Mottelson unified model in which the SU(3) states correspond to the intrinsic nucleon states of the unified model whereas the giant-resonance states correspond to those of the irrotational-flow Bohr model.

The  relationship between the symplectic model and the
Bohr-Mottelson unified  model  will be developed explicitly in the following sections.
It is obtained by considering the asymptotic (macroscopic) limits of symplectic-model representations that are rapidly approached in nuclei with relatively large numbers of nucleons.
In these limits, the SU(3) states are shown to approach those of a rigid rotor
and the giant-resonance excitations approach those of a harmonic vibrator.
The strong coupling between them then results in a coupled rotor-vibrator model with  many-nucleon wave functions.

In spite of this close correspondence with the standard unified model,
 there are notable differences.
In particular, the low-energy states are more like those of a triaxial rigid rotor with a truncated sequence of $K$ bands, than those of an axially-symmetric rotor with vibrational beta and gamma bands.
This is because the vibrational excitations of the microscopic symplectic model are basically those of the giant-resonance excitations which are expected to occur at higher energies.

A primary objective of this paper will be to show how the  symplectic model simplifies in its macroscopic limits so that it becomes  easy to 
apply to the analysis of experimental data and thereby identify the microscopic origins of what is observed.
Such simplifications will be particularly essential if one is to have hopes of understanding the transitions between different collective bands brought about by dynamical symmetry-breaking interactions.

Early applications of the symplectic model and its submodels 
were reviewed in 1985 \cite{Rowe85}.
These applications were with algebraic and schematic interactions, 
and, with a few exceptions, involved only a single symplectic-model irrep.
More recent calculations, with algebraic and schematic interactions, 
include those of Bahri \cite{RoweB00}, Dreyfuss \emph{et al.} 
\cite{DreyfussLDDB13}, and Tobin  \emph{et al.}  
\cite{TobinFLDDDB14}.
Symplectic model calculations with  general shell-model nucleon-nucleon interactions were initiated by Filippov and colleagues
\cite{FilippovO80,OkhrimenkoS81}
and developed by Vassanji \emph{et al.}\
\cite{VassanjiR82,VassanjiR84,CarvalhoVR87,CarvalhoRKB02}.

\section{The symplectic shell model}
The many-nucleon Hilbert space is infinite dimensional and in any calculation it is necessary to truncate it to a finite-dimensional subspace.
In the standard spherical shell model, the many-nucleon Hilbert space is expressed as a sum of subspaces ordered by increasing 
independent-particle-model energies as given, for example, by the Mayer-Jensen model \cite{MayerJ55}.
This makes it possible to truncate the Hilbert space to a shell-model subspace of states with lowest independent-particle-model energies.
However, for strongly-deformed nuclei, the Mayer-Jensen shell-model ordering is wildly inappropriate.
Already in light nuclei, one typically needs an effective charge of  approximately twice the mean nucleon charge to relate the quadrupole moments and E2 transition rates obtained in a Mayer-Jensen shell-model space
 to those observed.
For some of the more deformed states, such as those of the rotational band built on the first excited state of $^{16}$O 
\cite{BassichisR65,BrownG66,Suzuki76b,RoweTW06}
and on the so-called Hoyle state of $^{12}$C
\cite{DreyfussLDDB13}, one needs to consider shell-model states of 
 spherical harmonic-oscillator excitation energy $\sim\! 4\hbar\omega$,
 relative to the lowest spherical harmonic-oscillator  energy of the nucleus, as well as a large effective charge to describe their E2 properties \cite{HeydeW11}.

The inability of the spherical shell model to describe states of large deformation is explained in  the Nilsson extension of the Mayer-Jensen  model.
It is observed, for example  \cite{JarrioWR91} that
the Nilsson-model with a spheroidal potential 
determined by the observed deformation of a nucleus such as $^{168}$Er is able to give a lowest-energy state for this nucleus
with the observed deformation.
However,  because of  the crossing of spherical shell-model levels as they evolve into  Nilsson levels, such a Nilsson-model state has of the order of 12 harmonic-oscillator quanta more than
that  of the lowest-energy Mayer-Jensen states.

An important observation is that, in addition to being an algebraic collective model, the symplectic model also defines a coupling scheme for the complete many-nucleon Hilbert space and its expression as a sum of collective model subspaces \cite{Rowe16,Rowe16b}.
Thus, it gives a  decomposition of the many-nucleon Hilbert space, which differs from that of the Mayer-Jensen shell model 
and is more appropriate for the description of collective rotational states.
Moreover, as for the standard  shell model, the collective model subspaces can be ordered for any given nucleus, 
e.g., by the energies of their lowest-weight states with respect to a suitable model Hamiltonian; cf.\ discussion of  mean-field methods in Section \ref{sect:MFmodels}.

  The decomposition of a many-nucleon space into subspaces that carry representation of useful groups by means of so-called coupling schemes, is standard practice in traditional shell-model calculations
\cite{FrenchHMW69}.
Typical coupling schemes 
 are based on the use Wigner's U(4) supermultiplet group \cite{HechtPang69}
 which contains the ${\rm SU(2)}_S \times {\rm SU(2)}_T$
 spin and isospin  groups as subgroups, 
 the compact symplectic group \cite{Flowers52,FlowersS64a} which is the symmetry group of a pairing model, and the SU(3) group which is the dynamical group of the Elliott model \cite{Elliott58ab}.

For deformed rotational nuclei it is appropriate to consider a decomposition of the many-nucleon Hilbert space by means of a coupling scheme  based on the direct product group 
 ${\rm Sp}(3,\Rb) \times {\rm U}(4)$.
For such a coupling one could contemplate many-nucleon calculations in spaces classified by states labelled by the quantum numbers of, for example,  the groups in the chain
\beq
 {\rm Sp}(3,\Rb) \times 
 {\rm SU}(2)_{S_n}\times {\rm SU(2)}_{S_p} 
 \supset  {\rm U}(3) \times  {\rm SU(2)}_J ,
\eeq 
 where  SU(2)$_{S_n}$ and SU(2)$_{S_p}$ are the neutron and proton intrinsic spin groups, and SU(2)$_J$ is the total angular-momentum group.
 For convenience, we shall refer to the subspaces for the irreps of this coupling scheme as `collective subspaces'.
 
The symplectic shell model is then simply an expression of the Hilbert spaces of many-nucleon quantum mechanics as sums of collective subspaces of states that carry irreducible representations of the symplectic model together with other spin and isospin representations as required.

 For a calculation with a meaningful Hamiltonian in a space of many-nucleon states restricted to a collective subspace,
 one would  expect to see  a low-energy band of states emerging with properties close to those of a rotor model.
 However, as noted in Section \ref{sect:4.2}, one would not expect to see low-energy excited rotational bands such as beta or gamma vibrational bands.
 
Observe that,  because the full  quadrupole operator of a nucleus is an element of the ${\rm Sp}(3,\Rb)$ Lie algebra, there can be no isoscalar E2 transitions between the algebra's different representation spaces.
Thus, the observation of E2 transition between 
states of different rotational bands can
give information on the mixing of collective subspaces arising from symplectic model symmetry-breaking interactions.
Such mixing is expected to occur
and, indeed, a symmetry breaking interaction could very well result in a coherent mixing of collective spaces in the manner of a quasi-dynamical symmetry \cite{RoweCancun04}
and lead to unexpected results.
For example, it could be that strong pairing interactions could result in a mixing of lowest-grade SU(3) representations with the result that bands of states emerge with the properties of an axially symmetric rotor with vibrational excitations.
One can learn as much, perhaps more, from the limitations of a model as from its successes.

In concluding this section, it is emphasised that the primary focus of this paper is  on the interpretation of nuclear data in terms consistent with many-nucleon theory.
Thus, we make little mention of the very significant developments 
of the so-called $M$-scheme methods \cite{Whitehead72,WhiteheadWCM77}, 
that enable shell-model calculation in huge spaces.
The huge no-core shell-model calculations in  $M$-scheme basis have contributed enormously to establishing the foundations of nuclear structure theory in terms of many-nucleon quantum mechanics
which, in this paper, are taken as understood.
Also, because our concern is primarily with the interpretation of collective states in heavy nuclei, which are currently out of  the reach of detailed shell-model calculations, we make little mention of the technology associated with the implementation of shell-model calculations with general nucleon interactions in a symplectic-model basis.
We mention only that several authors, notably Suzuki and Hecht \cite{SuzukiH86,SuzukiH87}, 
and Escher and Draayer \cite{EscherD98},  contributed to its development.
Particularly important has been the facility to undertake multi-shell model calculations in a U(3)-coupled basis;
for this the most significant developments were no doubt:
the computer codes to determine U(3) coupling and recoupling coefficients of Akiyama, Draayer and Millener
\cite{AkiyamaD73,DraayerA73,Millener78}; 
the SU(3)-reduced matrix packages of Braunschweig, Bahri and Draayer 
\cite{Braunschweig78,Braunschweig78b, BahriD94};
and ultimately the impressively large calculations with realistic interactions in large multi-shell model spaces in a 
${\rm U(3)}\times {\rm SU}(2)_{S_n}\times {\rm SU(2)}_{S_p} 
 \supset  {\rm SU}(3) \times  {\rm SU(2)}_J$ coupled basis
 by Dytrych \emph{et al.}\
\cite{DytrychSBDV07a,DytrychSBDV07b,%
DytrychSBDV08,DytrychSDBV08b}.

\section{The representation of a collective model}
The collective models  have rich algebraic  structures
of which little use was made in the original formulations of the Bohr-Mottelson models. 
Indeed, they are often described as geometric models to distinguish them from purely algebraic models.
This is not inappropriate because their physical content is most readily understood in geometrical terms.
However, the theories of geometry, Lie groups and Lie algebras are very much intertwined.
What is  impressive is that, without explicit use of  their algebraic structures, the model representations anticipated  sophisticated developments in the theory of Lie group and Lie algebra representations which  subsequently became important in identifying their microscopic generalisations.

The most powerful procedure for constructing 
representations of a Lie group (or Lie algebra) is  by
extending a representation of a subgroup (subalgebra) by
the mathematical method of  induced representations
\cite{Mackey68}.
This procedure has a counterpart  in physics in terms of coherent state \cite{Perelomov85}
 and more general vector coherent state  (VCS) representations
 \cite{RoweR91}.
 A remarkable relationship between an induced representation and a unified collective model is that the irreps of the dynamical group for the model are induced from a smaller  group which defines the intrinsic states of the model's representations.

The standard theory of coherent-state representations was 
developed by many  
\cite{Klauder63,Sudarshan63,Perelomov72,Gilmore72,Onofri75} and is reviewed in refs.\,\cite{KlauderB-S85,Perelomov85}.
Its VCS extension was introduced \cite{RoweRC84,Rowe84}%
\footnote{Partial coherent-state representations were also defined by  Deenen and Quesne \cite{DeenenQ84}.}
specifically for the purpose of constructing irreps of the symplectic model but was quickly found to have a wide range of applications; cf.\
 \cite{Rowe12} and references therein.
VCS theory is intuitively natural from the perspective of collective models in which richer models are obtained by adding intrinsic degrees of freedom to a simpler model.
Thus, it is rewarding to discover its remarkable versatility in constructing representation of Lie algebras and Lie groups in general and
 that the collective models serve as prototypes for such applications.

A common property of collective models is that they are defined by  sets of intrinsic states and groups of transformations of these states.
Two types of representation are prevalent in collective models:
the first relates to boson-expansion methods appropriate for models with vibrational degrees of freedom; the second relates to models with rotational degrees of freedom.
The prototype  of a vibrational coherent-state representation  is  given by the Bargmann representation of a harmonic-oscillator
\cite{Bargmann61}.
The prototype of a rotational coherent-state representation is given by the rotor model.

A  valuable property of  coherent state and VCS methods is that they give expressions of  microscopic collective models in terms of the original collective models that they replace.
As a result, traditional practices for interpreting nuclear properties in terms of collective models continue to apply with some adjustment.

\subsection{Generalised Bargmann representations}
Recall that the states of a 1-dimensional harmonic oscillator span a unitary representation of a simple boson algebra with raising and lowering operators, $\hat c^\dag$, $\hat c$, that satisfy the commutation relation
\beq [\hat c,\hat c^\dag] = 1 ,\eeq
and that orthonormal basis states for the harmonic oscillator are  given by
\beq |n\rangle = \sqrt{\frac{1}{n!}} \, (\hat c^\dag)^n |0\rangle ,
\quad n = 0, 1, 2, \dots,  \label{eq:onHObasis}\eeq
where $|0\rangle$ is the boson vacuum state.
Thus, an arbitrary state of the  harmonic oscillator is expressed in the form
\beq |\psi\rangle = f(\hat c^\dag) |0\rangle ,\eeq
where $f(\hat c^\dag)$ is a function of the boson creation operator.
In the Bargmann representation, a state $|\psi\rangle$  is represented by a wave function, that is a function of a complex variable $z$, defined by the overlaps
\beq |\psi\rangle \to \Psi(z) = \langle 0| e^{z\hat c} |\psi\rangle.\eeq
The orthonormal basis states of equation (\ref{eq:onHObasis}) then have the wave functions
\beq |n\rangle \to \Psi_n(z) =  \sqrt{\frac{1}{n!}} \, z^n, \quad 
\quad n = 0, 1, 2, \dots . \eeq
The operation of a boson operator $\hat X$, 
i.e., $\hat c^\dag$ or $\hat c$,  on these wave functions is defined by the equation
\bal
 \hat\Gamma ( X) 
&\Psi_n(z) =  \langle 0| e^{z\hat c}\hat  X|n\rangle \nonumber\\
&= \langle 0|\big( \hat X + z [\hat c,\hat X] 
+ \tfrac12 z^2 [\hat c, [\hat c,\hat Z]]\big) e^{z\hat c}|n\rangle.
\label{eq:CSexp}
\end{align}
Thus, it is determined that
\beq \hat\Gamma ( c^\dag) = z, \quad 
\hat\Gamma ( c) = d/dz .  \eeq

The same methods can be used to derive exact boson expansions of  other Lie algebras with  raising and lowering operators 
\cite{Dobaczewski81,Dobaczewski81b,Dobaczewski82}.
The simplest example, is for the SU(2) Lie algebra, which has elements with commutation relations
\beq [S_0,S_\pm] = \pm S_\pm , \quad [\hat S_+,\hat S_+] = 2 S_0.
\eeq
If $|0\rangle$ is a lowest-weight state for an SU(2) irrep of angular momentum $j$, it is an eigenstate of $\hat S_0$ with eigenvalue $-j$. 
A state $|\psi\rangle$ of the irrep then has a coherent-state wave function defined as a function of a variable $z$ by
\beq \Psi(z) = \langle 0| e^{z\hat S_- }|\psi\rangle. \eeq
Thus the elements of the SU(2) Lie algebra  have coherent-state representations given, in accord with equation  (\ref{eq:CSexp}),
by
\bal
&\hat \Gamma(S_-) \Psi(z) =
\langle 0| \hat S_-  e^{z \hat S_-} |\psi\rangle = \frac{d}{dz}\Psi(z) \\
&\hat \Gamma(S_0) \Psi(z) =
\langle 0| \big(\hat S_- +z\hat S_0\big)  e^{z \hat S_- }|\psi\rangle 
\nonumber\\
&\phantom{\hat \Gamma(S_0) \Psi(z)}
  = \left(z\frac{d}{dz} - j\right)\Psi(z) \\
&\hat \Gamma(S_+) \Psi(z) =
\langle 0| \big(\hat S_+ -2 z\hat S_0  -z^2 \hat S_-   \big)  
e^{z \hat S_- }|\psi\rangle \nonumber\\
& \phantom{\hat \Gamma(S_+) \Psi(z)}
  = \left(2jz-z^2 \frac{d}{dz}\right) \Psi(z) .
\end{align}
This is the famous Dyson representation \cite{Dyson56}.
To make use of it, one needs an orthonormal set of basis wave functions relative to which the matrix elements of the SU(2) Lie algebra operators can be calculated.
A simple K-matrix procedure has been developed for this purpose
\cite{Rowe84,Rowe95}.

An orthonormal basis is defined
to within  normalisation factors as monomial functions 
\beq \Psi_{jm} (z) = k_m z^{j+m},
\eeq
where $m= -j$ for the lowest-weight state and
\beq \hat\Gamma(S_0) \Psi_{jm}(z) =m \Psi_{jm}(z) , \eeq
as desired.
Application of the raising  operator in the Dyson representation then gives
\bal   \hat\Gamma(S_+) \Psi_{jm}(z) 
&= k_m (j-m) z^{j+m+1}\nonumber \\
&= \frac{k_m}{k_{m+1}} (j-m)   \Psi_{j,m+1}(z) .
\end{align}
Similarly, application of the lowering operator gives
\bal   \hat\Gamma(S_-) \Psi_{j,m+1}(z) 
&= k_{m+1} (j+m+1)  z^{j+m} \nonumber\\
&=   \frac{k_{m+1}}{k_m} (j+m+1) \Psi_{jm}(z) .
\end{align}
Thus, the SU(2) matrix elements  are given by
\bal & \langle j,m+1 |\hat J_+ |jm\rangle =  
\frac{k_m}{k_{m+1}} (j-m) , \\
& \langle jm |\hat J_- |j,m+1\rangle =  
\frac{k_{m+1}}{k_m} (j+m+1) .
\end{align}
It follows that, to satisfy the Hermiticity relationships  
\beq 
\langle j,m+1 |\hat J_+ |jm\rangle =\langle jm |\hat J_- |j,m+1\rangle^* 
\eeq
of a unitary representation, the norm factors have ratios
\beq 
\left| \frac{k_{m+1}}{k_m}\right|^2 = \frac{j-m}{j+m+1} \eeq
and we obtain the  matrix elements
\beq \langle j,m\pm1 |\hat J_\pm |jm\rangle
= \sqrt{(j\mp m)(j\pm m+1)} \, .
\eeq

The coherent-state construction of representations works for many Lie algebras.  
For example, it works  as readily for the infinite-dimensional representations of the  non-compact SU(1,1) Lie algebra as for SU(2).
One can equally well start from a highest-weight state of a representation as from a lowest-weight state.

\subsection{Vector-coherent-state representations}
A serious limitation of the above coherent-state construction is that it apples
only to Lie algebras for which the raising (likewise the lowering) operators  commute with one another.
For example, it does not apply to representations of the SU(3) Lie algebra 
which comprises two commuting U(1) operators, three non-commuting raising operators and  three non-commuting lowering operators.

This obstacle is circumvented in a VCS representation by using, instead of a single intrinsic state, a set of intrinsic states that carry a multi-dimensional representation of an intrinsic-symmetry group.
A simple example is the extension of the Bargmann representation
to a  particle with intrinsic spin states.
A parallel collective-model example, could be the extension of the Bohr model  by the addition of intrinsic states of non-zero vorticity.
Indeed, the Bohr-Mottelson unified model  is seen from this perspective as a Bohr model with intrinsic nucleon states.

In a VCS representation, the SU(3) Lie algebra is seen as comprising the four elements of a U(2) subalgebra,  two commuting raising operators $\{ A_1,A_2\}$ and two commuting lowering operators 
$\{ B_1,B_2\}$.
One can then define an orthonormal basis of highest-grade states 
$\{ |\alpha\rangle\}$ for the irrep which are  a subset of states of the irrep that are annihilated by the two commuting raising operators
and transform as basis vectors $\{ \xi_\alpha\}$
for an irreducible U(2) representation.
The basis vectors $\{ \xi_\alpha\}$  then serve as intrinsic states for an irreducible VCS SU(3)  representation in which a VCS wave function 
$\Psi$ of a state $|\psi\rangle$ in an  irreducible SU(3) representation is   
a vector-valued function of  $\{ z_1, z_2\}$ coordinates defined by
\beq \Psi(z) = \sum_\alpha \xi_\alpha 
\langle \alpha |e^{\sum_i z_i \hat A_i} |\psi\rangle .
\eeq
The VCS construction has been shown to give explicit analytical expressions for the matrix elements of the SU(3) algebra 
\cite{RoweR91} and indeed for the known representations of this 
\cite{Pursey63} and other
 SU$(n)$  Lie algebras in canonical U$(n-1)$ bases \cite{BairdB63}.
However, for applications to nuclear collective models the representations of SU(3) are required in an SO(3)-coupled basis
\cite{RoweLeBR89,RoweVC89} for which the VCS representation of the rotor model provides a prototype.

\subsection{The representation of an asymmetric rotor}
\label{sect:rotor_rep}
The  Hilbert space $\Hb^{\rm ASR}$ for a completely asymmetric rigid rotor, with no intrinsic degrees of freedom, is the space of all  normalisable functions of the orientation angles of a set of axes fixed in the body of the rotor relative to a space-fixed set.
The group of all rotations of a  set of axes  is the group SO(3).
Thus,  $\Hb^{\rm ASR}$ is spanned by the  set of all wave functions of the form
\beq 
\psi_{KLM}(\Omega) = \sqrt{\frac{2L+1}{8\pi^2}}\,
 {\scrD}^L_{KM} (\Omega ) , \quad \Omega\in {\rm SO(3)}, 
 \label{eq:ASRwfn}
\eeq
in which  ${\scrD}^L_{KM}$, with $L\geq 0$ taking integer values,
is a Wigner rotation-matrix function, and the wave functions are
 normalised with respect to the standard SO(3) volume element
 $d\Omega$ such that
\bal
 \langle KLM | K'L'M'\rangle &= 
\int \psi^*_{KLM}(\Omega) \psi_{K'L'M'}(\Omega) \, d\Omega \nonumber\\
&= \delta_{K,K'}  \delta_{L,L'}  \delta_{M,M'} . \label{eq:ASRwfn2}
\end{align}

From a coherent-state perspective, the rotor is described by an intrinsic state $|\phi\rangle$ defined in the body-fixed axes of the rotor and the wave function $\psi_{KLM}$ is expressed as the overlap function
\beq \psi_{KLM} (\Omega) 
= \langle \phi | \hat R(\Omega) |KLM\rangle, \quad
\Omega \in {\rm SO(3)}.
\eeq
This expression is consistent with equation  (\ref{eq:ASRwfn}) as seen from the expansion
\beq \psi_{KLM} (\Omega) 
= \sum_N \langle \phi | KLN\rangle {\scrD}^L_{NM}(\Omega) 
\eeq
in which, if the intrinsic wave function $\phi$ is a delta function in $\Omega$,
\bal 
\langle \phi | KLN\rangle 
&=\int \phi^*(\Omega) \psi_{KLN}(\Omega) \, d\Omega \nonumber\\
&= \sqrt{\frac{2L+1}{8\pi^2}} \, \delta_{N,K} .
\end{align}

An advantage of the coherent state expression is that the action of an SO(3) tensor operator, such as a quadrupole moment operator  $\hat Q_{2\nu}$,
is immediately defined on a rotor-model wave function 
by the observation that, when the intrinsic axes of the rotor are the principal axes of the quadrupole tensor,
\bal \hat \Gamma(Q_{2\nu})\psi_{KLM}(\Omega) 
&=\langle \phi |\hat R(\Omega) \hat Q_{2\nu}|{KLM}\rangle 
\nonumber\\
&= \sum_\mu 
\langle \phi | \hat Q_{2\mu} \hat R(\Omega)|{KLM}\rangle 
{\scrD}^2_{\mu\nu} (\Omega ) \nonumber\\
&= \sum_\mu  \bar Q_\mu {\scrD}^2_{\mu\nu} (\Omega )
\psi_{KLM}(\Omega) .
\end{align}
where $\{ \bar Q_\mu\}$ are the quadrupole moments of the rotor in the principal axes frame and are such that
\beq  \langle \phi | \hat Q_{2\mu} |\psi\rangle =
[\delta_{\mu,0} \bar Q_0 + (\delta_{\mu,2} + \delta_{\mu,-2}) \bar Q_2]
 \langle \phi | \psi\rangle. \label{eq: intrQ}
\eeq
Thus, the coherent-state method provides a systematic  procedure for deriving the  standard rotor-model results
\cite{Rowebook70,BohrM75}.
This is particularly useful for rotors with intrinsic symmetries.

\subsection{Rotors with intrinsic symmetry}
In molecular physics,  molecules with relatively well-defined shapes are described  as symmetric tops, when two of their principal moments of inertia are equal, and as asymmetric tops, when all three are different.
In both cases, rotations can occur about all three principal axes.
However, rotations of the molecule that leave its principal axes invariant
are regarded as intrinsic motions and the set of such rotations form a group known as the intrinsic-symmetry group of the molecule.
Thus, the  intrinsic symmetry group of a symmetric top is the group 
D$_\infty$ which  comprises SO(2) $\subset$ SO(3)  
rotations about the symmetry axis and rotations through 
angle $\pi$ about  a perpendicular axis.
Likewise, an asymmetric top has an intrinsic symmetry
group D$_2$ which comprises rotations through multiples of $\pi$ 
about all of its principal axes.

To illustrate the significance of the intrinsic symmetry group,
 we consider the representations of Ui's rigid rotor model. 
The Lie algebra of Ui's model, discussed in Section \ref{sect:Ui_rotor}, contains 5 commuting quadrupole-moment and 3 angular-momentum operators.
An irreducible representation of this algebra is  induced by VCS methods starting from a representation  of the subalgebra containing the five quadrupole moments and the single angular-momentum component $L_0$,  
A basis for such a representation, $\hat\sigma$, is given by an infinite set of intrinsic states $\{ \xi_K\}$, with $K$ taking all integer values and for which
\begin{subequations}
\bal  
&\hat\sigma(L_0) \xi_{K} = K \xi_K , \quad
\hat\sigma(Q_{2\, 0}) \xi_{K} = \bar Q_0 \xi_K ,  \\
&\hat\sigma(Q_{2\, \pm 2}) \xi_{K} = \bar Q_2 \xi_{K\pm 2} .
\end{align}
\end{subequations}
However, to uniquely define an irreducible representation of the rotor model, it is also  necessary that the intrinsic states transform in a well-defined way under all elements of the intrinsic symmetry group D$_2$.
Transformation of the $\{ \xi_K\}$ basis under rotations through angle $\pi$ about the  axis of quantisation is already given by
\beq e^{{\rm i}\pi \sigma(L_0)} \xi_K = e^{{\rm i}\pi K} \xi_K .\eeq
Because there are two irreducible representations of
the discrete subgroup of SO(3) generated by rotations through an angle $\pi$ about an axis perpendicular to the axis of quantization,
we also specify that the intrinsic states  transform 
under  such a rotation $e^{{\rm i}\pi \hat\sigma(L_\perp)}$  
according to the equation
\beq  e^{{\rm i}\pi \hat\sigma(L_\perp)} \xi_K = \xi_{\bar K} , \quad
e^{{\rm i}\pi \hat\sigma( L_\perp)} \xi_{\bar K} = \xi_K,
\eeq
where the two irreducible representations are  characterised by the two
  possible sign relations
$\xi_{\bar K} = \pm \xi_{-K}$.
(Note that  for an odd-mass nucleus, $K$  takes half-odd integer values and 
changes sign under a $2\pi$ rotation about any axis;
thus $\xi_{\bar K} = \pm {\rm i}\xi_{-K}$, where ${\rm i} = \sqrt{-1}$.)

These intrinsic states now serve as basis vectors for a set of VCS wave functions 
 \bal 
 \psi_{\alpha LM} &= 
 \sum_K \xi_K \langle K | \hat R(\Omega)|{\alpha LM}\rangle 
 \nonumber\\
& =  \sum_{KN} \xi_K \langle K | \alpha L N\rangle
  {\scrD}^L_{NM} (\Omega ) , \quad \Omega\in {\rm SO(3)} ,
 \end{align}
for the rigid-rotor algebra, where the states $\{ |K\rangle\}$ are a subset of states 
of the rotor that transform in the same way under D$_2$ rotations as the
intrinsic states $\{ \xi_K\}$, i.e.,
\begin{subequations}
\bal 
&\hat L_0 |K\rangle = K |K\rangle , \quad
\hat Q_{2\, 0} |K\rangle = \bar Q_0 |K\rangle ,  \\
&\hat Q_{2\, \pm 2} |K\rangle = \bar Q_2 |K\pm 2\rangle .
\end{align}
\end{subequations}
It remains only to ensure that the consistency equations
\bal   
\sum_K  \hat\sigma(X)\xi_K 
&\langle K | \hat R(\Omega)|{\alpha LM}\rangle \nonumber\\
&=  \sum_K \xi_K \langle K |\hat X \hat R(\Omega)|{\alpha LM}\rangle
\end{align}
are satisfied for $X$ equal to any of the elements 
$\{ L_0, Q_{2\,0},Q_{2\,\pm 2}, e^{{\rm i} \pi  L_\perp}\}$ that define the intrinsic states.
With the standard normalisations given by equation  (\ref{eq:ASRwfn}),
we then obtain the  expression for the wave functions of an asymmetric top
\bal 
 \Psi_{KLM}
& (\Omega ) = \sqrt{\frac{2L+1}{16\pi^2 (1+ \delta_{K,0})}}  \label{eq:5.69}   \\
&  \times    \left\{ \xi_K {\scrD}^L_{KM}(\Omega )
    +(-1)^{L+K}  \xi_{ \bar K} {\scrD}^L_{-KM}(\Omega )\right\}  ,
\nonumber
\end{align}
in the familiar rotor-model form.

These expression are readily extended to nuclear rotational models with strongly-coupled spin degrees of freedom.
However, for present purposes, they will be used to provide a systematic procedure for the construction of SU(3) representations in an SO(3)-coupled basis and for their physical interpretation in rotor-model terms.

\section{\boldmath Representations of  SU(3)  in an SO(3) angular-momentum basis}
A fortuitous result that underlies the success of Elliott's model \cite{Elliott58ab}
 is that when the nuclear quadrupole moments of Ui's rotor model
 \cite{Ui70}
$\{ \hat Q_{2\nu}\}$ are restricted to the space of many-nucleon states of a single spherical harmonic-oscillator energy, 
they become elements  
$\{ \hat{\mathcal Q}_{2\nu}\}$ of an SU(3) Lie algebra.
This was shown in sect.\,\ref{sect:SU3<Sp3}.
It  follows  that the states of an irreducible SU(3) representation 
are the projected images of rigid-rotor model states.
The SU(3) representations in multi-shell spaces also acquire an enhanced significance as the building blocks of the microscopic collective theory.
 
Because of its use as a shell-model coupling scheme,
 the representation theory of SU(3) and associated technology for its use have been well developed 
(see the review article of Harvey \cite{Harvey68} for early references).
Most important was the development of programs to calculate
 SU(3) Clebsch-Gordan, Racah, and $9$-$\lambda\mu$  coefficients \cite{Hecht65,Akiyama66,Sebe68,AkiyamaD73,%
DraayerA73,Millener78};
which are freely available in both SU(2) and SO(3) bases \cite{AkiyamaD73}.

As defined in sect.\,\ref{sect:SU3<Sp3},  the SU(3) algebra 
is a subalgebra of the  U(3) algebra of operators
$\{ \hat{\mathcal C}_{ij}, i,j = 1,2,3\}$ that satisfy the commutation relations
\beq [\hat{\mc{C}}_{ij}, \hat{\mc{C}}_{kl}] 
= \delta_{j,k} \hat{\mc{C}}_{il} - \delta_{i,l} \hat{\mc{C}}_{jk}.
\label{eq:Cijcoms}
\eeq
It contains angular-momentum and quadrupole operators expressed in a spherical tensor basis by
\begin{subequations}\label{eq:6.L_k2}
\bal 
&\hat L_0 = \hat L_{23}  ,\quad 
\hat L_{\pm1} =\mp\tfrac{1}{\sqrt{2}}( \hat  L_{31} \pm {\rm i}  L_{12}),  \\
&\hat{\mc{Q}}_{2,0}  = 
    2  \hat{\mc{Q}}_{11} -  \hat{\mc{Q}}_{22}-  \hat{\mc{Q}}_{33}, \\
&   \hat{\mc{Q}}_{2,\pm 1} = 
  \mp \sqrt{6}\, ( \hat{\mc{Q}}_{12} \pm {\rm i}  \hat{\mc{Q}}_{13}), \\
&\hat{\mc{Q}}_{2,\pm 2} =  \textstyle\sqrt{\frac{3}{2}}\, 
 ( \hat{\mc{Q}}_{22}-  \hat{\mc{Q}}_{33} \pm 2{\rm i}  \hat{\mc{Q}}_{23}),
\end{align}
\end{subequations}
where 
\beq
\hat L_{ij}  =  -{\rm i} ( \hat {\mathcal{C}}_{ij} - \hat {\mathcal{C}}_{ji} ) ,
\quad \hat{\mathcal{Q}}_{ij} = 
\textstyle\frac12\big( \hat {\mathcal{C}}_{ij} 
+ \hat {\mathcal{C}}_{ji} \big) . \label{eq:LijQij}
\eeq
The SU(3) algebra is finite and semi-simple, and its irreducible representations are easily constructed.
They were determined  by numerical methods
\cite{Elliott58ab,ElliottH63} 
within the framework of the shell model for light nuclei, 
and subsequently  derived by the algebraic methods of VCS theory
\cite{RoweLeBR89}.

This section reviews a version \cite{Rowe12} 
of the VCS construction in basis states that relate  directly to those of Ui's rigid-rotor model.
In this construction, the SU(3) quadrupole moments 
$\{ \mc{Q}_\nu\}$ are represented as operators
\bal \hat\Gamma({\cal Q}_{2\nu}) 
& = (2\lambda+\mu+3) \hat\scrD^2_{0\nu} -
\tfrac12\,\big[ \hat{\bf L}\cdot \hat{\bf L}, \hat\scrD^2_{0\nu}\big]
\nonumber \\
&\quad + \sqrt{6}\big( \hat \sigma_+\hat\scrD^2_{2\nu} +
 \hat \sigma_-\hat\scrD^2_{-2,\nu}\big) , \label{eq:GammaQ}
\end{align}
on linear combinations of rotor-model wave functions of the form
 \bal \Phi^{(\lambda\mu)}_{KLM}&(\Omega) 
 =   \sqrt{ \frac{2L+1}{16\pi^2(1+\delta_{K,0})}}  \label{eq:6.phiKLM}  \\
& \times \big( \xi_{K}\scrD^L_{KM}(\Omega) +
(-1)^{\lambda+L+K} \xi_{-K} \scrD^L_{-K,M}(\Omega)\big) ,
\nonumber   
\end{align}
in which  $\hat\scrD^2_{\mu\nu}$ is an operator that acts
multiplicatively, i.e.,
\beq \hat\scrD^2_{\mu\nu} \Psi(\Omega)
= \scrD^2_{\mu\nu}(\Omega)\,\Psi(\Omega) ,
\eeq
and $\hat \sigma_{0}, \hat \sigma_{\pm}$ are operators that act on the intrinsic-spin states according to 
\begin{subequations}\label{eq:S+-ops.1}
\bal 
\hat \sigma_0 \xi_K &= \t\frac12 K \,\xi_K ,\\
\hat \sigma_\pm \xi_K &= \t\frac12 \sqrt{(\mu \mp K)(\mu\pm K+2)}\,\xi_{K\pm 2} .
\end{align}
\end{subequations}

A  significant property of equation (\ref{eq:GammaQ}) is that, 
if it were not for the term
$\tfrac12\,\big[ \hat{\bf L}\cdot \hat{\bf L}, \hat\scrD^2_{0\nu}\big]$ 
which is of negligible importance when 
$\hat\Gamma(\mc{Q}_{2,\nu})$ acts on states of angular momentum 
$L \ll 2\lambda+\mu+3$, 
the properties of the SU(3) states including their wave functions and matrix elements would be those of a rigid-rotor, albeit 
with $K$ limited to the range $K= \mu, \mu-2, \dots, 1 \text{\;or\;} 0$ of an SU(2) representation as defined by equation (\ref{eq:S+-ops.1}).
However, as the angular momentum increases and becomes non-negligible in comparison with $2\lambda+\mu+3$, the 
$\tfrac12\,\big[ \hat{\bf L}\cdot \hat{\bf L}, \hat\scrD^2_{0\nu}\big]$ term becomes of increasing importance and results in the termination of the rotation-like bands to those of a finite-dimensional SU(3) representation.
Thus, although the  $\hat\Gamma({\cal Q}_{2\nu})$ operators have well-defined actions on the infinite-dimensional space of the  rigid-rotor wave functions of equation  (\ref{eq:6.phiKLM}), 
they generate the  finite-dimensional subspace of an SU(3) irrep  when  applied repeatedly to any wave function that lies within the irreducible  SU(3) subspace of rotor-model wave functions.
This is illustrated in Subsection \ref{sect:mu=0or1} 
for $(\lambda\, 0)$ and $(\lambda\,1)$ representation for which the SO(3)-reduced matrix elements of the SU(3) quadrupole tensor have  simple analytical expressions.

For a generic representation with $\mu >1$, there are multiple states of a given angular momentum $L$ in the representation and
$K$ is no longer a precise quantum number.
However, one can construct an orthogonal basis within the space of states of a common  $L$ by first diagonalising the easily calculated matrix of SO(3)-reduced matrix elements of the SU(3) quadrupole operator restricted to this space.
It then only remains to renormalise these states as for multiplicity-free representations to obtain an orthonormal basis and the matrix elements of a unitary representation.

Note that a pair of so-called contragredient SU(3) representations 
$(\lambda\mu)$ and $(\mu\lambda)$ have complete sets of matrix elements that differ only in that their quadrupole matrix elements have opposite sign.
Thus, it is sufficient to determine the representations with 
$\lambda \geq \mu$ to obtain those for both $\lambda\geq\mu$ 
and $\lambda < \mu$.
 
 The following Subsections give derivations and details of the above results.

\subsection{\boldmath VCS wave functions for SU(3) in a rotor-model basis}
An irreducible representation of SU(3) is characterised by a highest-weight state $|\lambda\mu\rangle$, which is a state that is annihilated by the three raising operators 
$\hat{\mathcal C}_{12}, \hat{\mathcal C}_{13}$ 
and $\hat{\mathcal C}_{23}$,
and is an eigenstate of the operators
\beq \hat h_1 = \hat{\mc{C}}_{11} -\hat{\mc{C}}_{22}, \quad
\hat h_2 = \hat{\mc{C}}_{22} -\hat{\mc{C}}_{33};
\eeq
it satisfies the equations
\begin{subequations}
\bal
&\hat{\mathcal C}_{12}|\lambda\mu\rangle
= \hat{\mathcal C}_{13}|\lambda\mu\rangle
= \hat{\mathcal C}_{23}|\lambda\mu\rangle=0, \\
&\hat h_1|\lambda\mu\rangle = \lambda |\lambda\mu\rangle ,
\quad \hat h_2|\lambda\mu\rangle = \mu|\lambda\mu\rangle .
\end{align}
\end{subequations}

A first step in the construction of a VCS representation of SU(3) extends this highest-weight state $|\lambda\mu\rangle$ to a 
set of so-called highest-grade states.  
These are states that are annihilated by the commuting
$\hat{\mathcal C}_{12}$ and $\hat{\mathcal C}_{13}$  raising operators, but not by $\hat{\mathcal C}_{23}$, 
and are a basis for an irreducible representation of a 
U(2) $\subset$ SU(3) subalgebra.  
This U(2) sualgebra contains the U(1) element
\beq  \hat{\mc{Q}}_{2,0} = 2\hat h_1+ \hat h_2, \eeq
and elements $\hat L_0$ and $\hat{\mathcal{Q}}_{2,\pm 2}$ that commute with $\hat{\mc{Q}}_{2,0}$. 
The latter elements satisfy the SU(2) commutation relations
\beq [\hat L_0, \hat{\cal Q}_{2,\pm 2}] = \pm 2  \hat{\cal Q}_{2,\pm 2} ,\quad
[\hat{\cal Q}_{2,2},\hat{\cal Q}_{2,-2}] = 6 \hat L_0 ,
\label{eq:6.su2CRs}
\eeq
and can be identified with standard SU(2) spin operators
\beq \hat L_0 \equiv 2\hat\sigma_0 , \quad
\hat{\mc{Q}}_{2,\pm 2} \equiv \sqrt{6}\, \hat \sigma_\pm  , 
\label{eq:SU3intspin}
\eeq
with commutation relations
\beq [\hat \sigma_0, \hat \sigma_{\pm}] = \pm \hat \sigma_{\pm} , 
\quad [\hat \sigma_+, \hat \sigma_-] = 2\hat \sigma_0 . \eeq
Thus, the U(2) representation on the highest-grade states  follows directly from the well-known representations of the SU(2) spin algebra.

Let $\{ |(\lambda\mu)K\rangle\}$ denote an 
orthonormal  basis of highest-grade states for a U(3) representation defined by the equations
\bal 
&\hat{\mc{C}}_{1j}  |(\lambda\mu) K\rangle = 0, \quad
   \text{for}\; j=2,3 , \label{eq:U3Hgrade} \\
& \hat L_0 |(\lambda \mu) K\rangle = K |(\lambda\mu) K\rangle, 
\\
& \phantom{\hat{\mc{C}}_{1j}  |(\lambda\mu) K\rangle = 0,} \quad
   \text{for}\;  K = -\tfrac12 \mu,  -\tfrac12 \mu+1, \dots , \tfrac12 \mu ,
   \nonumber
\end{align}
and let $\{ \xi_K\}$ denote an equivalent set of  spin states.
The relationship (\ref{eq:SU3intspin}) then implies that these spin states  satisfy the equations
\begin{subequations}
\bal
& \hat \sigma_0 \xi_K = \t\frac12 K \,\xi_K ,\\
&\hat \sigma_\pm \xi_K 
= \t\frac12 \sqrt{(\mu \mp K)(\mu\pm K+2)}\,\xi_{K\pm 2} ,  \\
&\hat\sigma(\mc{Q}_{2,0}) \xi_K = (2\lambda+\mu) \xi_K ,        
\end{align}
\end{subequations}
and that the highest-grade states satisfy the corresponding equations
\begin{subequations}\label{eq:83}
\bal 
&\hat L_0 |(\lambda \mu) K\rangle = K |(\lambda\mu) K\rangle ,
\\
& \hat{\mc Q}_{2,\pm2} |(\lambda \mu) K\rangle \nonumber\\
& \quad\quad =  
\sqrt{\t\frac32(\mu \mp K)(\mu\pm K+2)}\,|(\lambda \mu) K\pm 2\rangle , 
 \\
&\hat{\mc Q}_{2,0} |(\lambda \mu) K\rangle 
= (2\lambda+\mu) |(\lambda \mu) K\rangle.  
\end{align} 
 \end{subequations}

Now recall that the states of an  irreducible SU(3) representation are linear combinations of the states obtained by rotating a suitable highest-weight state through all angles \cite{Elliott58ab}.
They are also obtained, with considerable redundancy, by rotation of the highest-grade states.
Thus, the highest-grade states serve as intrinsic states for  rotor-like VCS  representations of the U(3) model.
It follows that, an arbitrary  state  $|(\lambda\mu)\alpha LM\rangle$ of the desired SU(3) representation is  defined  by the  overlap functions
 \beq  \psi^{(\lambda\mu)}_{K\alpha LM}(\Omega) = 
 \langle (\lambda\mu)K |\hat R(\Omega ) 
 |(\lambda\mu)\alpha LM\rangle, 
 \label{eq:S+-}
\eeq
of an  SO(3) rotational angle $\Omega$.
In a VCS representation, these overlap functions are regarded as  components of a vector-valued function for which the elements of the set $\{ \xi_K\}$ are basis vectors.
The state $|(\lambda\mu)\alpha LM\rangle$ is then represented by a VCS wave function
\bal \Psi^{(\lambda\mu)}_{\alpha LM}(\Omega) 
&= \sum_K \xi_{K}
 \langle (\lambda\mu)K |\hat R(\Omega ) 
 |(\lambda\mu)\alpha LM\rangle \nonumber\\
& = \sum_K \xi_{K} 
\langle (\lambda\mu)K 
|(\lambda\mu)\alpha LK\rangle \scrD^L_{KM}(\Omega ) ,
  \label{eq:defnPsiOmega}
 \end{align}
which resembles  a rotor-model wave function
for which the wave functions $\{ \xi_K\}$ are interpreted as the intrinsic wave functions of rotational bands.
Moreover, in parallel with the rotor model, there are intrinsic symmetries, relating to the above-mentioned redundancies, that restrict the combinations of $\xi_{K} \scrD^L_{KM}(\Omega $) in equation  
(\ref{eq:defnPsiOmega}) to a linearly-independent set. 

The relevant intrinsic symmetry group, is the subset of SO(3) rotations that leave the space of highest-grade states invariant.
It is the group  D$_\infty$  generated by the rotations 
$ \{ e^{{\rm i} \alpha \hat L_0}; 0\leq \alpha < 2\pi\}$  and $e^{{\rm i} \pi \hat L_y}$,
where $\hat L_y$ is an angular-momentum operator perpendicular to $\hat L_0$, for which
\begin{subequations}\label{eq:6.rotHGstates}
\bal
&e^{-{\rm i}\alpha \hat L_0} |(\lambda\mu) K\rangle 
= e^{-{\rm i} \alpha K} |(\lambda\mu) K\rangle , \\
& e^{-{\rm i}\pi \hat L_y} |(\lambda\mu) K\rangle 
= (-1)^{\lambda} |(\lambda\mu),- K\rangle , \end{align}
\end{subequations}
(the second equation was determined by explicit construction of  the
 $\{|(\lambda\mu) K\rangle\}$ states).
Together with the identities
\begin{subequations}label{eq:6.Dfns} 
\bal
& \scrD^L_{KM}(\alpha,0,0) 
= e^{-{\rm i} \alpha K} \delta_{M,K} ,\\
&\scrD^L_{KM}(0,\pi,0) 
= (-1)^{L+K} \delta_{M,-K} ,
\end{align}
\end{subequations}
 these intrinsic symmetries imply that
\bal
&\langle (\lambda\mu),-K |(\lambda\mu)\alpha L,-K\rangle \nonumber \\
&\;\; = 
\langle (\lambda\mu),-K |e^{-{\rm i}\pi \hat L_y}   |(\lambda\mu)K\rangle
\langle (\lambda\mu)K | e^{{\rm i}\pi \hat L_y}  
  |(\lambda\mu)\alpha L,-K\rangle  \nonumber\\
&\;\; = (-1)^{\lambda + L+ K}
\langle (\lambda\mu)K |(\lambda\mu)\alpha LK\rangle .
\end{align}
It  follows that an orthonormal basis of VCS wave functions for the  irreducible SU(3) representation $(\lambda\mu)$ consists of linear combinations
 \beq \Psi^{(\lambda\mu)}_{\alpha LM}(\Omega) =  \sum_{K\geq 0} 
\Phi^{(\lambda\mu)}_{KLM}(\Omega)\, \mathcal{K}^{(L)}_{K \alpha} ,
 \label{eq:6.SO3ONwfns} \eeq
 where $\Phi^{(\lambda\mu)}_{KLM}$ is the rotor-model wave function given by equation (\ref{eq:6.phiKLM}), and the 
 $\mathcal{K}^{(L)}_{K \alpha}$ coefficients remain to be determined.

\subsection{\boldmath VCS representation of the SU(3) Lie algebra}
The VCS representation $\hat\Gamma(X)$ of an element $X$ in the SU(3) Lie algebra   is defined as  an operator on  VCS wave functions,  by the equation
\beq\hat \Gamma(X) \Psi^{(\lambda\mu)}_{\alpha LM} (\Omega)
=\sum_{K}\xi_{K} 
\langle (\lambda\mu)K | \hat R(\Omega) \hat X |(\lambda\mu)\alpha LM\rangle .
\label{eq:6.Gamma(X)}
\eeq
This immediately returns the expected expressions
\bal &  \hat\Gamma(L_0) \Psi^{(\lambda\mu)}_{\alpha LM} (\Omega) =
M\Psi^{(\lambda\mu)}_{\alpha LM}(\Omega) , \label{eq:5.L0}\\
& \hat\Gamma(L_\pm) \Psi^{(\lambda\mu)}_{\alpha LM} (\Omega) =
\sqrt{(L\mp M)(L\pm M+1)} \, \Psi^{(\lambda\mu)}_{\alpha L,M\pm 1}  (\Omega) ,
\label{eq:6.Lpm}
\end{align}
for the angular-momentum operators.
The VCS representation of the SU(3) quadrupole operators is  defined by the equation
\bal
&\Gamma({\cal Q}_{2,\nu})\Psi^{(\lambda\mu)}_{\alpha LM} (\Omega)
=\sum_K\xi_{ K} \langle (\lambda\mu)K|\hat R(\Omega)  
\hat {\cal Q}_{2,\nu} |(\lambda\mu)\alpha LM\rangle \nonumber \\
&\quad = \sum_{K\nu'}  \xi_{ K} 
\langle (\lambda\mu)K|\hat {\cal Q}_{2,\nu'} \hat R(\Omega)  
 |(\lambda\mu)\alpha LK\rangle\, \scrD^2_{\nu'\nu}(\Omega).
\label{eq:6.VCSQ}
\end{align}
By making the substitutions  
\begin{subequations}
\bal 
&\langle (\lambda\mu)K |\hat {\cal Q}_{2,0} 
= (2\lambda + \mu) \langle (\lambda\mu)K | ,\\
&\langle (\lambda\mu)K |\hat {\cal Q}_{2,\pm 1} = \textstyle
-\sqrt{\frac32}\langle (\lambda\mu)K |\hat L_\pm ,\\
&\sum_K\xi_{K} 
\langle (\lambda\mu)K |\hat {\cal Q}_{2,\pm 2}
=\sqrt{6}\, \hat \sigma_\pm \sum_K\xi_{K} \langle (\lambda\mu)K | ,
\end{align}
\end{subequations}
of which the first two are 
obtained from eqs.\,(\ref{eq:6.L_k2})--(\ref{eq:LijQij}) 
and the observation from (\ref{eq:U3Hgrade}) that 
\beq \langle (\lambda\mu)K | \hat C_{21}
= \langle (\lambda\mu)K | \hat C_{31} = 0 
\eeq
and the third is obtained from a comparison of equations  (\ref{eq:83}) and  (\ref{eq:S+-ops.1}),
 it follows that
\begin{subequations} \label{eq:6.95}
\bal 
&\sum_K\xi_{K} \langle (\lambda\mu)K |\hat {\cal Q}_{2,0}
\hat R(\Omega)  |(\lambda\mu)\alpha LM\rangle \nonumber\\
&\hspace{4cm} =(2\lambda+\mu) \Psi^{(\lambda\mu)}_{\alpha LM}(\Omega)  ,  
\\
&\sum_K\xi_{ K} \langle (\lambda\mu)K |\hat {\cal Q}_{2,\pm 1}
\hat R(\Omega)  |(\lambda\mu)\alpha LM\rangle \nonumber \\
&\hspace{4cm}=\textstyle -\sqrt{\frac32}\,  
\bar L_\pm \Psi^{(\lambda\mu)}_{\alpha LM}(\Omega) , 
\\ 
&\sum_K\xi_{ K} \langle (\lambda\mu)K |\hat {\cal Q}_{2,\pm 2}
\hat R(\Omega)  |\alpha LM\rangle\nonumber \\
&\hspace{4cm}=\sqrt{6}\, \hat \sigma_\pm \Psi^{(\lambda\mu)}_{\alpha LM}(\Omega) ,
\end{align}
\end{subequations}
where  $\bar L_\pm$ are  infinitesimal generators of left
rotations.
Their actions, defined by 
\bal
 \big[\bar L_k \scrD^L_{KM}\big](\Omega) 
 &= \langle LK|\hat L_k \hat R(\Omega) |LM\rangle \nonumber\\
 &= \sum_N \langle LK |\hat L_k |LN\rangle \scrD^L_{NM}(\Omega ) ,
\end{align}
give the expressions, familiar in the nuclear rotor model,
\begin{subequations}
\bal &\bar L_0 \scrD^L_{KM} = K \scrD^L_{KM} ,\\
& \bar L_\pm \scrD^L_{KM} = \sqrt{(L\pm K)(L\mp K+1)}\, 
\scrD^L_{K\mp 1,M}. 
\end{align}
\end{subequations}
Thus, equations (\ref{eq:6.VCSQ}) and (\ref{eq:6.95})
lead to the expression
\bal \hat\Gamma({\cal Q}_{2,\nu}) &= 
(2\lambda+\mu) \hat\scrD^2_{0\nu} -
\textstyle \sqrt{\frac32}\, \big( \hat\scrD^2_{1\nu}\bar L_+ +
\hat\scrD^2_{-1\nu} \bar L_-\big) \nonumber\\
&\quad+\sqrt{6}\,\big[ \hat \sigma_+\hat\scrD^2_{2\nu} 
+ \hat \sigma_-\hat\scrD^2_{-2\nu}\big]  \label{eq:6.express1}
\end{align}
which simplifies, by use of the identity
\beq \big[ \hat{\bf L}^2, \hat\scrD^2_{0\nu}\big] = 6\hat\scrD^2_{0\nu}+
\textstyle 2 \sqrt{\frac32}\, 
\big[ \hat\scrD^2_{1\nu}\bar L_+ +\hat\scrD^2_{-1\nu}\bar L_-\big],
\eeq
to equation  (\ref{eq:GammaQ}).

\subsection{\boldmath SU(3) transformations of rotor-model wave functions}
The transformations of the rotor-model wave functions 
(\ref{eq:6.phiKLM}) by the $\hat\Gamma({\cal Q}_{2\nu})$ 
operators given by equation  (\ref{eq:GammaQ}) are determined by use of the equations
\bal &\left[ \hat\scrD^2_0 \otimes \scrD^L_K \right]_{L'M'} =
(LK\, 20 |L'K)\, \scrD^{L'}_{KM'} , \label{eq:6.[DxD]}\\
& \left[ \hat \sigma_\pm \hat\scrD^2_{\pm2} 
\otimes \xi_{ K} \scrD^L_K \right]_{L'M'} \equiv
\hat \sigma_\pm \xi_{ K}  
\left[ \hat\scrD^2_{\pm2} \otimes \scrD^L_K \right]_{L'M'}   
\label{eq:6.[DxDxD]} \nonumber  \\
&\hspace{0.5cm} = \tfrac12 \sqrt{ (\mu\mp K)(\mu\pm K+2)} \, 
\nonumber\\
&\hspace{0.8cm}  \times(LK\, 2,\pm 2 |L',K\pm 2)\, 
\xi_{K\pm 2}\scrD^{L'}_{K\pm 2,M'} ,
\end{align}
from which it follows that
\bal 
[\hat \Gamma(\mc{Q}_2) \otimes &\Phi^{(\lambda\mu)}_{KL}]_{L'M'}
= \sqrt{\frac{2L+1}{2L'+1}}\sum_{K'\geq 0} 
\Phi^{(\lambda\mu)}_{K'L'M'} M^{L'L}_{K'K} \label{eq:QxPhi}
\end{align}
with
\begin{subequations}
\bal  &M^{L'L}_{KK} = \left[ (2\lambda\! +\!\mu+3) 
- \tfrac12 L'(L'+1) +\tfrac12 L(L+1)\right] \qquad \nonumber\\
&\quad\quad \times  (LK \,20|L'K ) \label{eq:6.M_KKmatrix}\\
&\quad\quad + \delta_{K,1}\,  (-1)^{\lambda +L+1}
\sqrt{\tfrac{3}{2}} (\mu+1)(L,-1\,22|L'1), \nonumber \\
&M^{L'L}_{K\pm 2,K} 
 = \sqrt{\tfrac32(\mu \mp K )(\mu \pm K+2) (1+\delta_{K,0)})}\ 
 \nonumber\\
&\quad\quad \times
(LK \,2,\pm 2|L',K\!\pm\!2). \label{eq:6.M_K+2,Kmatrix}
\end{align}
\end{subequations}

\subsection{\boldmath SO(3)-reduced matrix elements for SU(3) representations of type \boldmath $(\lambda0)$ and $(\lambda1)$}
\label{sect:mu=0or1}
SO(3)-reduced matrix elements are now obtained from the expression
\beq [\hat\Gamma({\cal Q}_2) \otimes 
\Psi^{(\lambda\mu)}_{\alpha L}]^{}_{L'M'}
= \sum_\beta \Psi^{(\lambda\mu)}_{\beta L'M'} 
\frac{\langle {\beta (\lambda\mu)} L' \| \hat {\mathcal{Q}}_2 \| {(\lambda\mu)} L\rangle} {\sqrt{2L'+1}}\label{eq:WEthm}
\eeq
of the Wigner-Eckart theorem  in an orthonormal basis.

For a $(\lambda\, 0)$ representation, 
for which the multiplicity index $\alpha$ of an SU(3) wave function
$\Psi^{(\lambda\,0)}_{\alpha LM}$ is not needed,
$K=0$ is a good quantum number and an orthonormal set of basis wave functions is of the form
\beq \Psi^{(\lambda0)}_{LM} = k_L \Phi^{(\lambda 0)}_{0LM} ,
\eeq
where $k_L$ is a  norm factor and $L$ is restricted to even or odd integers according as $\lambda$ is even or odd.
Equations (\ref{eq:GammaQ}) and (\ref{eq:QxPhi})
then lead to the reduced matrix elements
\bal
& \langle (\lambda\, 0) L' \| \hat {\mathcal{Q}}_2 \| (\lambda\, 0) L\rangle
=\frac{k_L}{k_{L'}} {\sqrt{2L+1}}\, (L0 \,20|L'0)  \nonumber\\
&\hspace {1.2cm} \times \left[ (2\lambda\! +\!3)  - \tfrac12 L'(L'+1) +\tfrac12 L(L+1)\right] 
\end{align}
and 
\begin{subequations}   \label{eq:6.(l0)Qmes1}
\bal  
&\langle(\lambda\, 0) L \| \hat {\mathcal{Q}}_2 \| (\lambda\, 0) L\rangle
= \sqrt{2L+1} \,    \nonumber\\
& \hspace{3.3cm} \times (2\lambda\! +\!3) (L0 \,20|L0),  \\
&\langle (\lambda\, 0) L\!+\!2 \| \hat {\mathcal{Q}}_2 \| (\lambda\, 0) L\rangle
= 2\frac{k_L}{k_{L\!+\!2}}
{\sqrt{2L+1}}\, (\lambda-L)  \nonumber\\ 
&\hspace{3.8cm} \times     (L0 \,20|L\!+\!2,0), \\
&{\langle (\lambda\, 0) L \| \hat {\mathcal{Q}}_2 \| (\lambda\, 0) L\!+\!2\rangle}
= 2\frac{k_{L\!+\!2}}{k_L} \sqrt{2L+5}\, (\lambda +L +3)
\nonumber\\
&\hspace{3.8cm}  \times(L\!+\!2,0 \,20|L0) \nonumber\\
&\hspace{0.4cm}
=2\frac{k_{L\!+\!2}}{k_L} \sqrt{2L+1}\, (\lambda +L +3) (L0 \,20|L\!+\!2,0) .
\end{align}
\end{subequations}
Thus, to satisfy the Hermiticity relations, as required for a unitary representation,
\beq \langle (\lambda\, 0) L' \| \hat {\mathcal{Q}}_2 
\| (\lambda\, 0) L\rangle
=  \langle (\lambda\, 0) L \| \hat {\mathcal{Q}}_2
 \| (\lambda\, 0) L'\rangle^* 
\eeq
the norm factors must have ratios 
\beq { \Big|\frac{k_{L+2}}{k_L}\Big|^2 =  \frac{\lambda-L}{\lambda+L+3}} \,.\eeq
It  follows that, for $L\leq \lambda$,
\bal
& \langle (\lambda\, 0) L+2 \| \hat {\mathcal{Q}}_2 \| (\lambda\, 0) L\rangle 
= \langle (\lambda\, 0) L \| \hat {\mathcal{Q}}_2 \| (\lambda\, 0) L+2\rangle  
\qquad \quad
\nonumber\\
&=\!\big[ 4(2L+1)(\lambda-L)(\lambda +L+3)\big]^\frac12 
(L0\, 20 | L+2,0) 
\label{eq:6.(l0)Qmes2b}
\end{align}
 and the sequence of increasing angular-momentum
states terminates at $L=\lambda$.

For a $(\lambda\, 1)$ representation, one similarly obtains
\begin{subequations}
\bal
&\langle (\lambda\, 1)L \| \hat{\mc{Q}}  \| (\lambda\,1)L\rangle = 
\sqrt{2L+1}\,  \\
&\quad \times
\left[ (2\lambda+4)(L 1\,2 0 |L 0) 
- (-1)^{\lambda +L} \sqrt{6} (L,\! -1 \, 2 2|L 1) \right], \nonumber\\
&\langle (\lambda\, 1)L\!+\!1 \| \hat{\mc{Q}}  \| (\lambda\,1)L\rangle
=  \sqrt{2L+1}\, (L 1 \, 2 0 | L\!+\!1, 1) \nonumber\\
& \quad\times
[2\lambda -L +3 +(-1)^{\lambda+L}(L+ 1) ]^\frac12 \nonumber\\
& \quad\times[2\lambda +L +5 -(-1)^{\lambda+L}(L+ 1)]^\frac12 ,\\
&\langle (\lambda\, 1)L\!+\!2  \| \hat{\mc{Q}}  \| (\lambda\,1)L\rangle
=  \sqrt{2L+1}\, (L 1 \, 2 0 | L\!+\!2, 1) \nonumber\\
&\quad \times
[2\lambda -2L +1 -(-1)^{\lambda+L}]^\frac12 \nonumber\\
&\quad \times[2\lambda +2L +7 -(-1)^{\lambda+L}]^\frac12  ,
\end{align}
\end{subequations}
consistent with the termination of the sequence of 
states at $L=\lambda+1$.

\subsection{\boldmath SO(3)-reduced matrix elements for SU(3) representations $(\lambda,\mu)$ with $\mu >1$} \label{sect:SU3mu>1}
The above multiplicity-free examples, highlight the fact that the Hermiticity relationships make it straightforward to renormalise an orthogonal basis to obtain  an orthonormal basis.  
An orthogonal basis for an irreducible  representation of SU(3) can be defined generally by the eigenstates  of the SO(3)-invariant operator 
$\hat X_3 =  [ \hat L \otimes \hat Q_2 \otimes \hat L]_0$
(or for a closer relationship with the rotor model, as discussed in the following section, by a particular linear combination of the SO(3)-invariants
$\hat X_3$ and $\hat X_4 =
[\hat L \otimes[\hat{\mc{Q}}\otimes\hat{\mc{Q}}]_2 \otimes \hat L]_0$).

For an orthonormal basis of  eigenstates of the operator
$\hat X_3 = [\hat L \otimes  \hat{\mathcal{Q}}_2 \otimes \hat L]_0$, the VCS wave functions  satisfy the equation
\bal \hat\Gamma(X_3) \Psi^{(\lambda\mu)}_{\alpha LM}
&= \Psi^{(\lambda\mu)}_{\alpha LM}
\frac{\langle{(\lambda\mu)}\alpha L \| 
[\hat L\otimes \hat{\mathcal{Q}}_2 \otimes \hat L]_0
 \| {(\lambda\mu)}\alpha L\rangle} {\sqrt{2L+1}}  \nonumber\\
&= \Psi^{(\lambda\mu)}_{\alpha LM} f(L)
\langle {(\lambda\mu)}\alpha L \|\hat{\mathcal{Q}}_2  
\| {(\lambda\mu)}\alpha L\rangle ,
\label{eq:6.X3redME}
\end{align}
with $f(L)= \sqrt{L(L+1)(2L-1)(2L+3)}$. 
 This implies that the reduced matrix elements 
 ${\langle{(\lambda\mu)}\beta L' \|  \hat{\mathcal{Q}}_2 
 \| {(\lambda\mu)}\alpha L\rangle}$,
 defined  by the Wigner-Eckart theorem (\ref{eq:WEthm}),
should be diagonal when $L'=L$, i.e., vanish if $L' =L$ but 
$\beta \not= \alpha$.

The first step of a so-called K-matrix  transformation 
\beq  \Psi^{(\lambda\mu)}_{\alpha LM}(\Omega) 
= \sum_K \Phi^{(\lambda\mu)}_{KLM}(\Omega)
\, \mc{K}^{L}_{K\alpha} , \label{eq:Kmat}
\eeq
to an orthonormal basis  
$\{\Psi^{(\lambda\mu)}_{\alpha LM}\}$
is  to construct an orthogonal set of functions
$\{ \Phi^{(\lambda\mu)}_{\alpha LM}\}$ (with Greek index $\alpha$) 
by a unitary transformation 
\beq   \Phi^{(\lambda\mu)}_{\alpha LM} =
\sum_K \Phi^{(\lambda\mu)}_{KLM} U^L_{K\alpha} ,
 \label{eq:Phi_alpha} 
\eeq
of the rotor-model wave functions $\{\Phi^{(\lambda\mu)}_{K LM}\}$
of equation (\ref{eq:6.phiKLM}),
where $U^L$ is the unitary matrix that diagonalises the matrix
$M^{LL}$  given by equations (\ref{eq:6.M_KKmatrix}) and
(\ref{eq:6.M_K+2,Kmatrix}).
It then only remains to renormalise this set to obtain orthonormal basis states, with VCS wave functions
\beq \Psi^{(\lambda\mu)}_{\alpha LM} 
= k_\alpha^L \Phi^{(\lambda\mu)}_{\alpha LM} ,
\eeq
that satisfy the Hermiticity relations
\bal 
\langle(\lambda\mu) \beta L' & \| \hat {\cal Q}_2 
 \|(\lambda\mu) \alpha L\rangle  \nonumber\\
& =  (-1)^{L'-L}   \langle(\lambda\mu)
 \alpha L\| \hat {\cal Q}_2 \|(\lambda\mu) \beta L'\rangle^*  \label{eq:hermiticity.eq}
 \end{align}
of a unitary representation.

An expansion of the left side of equation (\ref{eq:WEthm}) gives
\bal  
[\hat\Gamma({\cal Q}_2) \otimes &
\Psi^{(\lambda\mu)}_{\alpha L}]_{L'M'} \nonumber\\
&= k_\alpha^L  \sum_K 
[\hat\Gamma({\cal Q}_2) \otimes \Phi^{(\lambda\mu)}_{KL}]_{L'M'}
U^L_{K\alpha}
 \end{align}
and, with  (\ref{eq:QxPhi}) followed by the inverse of (\ref{eq:Phi_alpha}),
\bal 
&[\hat\Gamma({\cal Q}_2) \otimes \Psi^{(\lambda\mu)}_{\alpha L}]_{L'M'}
\nonumber\\
&\hspace{0.5cm} = k^L_\alpha \sqrt{\frac{2L+1}{2L'+1}} \sum_{KK'}
\Phi^{(\lambda\mu)}_{K'L'M'} M^{L'L}_{K'K} U^L_{K\alpha} \nonumber\\
&\hspace{0.5cm} =k^L_\alpha \sqrt{\frac{2L+1}{2L'+1}} \sum_{\beta KK'}
\Phi^{(\lambda\mu)}_{\beta L'M'} U^{L'*}_{K'\beta}
 M^{L'L}_{K'K} U^L_{K\alpha} .
\end{align}
This gives
 \bal
[\hat\Gamma({\cal Q}_2) \otimes 
\Psi^{(\lambda\mu)}_{\alpha L}]_{L'M'}
&=\sqrt{\frac{2L+1}{2L'+1}}  \sum_\beta \frac{k^L_\alpha}{k^{L'}_\beta}
\Psi^{(\lambda\mu)}_{\beta L'M'} \nonumber \\
&\quad \times
\sum_{KK'} U^{L'*}_{K'\beta} M^{L'L}_{K'K}  U^L_{K\alpha},
\end{align}
and, from the Wigner-Eckart theorem (\ref{eq:WEthm}), the reduced matrix element
\beq
\langle{(\lambda\mu)}\beta L' \|  \hat{\mathcal{Q}}_2 
 \| {(\lambda\mu)}\alpha L\rangle =
 \frac{k^L_\alpha}{k^{L'}_\beta}  \sqrt{2L+1}\,
{\cal M}_{\beta\alpha}^{L'L} .
\eeq
with 
\beq {\cal M}_{\beta\alpha}^{L'L} = 
\sum_{K,K'}  U^{L'*}_{K' \beta}  M^{L'L}_{K'K}  U^{L}_{K\alpha } .
\label{eq:calM}
\eeq
It likewise follows that
\beq
\langle{(\lambda\mu)}\alpha L\|  \hat{\mathcal{Q}}_2 
 \| {(\lambda\mu)}\beta L' \rangle =
 \frac{k^{L'}_\beta}{k^L_\alpha} \sqrt{2L'+1}
{\cal M}_{\beta\alpha}^{LL'} .
\eeq
Thus, to satisfy the Hermiticity equation (\ref{eq:hermiticity.eq}), it is required that
\beq  
\frac{k^L_\alpha}{k^{L'}_\beta} \sqrt{2L+1} \,
{\cal M}_{\beta\alpha}^{L'L} = (-1)^{L-L'}
 \frac{k^{L'}_\beta}{k^L_\alpha} \sqrt{2L'+1}\,
{\cal M}_{\beta\alpha}^{LL'^*} ,
\eeq
which means that the renormalisation coefficients have ratios given by
\beq 
\left(\frac{k^L_\alpha}{k^{L'}_\beta}\right)^2
= (-1)^{L-L'} \sqrt{\frac{2L'+1}{2L+1}}\,
\frac{\mathcal{M}^{LL'^*}_{\beta\alpha}}%
{\mathcal{M}^{L'L}_{\alpha\beta}}\, . \label{eq:kratios}
\eeq

The ratios found in this expression are
easily determined recursively, starting from the state of the
representation of angular momentum $L=0$ or $L=1$, which is always
multiplicity free and for which one can set $k^L = 1$.  This
expression yields normalisation factors $k^L_\alpha$ for those
states with
non-vanishing norms within the SU(3) representation under 
construction while those states for which $k^L_\alpha$ vanishes are
simply discarded.

\section{\boldmath Macroscopic limits of the SU(3) model}
\label{sect:Rotor_SU3}
Large-dimensional limits of the SU(3) model 
have been considered  by many authors
\cite{CarvalhoPhD84,LeBlancCVR86,LeschberD86,LeschberDR86,%
CastanosDL88,RoweVC89,NaqviD90}.
In such limits, the properties  
of an SU(3) representation asymptotically approach those of a rigid-rotor model.
We refer to these, and more precise limits brought to light by  VCS  representations \cite{RoweLeBR89,RoweVC89},  as macroscopic limits.
They are easily calculated and are valuable for assessing the relevance of particular SU(3) representations in  potential applications, 
e.g.,  as discussed in Section \ref{sect:CRV}.
This section examines the intimate relationship between the
rigid-rotor and the SU(3) model in  its macroscopic limits
and shows how closely accurately-computed results for finite-dimensional SU(3) representations approach their asymptotic limits.

It is first shown that, in the  limit in which the value of its Casimir invariant is  large, an irreducible representation of the SU(3) algebra contracts to that of a rigid-rotor model with intrinsic quadrupole moments
\beq \bar \iQ_0 = (2\lambda+\mu+3) , \quad 
\bar \iQ_2  = \sqrt{\tfrac32}\,(\mu +1). 
\label{eq:6.IntQmoments}\eeq 
It is then shown that an exceedingly close approach of SU(3) states to rotor-model states with integer-valued $K$ quantum numbers is given by eigenstates of the operator
\beq \hat {\mathcal{Z}}^{(\lambda\mu)} =
[\hat L \otimes[\hat{\mc{Q}}_2\otimes\hat{\mc{Q}}_2]_2 \otimes \hat L]_0 
-  \sqrt{\tfrac87}\, \bar \iQ_0
 [\hat L \otimes \hat{\mc{Q}}_2 \otimes \hat L]_0  .\label{eq:6.ZSU3op}\eeq 
It is also shown that precise SU(3) matrix elements are approached 
rapidly for $L/\lambda \lesssim 0.4$ by those of a rotor with the intrinsic quadrupole moments given by equation
(\ref{eq:6.IntQmoments}) and,  as shown in Subsection \ref{sect:Rotor_SU3.comparison}, even more rapidly by the asymptotic expressions
\begin{subequations} \label{eq:ASlimits}
\bal \label{eq:6.KLtoKL}
&\langle (\lambda\mu) K L\| \hat {\cal Q}_2 
\|(\lambda\mu) K L\rangle^{(AL)}  = \sqrt{2L+1}\,   \\
& \hspace{2.8cm} 
\times (LK\, 20 |LK)(\bar {\rm Q}_0 + \delta_{K,1} \Delta^{LL} ) , 
\nonumber  \\ 
& \langle (\lambda\mu) K,\! L\!+\!1\| \hat {\cal Q}_2 \|(\lambda\mu) K L\rangle^{(AL)}  = \sqrt{2L+1}  \label{eq:6.KLtoKL+1}\\
&\hspace{0.1cm} \times
 (LK\, 20 |L\!+\!1,\!K)  
\big[(\bar {\rm Q}_0 +\delta_{K,1}\Delta^{L\!+\!1,L})^2- (L+1)^2 \big]^{\tfrac12} ,
 \nonumber \\
&\langle (\lambda\mu) K,\! L\!+\! 2 \| \hat {\cal Q}_2 \|(\lambda\mu) K L\rangle^{(AL)} 
= \sqrt{2L+1}\, \\
&\hspace{0.1cm} \times (LK\, 20 |L\!+\!2,\!K)
\big[(\bar Q_0 +\delta_{K,1}\Delta^{L\!+\!2,L})^2- (2L+3)^2 \big]^{\tfrac12} ,
\nonumber \\
&\langle (\lambda\mu) K\!+\!2,\! L' \|\hat {\cal Q}_2 \|(\lambda\mu) K L\rangle^{(AL)} = \sqrt{2L+1} \\
&\hspace{0.1cm} \times (LK\, 22 |L',\!K\!+2) 
[\tfrac32 (\mu - K)(\mu\ + K+2)(1+\delta_{K,0})]^{\tfrac12} ,\nonumber
\end{align}
\end{subequations}
where $\bar \iQ_{0} = 2\lambda + \mu + 3$ and
\bal \label{eq:6,Delta2}
 \Delta^{L'L} &= (-1)^{\lambda+L+1} \sqrt{\tfrac32}\,(\mu+1)
 \frac{(L,\!-\!1\,22|L'1)}{(L1\,20|L'1)} \nonumber \\
 &= \Delta^{LL'}. 
\end{align}

\subsection{\boldmath Contraction of an SU(3) irrep}
\label{sect:SU3contraction}

A standard contraction \cite{InonuW53} of the SU(3) Lie algebra is obtained with renormalised quadrupole moments
\beq q_{2\nu} = \varepsilon {\cal Q}_{2\nu} ,\eeq
where $\varepsilon$ is a small parameter. 
In terms of these quadrupole moments, the SU(3) commutation relations become 
\begin{subequations}
\begin{align} 
&[ L_k,   L_{k'}] =  \sqrt{2}\, (1k'\,1k|1,k+k')\,   L_{k+k'} \, ,
\label{eq:6.LLcr.cont}\\
&{[}  L_k, q_{2,\nu}] = \sqrt{6}\, (2\nu\,1k |2,\nu+k)\,
 q_{2,\nu+k} \, . \\ 
& {[} q_{2\nu},  q_{2\mu}] =-3\sqrt{10}\,
 (2\mu\, 2\nu|1,\mu+\nu)\, \varepsilon^2 L_{\mu+\nu}  \nonumber\\
& \phantom{{[} q_{2\nu},  q_{2\mu}]} = {\rm O}(\varepsilon^2)  .
\end{align}
\end{subequations}
The circumstances under which this contraction is realised are obtained by setting  
 $\varepsilon^2 =1/ {\mathcal I}_2(\lambda\mu)$
\cite{CarvalhoPhD84,Rowe85},, 
where 
\bal  {\mathcal I}_2(\lambda\mu)
&=  \langle \hat {\mathcal{Q}}_2 \cdot \hat {\mathcal{Q}_2} 
+ 3 \hat {\bf L}\cdot  \hat {\bf L}\rangle  \nonumber\\
& =4(\lambda^2+ \lambda\mu +\mu^2 + 3\lambda+3\mu)
\end{align}
is the value of the SU(3) Casimir invariant for the representation
$(\lambda\mu)$, it follows that the $\{ q_{2\nu}\}$ quadrupole
moments of states of angular momentum $L$,
 for which $3 L(L+1)\ll {\cal I}_2(\lambda\mu)$,
are normalised such that
\beq \langle\hat q_2 \cdot \hat q_2\rangle =
\varepsilon^2\langle \hat {\mathcal{Q}}_2 
\cdot \hat {\mathcal{Q}}_2\rangle \approx 1 . \eeq
However, the commutators
$ [q_{2,\nu},  q_{2,\mu}] ={\rm O}(\varepsilon^2)$ become negligible.
Thus, as $\varepsilon \to 0$, the  SU(3) Lie algebra contracts to that
of the rotor algebra (see Section \ref{sect:Ui_rotor})
with commuting quadrupole operators.
The physical significance of this contraction is that  states of low angular momentum of a large-dimensional irreducible SU(3) representation, for which $\varepsilon \ll 1$, are expected to exhibit rotor-model properties,

The deformation parameters of the rotor model to which a given SU(3) representation contracts
\cite{LeschberDR86,LeschberD87,CastanosDL88}, 
are obtained by comparing the shape invariants 
$Q_2\cdot Q_2$ and $[Q_2\otimes Q_2\otimes Q_2]_0$
of a rotor with the corresponding SU(3) Casimir invariants.
For a rotor  with  intrinsic quadrupole moments
\beq \bar \iQ_0 = \beta\cos\gamma , \quad 
\bar \iQ_2 =\sqrt{\tfrac12}\, \beta\sin\gamma ,
\label{eq:6.IntQmomentsb}\eeq 
these invariant are given by
\beq   Q_2\cdot Q_2 = \beta^2, \quad 
\sqrt{70}\,[Q_2\otimes Q_2\otimes Q_2]_0 
=-2\beta^3\cos3\gamma .
\label{eq:6.shapeinvariants}  \eeq
It is then determined that, with the intrinsic quadrupole moments
\beq \bar \iQ_0 = (2\lambda+\mu+3) , \quad 
\bar \iQ_2  = \sqrt{\tfrac32}\,(\mu +1),
\label{eq:6.IntQmomentsc}\eeq 
the rotor-model invariants take the values
\bal 
 Q_2\cdot Q_2 
   &= ( 2\lambda + \mu +3)^2 + 3\mu(\mu+2) \nonumber\\
   &= 4(\lambda^2 + \lambda\mu + \mu^2 + 3\lambda + 3\mu + 3),
    \label{eq:6.Q.Q}\\
\sqrt{70} \, [Q_2 &\otimes Q_2\otimes Q_2]_0 \nonumber\\
  &= - 8(2\lambda+\mu+3)(\lambda - \mu)(\lambda + 2\mu +3) ,
\label{eq:6,Q^3SU3}
\end{align}
which, with neglect of the constant 3 on the right side of 
equation  (\ref{eq:6.Q.Q}), are  the values of the SU(3) Casimir invariants 
$\hat{\mathcal{I}}_2$ and $\hat {\mathcal{I}}_3$.

\subsection{\boldmath The $K$ quantum number}
\label{sect:6.Kqno}
Given the close relationship between the SU(3) and rotor models, 
it was natural to seek an optimal one-to-one correspondence between states of an irreducible SU(3) representation and a subset of
rotor model states with good $K$ quantum numbers.
Rosensteel, Leschber and colleagues \cite{RosensteelR77b,LeschberDR86} suggested
that the SU(3) states of the corresponding pairs should be eigenstates of some linear combination of the SO(3)-invariant operators
$ [\hat L \otimes \hat{\mc{Q}}_2 \otimes \hat L]_0$ and
$[\hat L \otimes[\hat{\mc{Q}}\otimes\hat{\mc{Q}}]_2 
\otimes \hat L]_0$. 
This proved to be correct \cite{RoweT08} and, as this section shows,  gives useful canonical  SO(3)-coupled bases for irreducible SU(3) representations with extraordinarily good $K$ quantum numbers.

It has been shown  \cite{RoweT08}  that  rotor model states with good $K$ quantum numbers and for which the  intrinsic quadrupole moments take the constant $K$-independent values
\beq \langle\xi_K |\hat Q_{2,0} |\xi_K\rangle = \bar \iQ_0, \quad
\langle\xi_{K\pm 2} |\hat Q_{2,\pm 2} |\xi_K\rangle = \bar \iQ_2 ,\eeq
are eigenstates of the operator 
\beq \label{eq:Rotor_SU3.Zrot}
 \hat {\mathcal{Z}}^{({\rm rot})} =
[\hat L \otimes[\hat{{Q}}_2 \otimes\hat{{Q}}_2 ]_2 \otimes \hat L]_0 
-  \sqrt{\tfrac87}\, \bar Q_0 
[\hat L \otimes \hat{{Q}}_2 \otimes \hat L]_0  .
\eeq
This was shown by observing that, for such a generally triaxial rotor, the ratios
\beq R(K',L,K) = 
\frac{\langle K'L \| [\hat{{Q}}_2\otimes\hat{{Q}}_2]_2 \| KL\rangle}
{\langle K'L \| \hat{{Q}}_2  \| KL\rangle} 
\eeq
 have the  $L$- and $K$-independent off-diagonal values 
\beq  R(K+2,L,K) = \sqrt{\tfrac87}\, \bar \iQ_0 ,\eeq
for all $L$ and $K$.
This implies that standard rotor-model  states $\{ |KLM\rangle\}$, 
with good $K$ quantum numbers, are eigenstates of the operator 
$\hat {\mathcal{Z}}^{({\rm rot})}$ of (\ref{eq:Rotor_SU3.Zrot}).
Thus, an approach of an SU(3) state of angular momentum $L$ to a state with a good $K$ quantum number, which becomes precisely that of the rotor model in a $\lambda/L\to \infty$ asymptotic limit, can  be defined as an eigenstate of the operator
$\hat {\mathcal{Z}}^{(\lambda\mu)}$ of equation  (\ref{eq:6.ZSU3op}).

Recall that  SU(3) states with the same $L,M$ angular-momentum  values were distinguished in the VCS construction of Section \ref{sect:SU3mu>1} by choosing them to be eigenstates of the third-order  SO(3)-invariant operator
$[\hat L \otimes \hat{\mc{Q}}_2 \otimes \hat L]_0$.
Thus, to obtain SU(3) states corresponding  to rotor-model states with good $K$ quantum numbers, one has simply to adjust the VCS construction by replacing this operator with the linear combination 
$\hat {\mathcal{Z}}^{(\lambda\mu)}$
of the third- and fourth-order SO(3) invariants.
There should then be far less mixing of rotor model wave functions in the construction of orthonormal SU(3) states.
 The mean values of the $K$ quantum numbers, 
\beq \langle K\rangle_{\alpha L} 
= \sum_K K\, |\mc{K}^{L}_{K\alpha}|^2 , \label{eq:6.Kave} \eeq
where $\{ \mc{K}^{L}_{K\alpha}\}$ are the coefficients in the expansion (\ref{eq:Kmat}) for a range of SU(3) eigenstates of 
$\hat {\mathcal{Z}}^{(\lambda\mu)}$ are shown in 
Table \ref{tab:6.Kqno} for two SU(3) representations, 
$(70\,6)$ and $(70\,7)$, typical of those appropriate for rare-earth nuclei and one, $(10\,4)$, that is more typical for a light nucleus.
They are seen to be  astonishingly close to corresponding integer rotor-model values with the exception of those for a few states
in the upper angular-momentum reaches of the finite-dimensional SU(3) bands.
Also shown are the quadrupole moments of these states
in comparison with their rotor-model values.

It will be noted that the rotor contraction limit is expected to be valid when both $\lambda/ L$ and $\mu/L$ take large values.
However, assuming that $\lambda \geq \mu$, the asymptotic limit only requires  $\lambda/L$ to be large.
Results are shown for both even and odd values of $\mu$ because
 it is known that in the  rotor-model description of a $K=1$ band, which occurs only in SU(3) for odd $\mu$, the matrix elements of the quadrupole operators connect intrinsic states with $K=\pm 1$ and, as a consequence, show different properties to those of other bands.
It is in fact necessary to include the cross-over intrinsic terms in both the precise and asymptotic limits to get the good agreement 
shown in  Table \ref{tab:6.Kqno}.

Results are shown for large values of $\lambda$ 
and  small values of $\mu$, typical of the values appropriate for applications in rare-earth nuclei.
They are also shown for a $(10\,4)$ representation for which 
one would not expect the results of the contraction limit to be accurate
for any but the lowest values of $L$.
Thus, it is quite remarkable to discover the  extent to which 
$K$ remains a good quantum number for this representation until the point at which the bands reach their upper bounds; i.e., for 
$L\leq10$ for the $(10\,4)$ representation.

\begin{table}[pth]
\caption{\label{tab:6.Kqno} 
Comparison of the $K$ quantum number and diagonal quadrupole matrix element 
$ \langle (\lambda\mu)\alpha L \| \hat{\mathcal{Q}}_2 
\|(\lambda\mu)\alpha L\rangle$ computed precisely and in their rotor-model limits.
The $K$ quantum number is denoted by 
$K^{(\rm rot)}_{\alpha L}$ in the rotor-model limit for a state 
$|(\lambda\mu)\alpha LM\rangle$ and its mean value in the corresponding SU(3) state is denoted by
$\langle K\rangle_{\alpha L}$.
The quadrupole matrix elements
are denoted by $\langle \hat Q\rangle^{({\rm rot})}_{\alpha L}$
for the contraction (rotor) limit given by equation  (\ref{eq:6.KLtoKL}), 
and by $\langle \hat Q\rangle^{({\rm su3})}_{\alpha L}$
when computed precisely.  }\vspace{0.2cm}
$ \begin{array}{|c|c|c|r|r|c|c|} \hline
(\lambda\;\mu)& \alpha &L& 
\langle \hat Q\rangle^{({\rm rot.})}_{\alpha L}\;
& 
\langle \hat Q\rangle^{({\rm su3})}_{\alpha L}\;
& K^{({\rm rot})}_{\alpha L} &  \langle K\rangle_{\alpha L}\;
\\   \hline
(70\, 6) & 1  & 0 & 0   \;\;\;\;\;\;  & 0 \;\;\;\;\;\;  &  0    &  0.000\\
             & 2  & 2 & 178.089     &     178.089 &  2    & 2.000\\
             & 3 &  4 & 318.937     &     318.937 & 4     &  4.000\\
             & 4 &  6 & 425.927     &     425.927 & 6     &  6.000\\
             & 1 &  30 &-582.098 & -582.193  &0        & 0.000\\
             & 2 &  30 &-574.587 &-574.609  & 2        & 2.000\\
             & 3 &  30 &-552.055 &-552.050  & 4        & 4.000\\
             & 4 &  30 &-514.500 &-514.388  & 6        & 6.000\\
  \hline
(70\, 7) & 1 & 1 &     95.304 &     95.304 & 1   &  1.000\\
             & 2 & 3 &   256.174 &  256.174 & 3   &  3.000\\
             & 3 & 5 &   377.874 &  377.874 & 5   &  5.000\\
             & 4 & 7 &   475.213 &  475.213 & 7   &  7.000\\
             & 1 & 31 & -546.077 &  -545.784 & 1   &  1.000\\
             & 2 & 31 & -579.311 &  -545.784 & 3   &  3.000\\
             & 3 & 31 & -550.495 &  -550.462 & 5   &  5.000\\
             & 4 & 31 & -507.272 &  -507.108 & 7   &  7.000\\
\hline
(10\, 4)  & 1 & 2 &  -32.271 &  -32.281  & 0  & 0.000\\
             & 2 & 2 &    32.271 &   32.281 &  2  & 2.000\\
             & 3 & 4 &    57.794 &   57.811 &  4  & 4.000\\
             & 1 & 10 & -62.077   &  - 63.297  & 0   &  0.003\\
             & 2 & 10 & -55.305   &   -54.896  & 2   &  2.001\\
             & 3 & 10 & -34.989    &  -34.177  &  4   &  3.996\\
             & 1 & 12 & -67.663     &  -39.409 &  -   &  1.012\\
             & 2 & 12 & -62.458    &  -44.447  & -    & 3.872\\
             & 1 & 14 & -72.830    &  -35.837  & -    & 2.010\\
\hline   
\end{array}$%
\end{table}

\subsection{\boldmath Comparison of asymptotic and exact SU(3) matrix elements} \label{sect:Rotor_SU3.comparison}
The results of Table \ref{tab:6.Kqno} show that, when orthonormal basis states for an  SU(3) irrep are given by eigenstates of the Hermitian operator $\hat Z^{(\lambda\mu)}$,
the $K$ quantum numbers are defined and rapidly approach the integer values matching those of the rotor model for even modestly large values of $\lambda/ L $.
Thus, one obtains asymptotic limits, that should be more accurate than the contraction limits, from the observation that the SU(3) wave functions are given accurately to within normalisation factors by good $K$ rotor-model wave functions.  
The asymptotic limits that result are given by equations 
(\ref{eq:ASlimits})-(\ref{eq:6,Delta2}).
In fact, the contraction and  asymptotic limits of 
 the diagonal matrix elements
$ \langle (\lambda\mu)K L \| \hat{\mathcal{Q}}_2 
\|(\lambda\mu)K L\rangle$ coincide
and both are given by equation  (\ref{eq:6.KLtoKL}).

Transition matrix elements, given by the rotor contraction and 
the higher-order asymptotic expressions 
(\ref{eq:6.KLtoKL+1})-(\ref{eq:6,Delta2}),
are compared with accurately computed matrix elements
in Table \ref{tab:6.QME}, again for the irreps $(70\,6)$, $(70\,7)$ and $(10\,4)$.
For the large in-band transitions, the matrix element of the rotor contraction limit are seen to be remarkably accurate and the higher-order asymptotic limit are as good as exact for all matrix elements shown with the only exceptions being for the $\Delta L = 1$ in-band transitions of the $K=1$ band; note the comparisons for the 
$L=29, K=1$ to $L=30, K=1$ matrix elements
for the $(70\,7)$ representation.

$K=1$ bands are special both in SU(3) and in the rotor model
because it is only for such a band that there is a matrix element of a quadrupole operator between the $K=\pm1$ intrinsic states; it is given by the $\Delta^{L'L}$ terms in equations
(\ref{eq:ASlimits})-(\ref{eq:6,Delta2}).
Thus, for an in-band $K=1$ matrix element there are two contributing terms which for a $\Delta L =1$ matrix element are of opposite sign.
In a situation in which they of near equal magnitude, the error in their difference, as given by their asymptotic values, can then be relatively large.
This cancellation of terms of opposite sign is exhibited for
$\Delta L = 1$ in-band $K=1$ transition matrix elements  in more detail in the lower part of Table \ref{tab:6.QME}.
Note that, in contrast, the two terms contributing to $\Delta L=2$ transition matrix between $K=1$ states are of the same sign and their sum is large and given to a high level of accuracy.

There is also a second reason for larger differences between the precise and asymptotic values of interband transition matrix elements for SU(3) irreps with $\mu \ll \lambda$.
It arises because, although the mixing of states of different 
integer-$K$ values is exceedingly small in appropriately defined SU(3) states, the huge difference in the magnitudes of the $K$-conserving matrix elements relative to the $K$-changing matrix elements, means that the mixing has  an exaggerated effect on the magnitude of the interband matrix elements.
Thus, in general, the smaller inter-band matrix elements are expected to be given considerably less accurately by their asymptotic limits than the large matrix elements as is seen to be the case.

\begin{table*}[p]
\caption{\label{tab:6.QME}
Selected SU(3) quadrupole matrix elements 
$\langle (\lambda\mu)\alpha_f L_f \| \hat{\mathcal{Q}}_2 
\|(\lambda\mu)\alpha_i L_i\rangle$ 
calculated in the rotor contraction (RC) and the $\lambda/L\to\infty$ asymptotic limit (AL) compared with precisely computed values for three irreps.
The matrix elements are denoted by
$\langle \hat Q\rangle^{({\rm RC})}_{fi}$, 
when evaluated in the rotor contraction limit, by 
$\langle \hat Q\rangle^{({\rm AL})}_{fi}$
when evaluated in the higher-order asymptotic limit, and by 
$\langle \hat Q\rangle^{({\rm su3})}_{fi}$
when evaluated precisely. 
Matrix elements that do not satisfy the hermiticity relationships of a unitary representations are shown as asterisks.
 (A similar comparison  of such matrix elements for $(32\, 5)$ and $(10\, 4)$ representations was given in Ref.\ \cite{RoweT08}.)
} \vspace{0.2cm}
$ \begin{array}{|c|c|c|c|c|c|r|r|r| } \hline
(\lambda\;\mu) &K^{({\rm rot})}_{\alpha_f L_f} & L_f 
& K^{({\rm rot})}_{\alpha_i L_i} & L_i &L_f/\lambda
&\langle \hat Q\rangle^{({\rm RC})}_{fi} 
& \langle \hat Q\rangle^{({\rm AL})}_{fi} 
& \langle \hat Q\rangle^{({\rm su3})}_{fi}   \\ 
\hline
(70\; 6)& 0 & 10 & 0 & 8 & 0.14 & 397.170 & 393.928 & 393.919  \\
           & 0 & 10 & 2 & 8 & 0.14 & 10.407   & 10.301   & 10.385  \\
           & 2 & 10 & 2 & 8 & 0.14 & 379.415 & 376.318 & 376.311  \\
           & 2 & 10 & 4 & 8 & 0.14 & 4.249     & 3.839     & 3.867  \\
           & 4 & 10 & 6 & 8 & 0.14 & 1.967     & 1.376     & 1.385 \\
           & 0 & 30 & 2 & 29 & 0.43 & 45.365  & 44.900 & 45.808 \\  
           & 2 & 30 & 2 & 29 & 0.43 & 94.078  & 92.152 & 92.111 \\   
           & 2 & 30 & 4 & 29 & 0.43 & 29.730  & 26.861 & 27.402 \\ 
           & 4 & 30 & 6 & 29 & 0.43 & 27.276  & 19.089 & 19.460 \\ 
           & 0 & 30 & 0 & 28 & 0.43 & 700.754 & 643.476 & 643.290 \\
           & 0 & 30 & 2 & 28 & 0.43 &   21.700 &   21.478 & 23.425 \\
           & 2 & 30 & 2 & 28 & 0.43 & 697.530 & 640.516 &640.379 \\
           & 2 & 30 & 4 & 28 & 0.43 &   13.263 &   11.983 & 12.996 \\
           & 4 & 30 & 6 & 28 & 0.43 &   11.333 &     7.932 & 8.593 \\ 
\hline%
(70\;7)
           & 1 & 20 & 1 & 19 & 0.29 & 27.111 & 25.981 & 26.340 \\
           & 1 & 22 & 1 & 21 & 0.31 & 22.895 & 21.405 & 21.984 \\
           & 1 & 24 & 1 & 23 & 0.34 & 19.092 & 17.103 & 18.032 \\
           & 1 & 30 & 1 & 29 & 0.43 &   9.487  &   0.000 & 7.948 \\
           & 1 & 30 & 3 & 29 & 0.43 & 35.336  & 34.214 & 34.271 \\
           & 3 & 30 & 3 & 29 & 0.43 &141.668 &138.806 & 138.967 \\
           & 3 & 30 & 5 & 29 & 0.43 &  32.588 &  28.222 & 28.799 \\
           & 1 & 30 & 1 & 28 & 0.43 & 685.855 & 627.359 & 627.034 \\
           & 1 & 30 & 3 & 28 & 0.43 &   16.325 &   15.807 & 16.493 \\
           & 3 & 30 & 3 & 28 & 0.43 & 698.155 & 641.881 & 641.850 \\
           & 3 & 30 & 5 & 28 & 0.43 &     14.03 &   12.153 & 13.173 \\
\hline%
(10\;4) 
            & 0& 2 & 0 & 0 & 0.00 & 27.000 & 26.833 & 26.823  \\
            & 2& 2 & 0 & 0 & 0.20 &   8.660 &   8.485 & 8.516  \\
            & 0& 4 & 0 & 2 & 0.40 &  43.296 & 41.816 &  41.743  \\
           & 2 & 4 & 0 & 2 & 0.40 &    8.964 & 8.783   &   9.081 \\
           & 4 & 4 & 0 & 2 & 0.40 &    0        & 0          &   0.003  \\
           & 0 & 8 & 0 & 6 & 0.80 & 63.894 & 53.126 & 52.406  \\
           & 2 & 8 & 0 & 6 & 0.80 & 10.607 & 10.392 & 12.629\; \\
           & 4 & 8 & 0 & 6 & 0.80 &   0        &   0        &  0.158 \\
           & 2 & 8 & 2 & 6 & 0.80 & 59.286 & 49.295 &  48.911 \\
\hline%
(70\;7)
           & 1 & 26 & 1 & 25 & 0.37 & 15.625 &  12.890 & 14.409 \\
           & 1 & 28 & 1 & 27 & 0.40 & 12.438 &    8.409 & 11.062 \\
           & 1 & 30 & 1 & 29 & 0.43 &   9.487 &    0.000 & 7.948 \\
           & 1 & 32 & 1 & 31 & 0.46 &   6.736 &   
           * \;\;\;\;\;\;  & 5.036  \\         
           & 1 & 34 & 1 & 33 & 0.49 &   4.159 &   
           *\;\;\;\;\;\; & -2.304  \\
           & 1 & 36 & 1 & 35 & 0.51 &   1.732 &  
           *\;\;\;\;\;\; &- 0.266 \\
           & 1 & 38 & 1 & 37 & 0.54 &  -0.562 &  
           *\;\;\;\;\;\; & -2.684\\           
\hline
   \end{array}$%
\end{table*}

\section{\boldmath Representations of Sp$(3,\Rb)$ in a U(3) basis} 
\label{sect:Sp3Rreps}
As defined in Section \ref{sect:Sp3Rmodel}, the Sp$(3,\Rb)$ 
Lie algebra comprises elements of a U(3) subalgebra and 
giant-monopole and giant-quadrupole raising and lowering operators.
Thus,  an irreducible Sp$(3,\Rb)$ representation is defined by a set of 
so-called lowest-grade states that are annihilated by the giant-resonance lowering operators and are a basis for an irreducible U(3) representation.
Basis states for the Sp$(3,\Rb)$ representation are then generated by acting 
repeatedly on these lowest-grade U(3) states with the giant-resonance raising operators $\{ \hat {\mathcal A}_{ij}\}$.
Matrix elements for such a representation were first derived by use of recursion relations \cite{Rosensteel80} and subsequently by VCS methods  \cite {RoweRC84,Rowe84}.

This section outlines the VCS 
construction. 
It starts with a 
 basis of lowest-grade 
states $\{|w_0\alpha\rangle\}$ for the desired Sp$(3,\Rb)$ 
representation.  These states transform under U(3)
as a  set of harmonic-oscillator boson vacuum states  $\{|w_0\alpha)\}$
with $z$-independent  coherent-state wave functions $\{ \xi^{w_0}_\alpha\}$.
States of  the Sp$(3,\Rb)$ representation are then defined with 
VCS 
wave functions 
\beq \Psi(z) =\sum_\alpha  \xi^{w_0}_\alpha 
\langle w_0 \alpha |\exp \left[ \tfrac12 
{\textstyle\sum}_{ij} z_{ij} \hat {\mathcal B}_{ij}\right] |\psi\rangle , 
\label{eq:7.VCSwfn}\eeq
where $z=\{ z_{ij}= z_{ji}\}$ is a set of 6 complex variables.

Section \ref{sect:Sp3R_VCS}  shows that, in this VCS representation,
 the elements of the Sp$(3,\Rb)$ Lie algebra, as defined in Section \ref{sect:SU3<Sp3}, are realised as operators of the form
\begin{subequations}\label{eq:8.VCS}
\bal
&\hat\Gamma({\mathcal{C}}_{ij}) =\hat \Cb_{ij} + (z\nabla)_{ij} ,\\
&\hat\Gamma({\mathcal{B}}_{ij}) = \nabla_{ij} ,\\
&\hat\Gamma({\mathcal{A}}_{ij}) = [ \hat\Lambda,z_{ij}],
\end{align}
\end{subequations}
where ${\hat\Cb}_{ij}$ is the representation of the U(3) element 
$\hat{\mc{C}}_{ij}$ on the space of lowest-grade states,
$\{\nabla_{ij} =  (1+\delta_{i,j}){\partial}/{\partial z_{ij}}\}$
and $\{ z_{kl}\}$  satisfy the  commutation relations
\beq   [\nabla_{ij}, z_{kl} ] 
=  \delta_{i,k} \delta_{j,l} + \delta_{i,l} \delta_{j,k},  \eeq
$\hat\Lambda$ is a U(3) scalar operator
\beq \hat\Lambda 
= \tfrac12 ({\hat\Cb}+z\nabla) \cdot  ({\hat\Cb}+z\nabla)
-\tfrac14 z\nabla \cdot z\nabla - z\cdot \nabla 
\label{eq:8.VCS_A2}\eeq
 that is diagonal in a U(3)-coupled basis for the representation,
$z\nabla$ is a tensor with components 
$(z\nabla)_{ij}  = \sum_k z_{ik}\nabla_{kj}$ 
and $z\cdot\nabla= \sum_{i} (z\nabla)_{ii}$.

The six variables $z_{ij}$ and the corresponding derivative operators
$\nabla_{ij}$ in these expressions can be regarded as coherent-state representations of boson operators,
$a^\dag_{ij}$ and $a_{ij}$, with commutation relations
\beq [a_{ij}, a^\dag_{kl} ] 
=  \delta_{i,k} \delta_{j,l} + \delta_{i,l} \delta_{j,k} . 
\label{eq:aijbosonops}\eeq
The linear combinations 
\begin{subequations}
\bal
&s^\dag = \sqrt{\tfrac16}\sum_i a^\dag_{ii} , 
\quad s = \sqrt{\tfrac16}\sum_i a_{ii} \\
&d^\dag_{0}  
   = \sqrt{\tfrac{1}{12}}\, (2 a^\dag_{11} - a^\dag_{22} - a^\dag_{33}),\\
& d_{0} 
 = \sqrt{\tfrac{1}{12}}\, (2a_{11} - a_{22} -a_{33}),  \\
&d^\dag_{\pm 1} 
  =\mp \sqrt{\tfrac{1}{2}}\, (a^\dag_{12} \pm {\rm i} a^\dag_{13}),\\
&d_{\pm 1}
 =\mp \sqrt{\tfrac{1}{2}}\, a_{12} \mp {\rm i} a_{13}), \\
& d^\dag_{\pm 2}   = \sqrt{\tfrac{1}{8}}\, 
  (a^\dag_{22} - a^\dag_{33}  \pm 2{\rm i} a^\dag_{23}) ,  \\
& d_{\pm 2}  = \sqrt{\tfrac{1}{8}}\, (a_{22} - a_{33}   \mp 2{\rm i} a_{23}) , 
\end{align}
\end{subequations}
are then interpreted as  the $s$ ($L=0$) and $d$ ($L=2$) boson 
raising and lowering operators of a 
Bohr model of monopole and quadrupole vibrations. 
The boson operators are also recognised 
as components of a U(3) $\{ 2\,0\,0\}$ tensor.
Thus, the VCS representation (\ref{eq:8.VCS}a)-(\ref{eq:8.VCS}c) of the Sp$(3,\Rb)$ Lie algebra is understood in terms of  representations of two simpler commuting algebras:  the lowest-grade representation of U(3) and the representation of the boson algebra of the  Bohr model.
The combination of these two algebras is known as a U(3)-boson algebra \cite{RosensteelR81}.

The operators of equation (\ref{eq:8.VCS}) define a representation of the Sp$(3,\Rb)$ Lie algebra in terms of a  U(3)-boson algebra in which a set of  boson vacuum state
$\{ |w_0 \alpha)\}$, which satisfy the equations 
\beq  a_{ij} |w_0 \alpha) = 0 , \quad i,j=1,2,3 ,\eeq
are the lowest-grade states of a set of U(3)-coupled states
\beq \{ |w_0 n \rho w \alpha) 
= [ P_n(a^\dag) \otimes |w_0) ]_{\rho w \alpha} \} ,  
\label{eq:8,U3bosonbasis} \eeq
in which 
$ P_{n}(a^\dag)\propto 
[a^\dag \otimes a^\dag \otimes\cdots a^\dag ]_{n\alpha}$  
is a U(3) tensor with components 
that transform as a basis for a U(3) irrep
$n = \{ n_1\, n_2\, n_3\}$ for which $n_1\geq n_2\geq n_3 \geq 0$ 
are even integers and
$\rho$ indexes the multiplicity of the  U(3) irrep 
$\{w\}$ in the tensor product $\{n\}\otimes \{w_0\}$.

In this Sp$(3,\Rb)$ representation the matrix elements of 
the U(3) operators $\hat\Gamma({\mathcal{C}}_{ij})$ are
identical to those of a standard unitary representation of U(3) in an
orthonormal U(3)-coupled basis and the matrix elements of 
$\hat \Lambda$ are easily determined; 
see equation (\ref{eq:7.Omega3}).
Thus, it only remains  to determine the SU(3)-reduced matrix elements
$(w_0 n' \rho' w' \trib a^\dag \trib w_0 n \rho w)$ and 
 the related matrix elements
$(w_0 n \rho w \trib a \trib w_0 n' \rho' w')$.
It is shown in section \ref{sect:U3bosonreps}
that
\bal 
&(w_0 n' \rho' w' \trib a^\dag \trib w_0 n \rho w) 
= ({n'}\trib \hat a^\dag \trib n)    \label{eq:8.U3bosonrme} \\
&\times U((\lambda_0\mu_0) (\lambda_n\mu_n)  (\lambda_{ w'}\mu_{ w'}) (2\,0);
\rho(\lambda_ w\mu_ w) \, \rho'(\lambda_{n'}\mu_{n'})) , \nonumber
\end{align}
where $U((\lambda_0\mu_0) (\lambda_n\mu_n)  (\lambda_{ w'}\mu_{ w'}) (2\,0);
\rho(\lambda_ w\mu_ w) \, \rho'(\lambda_{n'}\mu_{n'}))$ is an SU(3) Racah coefficient with
\begin{subequations}  \label{eq:su3qnos}
\bal
&\lambda_n = n_1-n_2, && \mu_n = n_2-n_3 ,\\
&\lambda_w = w_1-w_2, & &\mu_w = w_2-w_3 ,
\end{align}
\end{subequations}
and
\bal
({n'} \trib a^\dag& \trib n) = 
\delta_{n'_1,n_1+2}\delta_{n'_2,n_2} \delta_{n'_3,n_3} 
 \nonumber \\
& \times \left( \frac{(n_1+4)(n_1-n_2+2)(n_1-n_3+3)}
{2(n_1-n_2+3)(n_1-n_3+4)}\right)^{\frac12}  \nonumber \\
 &+ \delta_{n'_1,n_1} \delta_{n'_2,n_2+2} \delta_{n'_3,n_3}   
\nonumber \\
&\times  \left( \frac{(n_2+3)(n_1-n_2)(n_2-n_3+2)}
{2(n_1-n_2-1)(n_2-n_3+3)}\right)^{\frac12} \nonumber\\
&+\delta_{n'_1,n_1} \delta_{n'_2,n_2}\delta_{n'_3,n_3+2}   \nonumber\\
&\times \left( \frac{(n_3+2)(n_2-n_3)(n_1-n_3+1)}
{2(n_1-n_3)(n_2-n_3-1)}\right)^{\frac12}  .  \label{eq:8.bosonmes}
\end{align}

The above equations satisfy the  Hermiticity requirements of a unitary representation of the U(3)-boson algebra.
They also give explicit expressions of the matrix elements of the VCS operators of equations 
(\ref{eq:8.VCS}a)-(\ref{eq:8.VCS}c)  in this U(3)-boson basis.
 However, the representations of the Sp$(3,\Rb)$ algebra obtained in this way do not satisfy  the Hermiticity requirements of a unitary representation.  
 This is because  the orthonormal U(3)-boson basis is not an orthonormal basis for  a unitary representation of the Sp$(3,\Rb)$ Lie algebra.
 This is to be expected because,  with respect to the boson inner product, 
 the Hermitian adjoint of $\nabla_{ij}$ 
 is $\nabla_{ij}^\dag = z_{ij}$ whereas 
 $\hat\Gamma({\mathcal{B}}_{ij})^\dag = z_{ij} 
\not=\hat\Gamma({\mathcal{A}}_{ij})$.
Thus, a transformation to an appropriate orthonormal basis is 
required and, as reviewed in Section \ref{sect:Kmat_theory}, it is
is obtained by a K-matrix algorithm.

Special cases are the multiplicity-free representations founded on one-dimensional U(3) lowest-grade representations.
These are the representations corresponding to multiple giant-resonance excitations of a  spherical-harmonic oscillator  closed-shell nucleus 
which relates to a 6-dimensional Bohr model with only vibrational degrees of freedom \cite{CastanosCM84}.
Analytical expressions are obtained for the matrix elements of these representations in Section \ref{sect:MFreps}.

 \subsection{\boldmath VCS  representation of the Sp$(3,\Rb)$ Lie algebra}
 \label{sect:Sp3R_VCS}
The Hilbert space $\Hb^{w_0}$ of an irreducible Sp$(3,\Rb)$ representation is  spanned by sets of coherent states
\beq  
\big\{ |w_0 \alpha ; z\rangle = 
\exp \big( \tfrac12 \sum_{ij} z^*_{ij} \hat {\mathcal{A}}_{ij}\big) |w_0\alpha\rangle \big\}, 
\eeq
where $\{ |{w_0}\alpha \rangle\}$ is a basis of lowest-grade states
for a U(3) representation $w_0$ and $\{ z_{ij}\}$ is a set of complex variables.
A state vector $|\psi\rangle\in\Hb^{w_0}$ is then characterised by the set of overlap functions
\beq \big\{\Psi_\alpha(z) =  \langle w_0 \alpha ; z|\psi\rangle
=  \langle w_0 \alpha | e^{\hat Z(z)} |\psi\rangle \big\} , 
\eeq
with  
$\hat Z(z)  = \tfrac12 \sum_{ij} z_{ij} \hat {\mathcal{B}}_{ij}$.
In a VCS representation, these overlaps are regarded as  components of the  vector-valued  wave function $\Psi$  
defined by  equation (\ref{eq:7.VCSwfn}).

 The representation of state vectors in $ \Hb^{w_0}$ by
VCS wave functions  also defines a representation 
$\hat X\to \hat\Gamma (X)$ of  a linear operator  $\hat X$ on state vectors by
\beq \hat\Gamma(X)\Psi(z)  =   \sum_\alpha \xi^{w_0}_\alpha  
\langle w_0 \alpha | e^{\hat Z(z)}\hat X |\psi\rangle .\label{eq:VCSops}
\eeq
A special case is when $\hat X= \hat {\cal C}_{ij}$ and 
$|\psi\rangle = |w_0\alpha\rangle$ is a lowest-grade state.
In this situation, 
\beq \hat\Gamma({\cal C}_{ij}) \xi^{w_0}_\alpha 
= \hat \Cb_{ij} \xi^{w_0}_\alpha ,\eeq
where 
\bal \hat \Cb_{ij} \xi^{w_0}_\alpha &= 
\sum_\beta \xi^{w_0}_\beta  
\langle w_0 \beta | e^{\hat Z(z)}\hat {\cal C}_{ij} |w_0\alpha\rangle   
\nonumber\\
&= \sum_\beta \xi^{w_0}_\beta  
\langle w_0 \beta| \hat {\cal C}_{ij} |w_0\alpha\rangle . \label{eq:7.Cbij}
\end{align}
With the definition (\ref{eq:VCSops}), the VCS  operators representing  elements  of the 
Sp$(3,\Rb)$ Lie algebra are  determined by use of the  expansion
\beq e^{\hat Z} \hat X = \big( \hat X + [ \hat Z,\hat X] 
+ \tfrac{1}{2!} [\hat Z, [\hat Z,\hat X]] +
 \cdots \big) e^{\hat Z} ,  \label{eq:7.BCH}\eeq
which, because $\hat Z$ is a sum of Sp$(3,\Rb)$ lowering operators,  terminates at or before the third  term.
The identities
\begin{subequations}
\bal
& \sum_\alpha \xi^{\{w_0\}}_\alpha  
\langle w_0 \alpha |\hat{\cal C}_{ij} e^{\hat Z(z)} |\psi\rangle
= \hat \Cb_{ij}\Psi(z) , \\
&\sum_\alpha \xi^{\{w_0\}}_\alpha  
\langle w_0 \alpha |\hat{\mathcal{B}} _{ij} e^{\hat Z(z)} |\psi\rangle
=\nabla_{ij}\Psi(z) ,\\
&\sum_\alpha \xi^{\{w_0\}}_\alpha  
\langle w_0 \alpha |\hat{\mathcal{A}} _{ij} e^{\hat Z(z)} |\psi\rangle
=0 ,
\end{align}
\end{subequations}
and the observation that 
$ \hat{\mathcal{B}} _{ij} e^{\hat Z(z)} =\nabla_{ij} e^{\hat Z(z)}$
then leads to the VCS representation 
\begin{subequations}\label{eq:7.VCS}
\bal
&\hat\Gamma({\mathcal{C}}_{ij}) =\hat \Cb_{ij} + (z\nabla)_{ij} , \\
&\hat\Gamma({\mathcal{B}}_{ij}) = \nabla_{ij} ,  \\
&\hat\Gamma({\mathcal{A}}_{ij}) = (\hat \Cb z)_{ij} + (\hat \Cb z)_{ji} +
 (z\nabla z)_{ij} -4 z_{ij} .
\end{align}
\end{subequations}
The expressions $\hat\Gamma({\mathcal{C}} _{ij})$ and $\hat\Gamma({\mathcal{B}}_{ij})$  are  simple.  
The operator $\hat\Gamma({\mathcal{A}}_{ij})$ is not so simple.  
However, because $\hat {\mathcal{A}}_{ij}$ and $z_{ij}$ are both components of U(3) 
$\{ 2\,0\,0\}$ tensors, it follows that, if
$\hat\Gamma({\mathcal{A}}_{ij})$ is expressed as a commutator
\beq \hat\Gamma({\mathcal{A}}_{ij}) = [ \hat\Lambda,z_{ij}] , 
\label{eq:7.ALamz}\eeq
the operator $\hat\Lambda$ must be a U(3) scalar operator.  
A U(3) scalar $\hat\Lambda$ operator that satisfies these requirements is given (to within terms that commute with $z_{ij}$ by equation (\ref{eq:8.VCS_A2}).

\subsection{\boldmath Representations  of the U(3)-boson algebra}
\label{sect:U3bosonreps}
The operators $\{ a^\dag_{ij}, a_{ij}\}$ of equation  (\ref{eq:aijbosonops}),  when expressed as linear combinations of $s$ and $d$ boson operators, are seen as components of a U(3)  $\{2\,0\,0\}$ tensor
with highest-weight component $a^\dag_{11}/\sqrt{2}$.
Orthonormal U(3)-coupled basis states for this harmonic oscillator
are then expressible in the form
\beq \{ |n\alpha ) = P_{n\alpha}(a^\dag) |0) \} , \label{eq:8.59}\eeq
where $|0)$ is the boson vacuum state and 
$P_{n\alpha}(a^\dag)$ is a polynomial in the boson creation operators.  
However, their inner products are expressed with round brackets
\beq (n \alpha | n'\beta) = (0|P_{n\alpha}(a) P_{n'\beta} (a^\dag) |0) =
\delta_{n,n'} \delta_{\alpha,\beta} \eeq
 to distinguish them from the inner products for an orthonormal basis for the related states of an Sp$(3,\Rb)$ representation with angle brackets.
The required polynomials are U(3)-coupled tensor products
\beq P_{n\alpha}(a^\dag) \propto[ a^\dag \otimes \dots \otimes a^\dag]_{n\alpha} , \eeq
where $n = \{ n_1\,n_2\,n_3\}$ is a triple of even integers with $n_1\geq n_2 \geq n_3 \geq 0$.
SU(3)-reduced matrix elements $({n'} \trib a^\dag \trib n)$
 of the boson operators in the extended Bohr model
have been derived  \cite{Quesne81,RosensteelR83,LeBlancR87} 
and are given by equation (\ref{eq:8.bosonmes}).

The extension of the  6-dimensional harmonic-oscillator space
to the space of  an irreducible representation of the U(3)-boson algebra is  achieved by replacing the single vacuum state of the oscillator with a set of vacuum states that are in one-to-one correspondence with the lowest-grade U(3) states of the desired 
Sp$(3,\Rb)$ representation.
An orthonormal basis for the U(3)-boson model is then given by the U(3)-coupled states
\beq \{ |w_0 n \rho w \alpha) 
= [ P_n(a^\dag) \otimes |w_0) ]_{\rho w \alpha}  \} , \eeq
where $\{|w_0\alpha\rangle\}$ is 
a basis of boson vacuum states 
for a 
U(3) irrep $\{w_0 \} \equiv \{ N_0 (\lambda_0\mu_0)\}$ 
that is isomorphic
to the lowest-grade U(3)  irrep of the
Sp$(3,\Rb)$ irrep $\langle w_0 \rangle \equiv \langle N_0 (\lambda_0\mu_0)\rangle$; 
$\rho$ indexes the multiplicity of the U(3) irrep $\{w\}$ in the tensor product $\{n\}\otimes \{w_0\}$.
One then obtains the SU(3)-reduced matrix elements
(\ref{eq:8.U3bosonrme}) of the boson operators
 by standard Racah recoupling methods. 

Matrix elements of operators in the SU(3) Lie algebra are determined in the U(3)-boson representation, as in any representation with an SU(3)-coupled basis, from the reduced matrix elements of the SU(3) tensor with components
\beq
\hat X^{(1\,1)}_{lm} 
= \delta_{l,2} \hat{\mathcal{Q}}_{2,m} + \delta_{l,1}\sqrt{3}\,  L_m .
\eeq
SU(3)-reduced matrix elements of this tensor are obtained from the observation that its scalar product  
\beq   \hat X^{(1\,1)}\cdot \hat X^{(1\,1)} = 
 \sum_q \big(\hat X^{(1\,1)}_q\big)^\dag \hat  X^{(1\,1)}_q ,\eeq
 is the  SU(3) Casimir invariant
 \beq 
\hat {\mathcal I}_2 
=\hat {\mathcal{Q}} \cdot \hat{\mathcal{Q}}+ 3\hat L\cdot\hat L ,
 \eeq
 which, for an SU(3) representation $(\lambda\mu)$, has values 
  \beq 
  {\mathcal I}_2(\lambda\mu) =  
  4 (\lambda^2+\lambda\mu+\mu^2 + 3\lambda+3\mu) .
  \eeq
It follows that
\bal | \langle\lambda'\mu' \trib \hat X^{(1\,1)} 
&\trib \lambda\mu \rangle |^2
 = \delta_{\lambda',\lambda} \delta_{\mu',\mu} \nonumber\\
&\times 4 (\lambda^2+\lambda\mu+\mu^2 + 3\lambda+3\mu)
 . \label{eq:106}
\end{align}
The sign of the matrix element 
\beq \langle\lambda\mu \trib \hat X^{(1\,1)} \trib \lambda\mu \rangle 
= \pm 2 (\lambda^2+\lambda\mu+\mu^2 + 3\lambda+3\mu)^{\frac12} \eeq
is undetermined by this expression but is fixed by the phase convention used in the determination of SU(3) Clebsch-Gordan coefficients to be the  same  as that of the coefficient 
$\big( (\lambda\mu) \alpha L, (11) 1 \| (\lambda\mu) \alpha L\big)$.

Because the operator $\hat\Lambda$ appearing in equation
(\ref{eq:7.ALamz})  is diagonal in the  U(3)-boson  basis, 
\beq \hat\Lambda |w_0 n \rho w \nu) 
= \Omega_{n w} |w_0 n \rho w \nu) , \eeq
and has eigenvalues given by 
\beq \Omega_{n w} = \tfrac14 \sum_{j=1}^3
\big[2 w_j^2-n_j^2 + 8( w_j-n_j) - 2j(2 w_j-n_j) \big] ,
\label{eq:7.Omega3}
\eeq
the matrix elements of the VCS representations of the Sp$(3,\Rb)$ algebra have the explicit expressions
\begin{subequations} \label{eq:bGamma}
\bal
&(w_0 n \rho w \trib \hat\Gamma (\mathcal{C}) \trib w_0 n' \rho' w') 
=\pm 2 \delta_{n,n'} \delta_{\rho,\rho'}\delta_{w,w'}  \nonumber\\
& \hspace{2.5cm} \times
(\lambda^2+\lambda\mu+\mu^2 + 3\lambda+3\mu)^{\frac12} , \\
&(w_0 n \rho w \trib \hat\Gamma (\mathcal{B}) \trib w_0 n' \rho' w') 
 \nonumber\\
& \hspace{3.2cm} =(w_0 n \rho w \trib a \trib  w_0 n' \rho' w') , \\
&(w_0 n' \rho' w' \trib \hat\Gamma (\mathcal{A}) \trib w_0 n \rho w) 
 = (\Omega_{n' w'}-\Omega_{n w}) \nonumber\\
& \hspace{3.2cm}\times  (w_0 n' \rho' w' 
\trib a^\dag \trib w_0 n \rho w),  
\end{align}
\end{subequations}
in  the above-defined  U(3)-boson basis.
However, while this basis is  orthonormal with respect to the 
(round bracket) inner product of the U(3)-boson representation,
it is not orthonormal with respect to the appropriate
(angle bracket) inner product for a unitary Sp$(3,\Rb)$ representation.
 As a consequence the matrix elements  (\ref{eq:bGamma})
are inconsistent with the Hermiticity relationship 
 $\hat\Gamma(\mc{A}_{ij}) = \hat\Gamma(\mc{B}_{ij})^\dag$ that should be satisfied in an appropriate orthonormal basis.
The following subsection uses K-matrix methods to derive an appropriate orthonormal basis.
But first we consider the simple cases in which 
the multiplicity index $\rho$ is not needed
and for which analytical results are simply obtained \cite{RoweRC84}.

\subsection{\boldmath Multiplicity-free representations of Sp$(3,\Rb)$} \label{sect:MFreps}
Multiplicity-free representations are representations for which the basis states are uniquely defined by their quantum numbers without a need of multiplicity indices.
For Sp$(3,\Rb)$, the multiplicity-free representations  
$\langle w_0\rangle \equiv \langle N_0 (\lambda_0\mu_0)\rangle$ 
are those with $\lambda_0=\mu_0=0$.
The significant property of multiplicity-free  representations is that the U(3)-boson basis states are already an orthogonal set with respect to the Sp$(3,\Rb)$ inner product and it only remains to renormalise them to obtain an orthonormal basis.
It follows immediately  from the equations (\ref{eq:bGamma}) 
that the appropriate renormalisation factors should be such as to give the matrix elements in the corresponding orthonormal basis 
(with angle brackets) for these multiplicity-free representations
\begin{subequations} \label{eq:Gamma}
\bal
\langle w_0 n  w \trib \hat{\mathcal{C}} \trib w_0 n'  w' &\rangle 
= 2 \delta_{n,n'} \delta_{w,w'} \nonumber\\
&\times(\lambda^2 + \lambda\mu + \mu^2 +3\lambda + 3\mu)^{\frac12} , \\
\langle w_0 n  w \trib \hat{\mathcal{B}} \trib w_0 n'  w' &\rangle 
=(\Omega_{n' w'}-\Omega_{n w})^{\frac12} \nonumber\\
&\hspace{1.1cm} \times (w_0 n  w \trib a \trib w_0 n'  w')  , \\
\langle w_0 n'  w' \trib \hat{\mathcal{A}} \trib w_0 n  w &\rangle 
= (\Omega_{n' w'}-\Omega_{n w})^{\frac12} \nonumber\\
&\hspace{1.05cm} \times  (w_0 n'  w' \trib a^\dag \trib w_0 n  w).
\end{align}
\end{subequations}
These expressions, given here as a special simple case of the  VCS construction \cite{RoweRC84},
were first derived in an equivalent form by 
Casta{\~n}os \emph{et al.}\ \cite{CastanosCM84}.

\subsection{\boldmath K-matrix theory and generic  representations of 
Sp$(3,\Rb)$ with lowest-weight states} \label{sect:Kmat_theory}
All the Sp$(3,\Rb)$ representations  that arise in nuclear physics
have lowest-weight states and are  unitary when expressed in  appropriate orthonormal bases.  
However, as shown above, they are not unitary in  bases for which the U(3)-boson algebra is unitary.
Thus, to make use of the simple results obtained in a 
U(3)-boson model basis, an orthonormal basis of 
VCS wave functions  $\{ \Psi_\sigma\}$ is expanded,
\beq \Psi_\sigma
= \sum_s \phi_s {\mc{K}}_{s\sigma} , \label{eq:^K}\eeq
in terms of an orthonormal  basis $\{ \phi_s\}$ for a corresponding U(3)-boson model representation.
The matrix elements of $\mc{K}$, expressed as overlaps  in terms of the U(3)-boson inner product by 
\beq {\mc{K}}_{s\sigma} = (\phi_s, \Psi_\sigma ) ,\eeq
 are  determined by the K-matrix methods of \cite{Rowe84,Rowe95}.

From the expansion
\beq \hat\Gamma(X) \Psi_\sigma 
= \sum_\tau \Psi_\tau \langle\tau |\hat X |\sigma\rangle ,
\eeq
for an element $X$ of the Sp$(3,\Rb)$ Lie algebra,
we  obtain the relationship 
\beq \sum_s (\phi_t, \hat\Gamma(X) \phi_s) \mc{K}_{s\sigma}
= \sum_\tau \mc{K}_{t\tau} \langle\tau |\hat X |\sigma\rangle .
\eeq
This  and a corresponding equation for the operator $\hat X^\dag$ are
conveniently expressed as matrix equations
\beq \Gamma(X) \mc{K} = \mc{K} M(X) , \quad
\Gamma(X^\dag) \mc{K} =\mc{K} M(X^\dag). \label{eq:210}
\eeq
where
\beq \Gamma_{st}(X) = (\phi_s,\hat\Gamma(X) \phi_t) , \quad
M_{\sigma\tau} (X) = \langle \sigma| \hat X | \tau\rangle .
\eeq

We now seek a set of $\mc{K}$ matrices such that the $M$ matrices satisfy the Hermiticity relationships 
\beq M_{\sigma\tau}(X^\dag) = M_{\tau\sigma}^* (X) 
\label{eq:M(X)b} \eeq
of a unitary representation.
Multiplying the right-hand side of each equation in (\ref{eq:210}) by 
$\mc{K}^\dag$ gives the identities
\beq
\Gamma(X) S = \mc{K} M(X) \mc{K}^\dag, \quad
\Gamma(X^\dag) S =\mc{K} M(X^\dag) \mc{K}^\dag, \label{eq:210b}
\eeq
where $S$ is the Hermitian matrix
\beq S =  \mc{K} \mc{K}^\dag .
\eeq
Taking the Hermitian adjoint of the first equation of (\ref{eq:210b}) and imposing the required Hermiticity condition  (\ref{eq:M(X)b}), then gives the equation
\beq 
S\Gamma(X)^\dag = \mc{K} M(X)^\dag  \mc{K}^\dag
= \Gamma(X^\dag) S \label{eq:GW=SG}
\eeq
for $X$ in the Sp$(3,\Rb)$ Lie algebra.
This equation is equivalently expressed as an operator identity
\beq 
\hat S\hat \Gamma(X)^\dag = \hat\Gamma(X^\dag) \hat S ,
\label{eq:GW=SGb}
\eeq

A result, confirmed by this identity, is that
\beq \hat\Gamma(\mc{C}_{ij}) \hat{\mc{S}}
= \hat{\mc{S}} \hat\Gamma(\mc{C}_{ij}) ,    \eeq
for every element $\mc{C}_{ij}$ of the 
U(3) $\subset$ Sp$(3,\Rb)$ subalgebra.
Thus,  $\hat{\mc{S}}$ commutes with the VCS representation of elements of the ${\rm U(3)} \subset {\rm Sp}(3,\Rb)$ subalgebra
and it follows that the $\mc{S}$ matrices are block diagonal, i.e.,
they have no off-diagonal elements between states of inequivalent U(3) representations.
Also, if restricted to a single irreducible U(3) representation, 
${\mc{S}}$ becomes a multiple of the identity (by Schur's lemma).
This is consistent with the observation that the U(3) subrepresentations of both the Sp$(3,\Rb)$ and U(3)-boson representations are already unitary.

It is then appropriate to augment the labelling of basis states to include their U(3) quantum numbers. 
Thus, we now label basis states for the Hilbert space of the Sp$(3,\Rb)$ representation by
$\{ |\sigma w\alpha\rangle\}$, where $\alpha$ labels basis states for an irreducible U(3) representation $w$, and $\sigma$ distinguishes multiple occurences of this representation.
Basis states $\{ |sw\alpha)\}$  for the U(3)-boson space are similarity labelled.
Equation (\ref{eq:^K}) is then expressed in the form
\beq \Psi_{\sigma w \alpha} = \sum_s \phi_{sw\alpha} 
\mc{K}^w_{s \sigma} 
\eeq
and the $\mc{S}^w$ block has matrix element
\beq  \mc{S}^w_{st}
= \sum_\sigma \mc{K}^w_{s\sigma} \mc{K}^{w*}_{t\sigma} .
\eeq

The following shows that the $\mc{S}^w$ matrices can be determined recursively starting from that for the multiplicity-free lowest-grade U(3) representation, $\mc{S}^{w_0}$, which can be set equal to the identity matrix.
The $S^w$ matrices for successively higher-grade U(3)
representations are then obtained from equation  (\ref{eq:GW=SGb}) with $X= \mc{B}_{ij}$ which gives 
\beq \hat{\mc{S}} a^\dag_{ij} = [\hat\Lambda,a_{ij}^\dag] \hat{\mc{S}} \eeq
and hence
\beq  \hat{\mc{S}} a^\dag\cdot a = \sum_{ij} [\hat\Lambda,a_{ij}^\dag ]\hat{\mc{S}} a_{ij},
\label{eq:Seqn}\eeq
where $a^\dag\cdot a = \sum_{ij} a_{ij}^\dag a_{ij}$.
Taking matrix elements of both sides of this expression with
\beq {\mc{S}}^w_{ts} = (tw\alpha | \hat {\mc{S}} |sw\alpha), \eeq
and noting that $a^\dag \cdot a$ and $\hat\Lambda$ are diagonal, gives the equation
\bal {\mc{S}} ^{w'}_{st} (tw'\beta |a^\dag \cdot a |tw'\beta) =
 &\sum_{ijuvw\alpha} 
 (sw'\beta  [\hat\Lambda,a_{ij}^\dag |uw\alpha) \mc{S}^w_{uv} 
 \nonumber\\
 &\times(v w\alpha | a_{ij} |tw'\beta)  . \label{eq:220}
 \end{align}
From the observation that $[a^\dag \cdot a,a^\dag_{ij}] = 2a^\dag_{ij}$ and, because 
$a^\dag_{ij}$ increases the harmonic-oscillator energy of a state by 
$2\hbar\omega$ implying that
 $[\hat H_{\rm HO}, a^\dag_{ij}] = 2a^\dag_{ij}$,  it follows that
\beq a^\dag \cdot a\, |sw\alpha) = (N_w - N_{w_0}) |sw\alpha) ,
\eeq
where
\beq N_w = w_1+w_2+w_3 .\eeq
Equation (\ref{eq:220}) then becomes
\bal   {\mc{S}} ^{w'}_{st} = \frac{1}{N_{w'}-N_0} 
& \sum_{ijuvwr} (sw'\beta  [\hat\Lambda,a_{ij}^\dag |uw\alpha) 
 \nonumber\\
&\times  \mc{S}^w_{uv} (v w\alpha | a_{ij} |tw'\beta)
\nonumber \\
= \frac{2}{N_{w'}-N_0} 
&\sum_{w uv} \big( \Omega_{n(s)w'} - \Omega_{n(u)w} \big) 
\nonumber\\
&(s w' \trib a^\dag \trib  u \!\,  w) 
 (t  w' \trib a^\dag \trib v w)^* {\mc{S}} ^{w}_{uv}, \label{eq:S}
\end{align}
where the matrix elements on the right side of this equation are reduced with respect to SU(3) and $n(s)$ is the value of $n$ in the multiplicity label 
$s\equiv n\rho$.
Thus, from $S^{w}$,  one obtains a matrix $S^{w'}$ matrix with 
$N_{w'} = N_{w}+2$, etc.

Once a Hermitian matrix ${\mc{S}} ^w$, with elements 
${\mc{S}}^w_{st} 
= \sum_\alpha \mc{K}^w_{s\sigma} \mc{K}^{w *}_{t\sigma}$,
 has been determined, it can diagonalised by a unitary transformation.   
 Also, because the ${\mc{S}} ^w$ matrices are non-negative by construction, their eigenvalues are real and non-negative. 
 Thus, the ${\mc{S}}^w$ matrices are expressible  in the form
\beq {\mc{S}}^w_{st} =  \sum_\sigma 
U^w_{s\sigma} \big( k^w_\sigma\big)^2 U^{w *}_{t\sigma} ,\label{eq:7.Sdiag}
\eeq
where $U^w$ is unitary, from which one obtains the $\mc{K}^w$
matrix with elements
\beq \mc{K}^w_{s\sigma} = 
U^w_{s\sigma} k^w_\sigma .\label{eq:7.Ksoln}\eeq
The VCS wave functions $\{\Psi_{\sigma w \alpha}\}$ for
 an orthonormal basis $\{ | \sigma w \alpha\rangle\}$
for the Hilbert space $\Hb^{w_0}$ of an Sp$(3,\Rb)$ irrep 
 are then  expressed by the equation
 \beq \Psi_{\sigma w \alpha}=
\hat{\mc{K}}  \phi_{\sigma w \alpha}   
= \sum_s \phi_{s w\alpha}\, U^w_{s\sigma} k^w_\sigma .\eeq

It is noted from this expression that any state 
$ | \sigma w \alpha\rangle$,
with VCS wave function $\Psi_{\sigma w \alpha}$,
 for which $k^w_\sigma$ is zero will vanish.  
This can happen if there are normalisable states in the U(3)-boson Hilbert space that have no counterparts in the Sp$(3,\Rb)$ 
 space $\Hb^{w_0}$.  
The ${\rm Sp(3,\Rb)} \to {\rm U(3)}$ branching rules \cite{RoweWB85} show precisely when this can happen which, in fact, is only for Sp$(3,\Rb)$ irreps in nuclei of nucleon number $A\leq 6$.
The non-vanishing matrix elements of any element  
$X$ in the Sp$(3,\Rb)$ Lie algebra
 are then obtained from equation  (\ref{eq:210}) and given by
 \beq \langle \sigma  w\, \trib \hat X \trib \tau  w'\rangle 
= \frac{1}{k^w_\sigma }    \sum_{st} U^{w*}_{s\sigma} 
({s w} \trib\hat \Gamma(X) \trib {t w' }) U^{w'}_{t\tau} k^{w'}_\tau .
\label{eq:SpME}
\eeq

\section{Macroscopic limits of the symplectic model}
In this section, we consider the macroscopic limits of the symplectic model that are approached in heavy deformed nuclei and which provide the means to interpret their properties  quickly and easily in a manner that relates to a microscopic many-nucleon interpretation.
 
Of particular interest is the evolution of the symplectic model of 
nuclear collective dynamics to that of a rigid (generally triaxial) rotor strongly coupled to high-energy giant-resonance vibrations.
The macroscopic model that emerges is shown to be
 much as envisaged in the phenomenological Bohr-Mottelson 
 \cite{BohrM53} and Davydov-Filippov 
 \cite{DavydovF58,DavydovF59}
 models, albeit with the low-energy beta- and gamma-vibrational excitations replaced with the higher-energy giant resonance excitations of a deformed nucleus.

The macroscopic limits of the SU(3) matrix elements to those of a rigid rotor have been given in section \ref{sect:Rotor_SU3}.
Thus, for representations with non-zero U(2)$_S$ spin, the close relationship of the symplectic model to the phenomenological collective models suggests that, because of the adiabaticity of the low-lying SU(3)-rotational states of the symplectic model, a strong spin-orbit interaction will result in a corresponding strong coupling of the rotational and spin degrees of freedom of the model.
It could also result in rotationally-aligned coupling and to a mixing of rotational bands from different Sp$(3,\Rb)$ representations.
These are  properties of the model that need to be explored, to see  the extent to which the insights gained from the many analyses in terms of the unified model are consistent with its microscopic realisation.

The U(1) $\times$ SU(3) labelling $\{N(\lambda\mu)\}$ of an irreducible U(3) representation $\{w\}= \{w_1,w_2,w_3\}$ is given by setting
\beq N_w= w_1+w_2+w_3, \quad
\lambda_w = w_2-w_3, \quad \mu_w = w_2-w_3 .\eeq
An irreducible Sp$(3,\Rb)$ representation $\langle w_0\rangle$
with lowest-grade U(3) representation $\{w_0\}$ is then similarly labelled by the U(1) $\times$ SU(3) quantum numbers
$\langle N_0(\lambda_0\mu_0)\rangle$ with
$N_0 = N_{w_0}$,  $\lambda_0 = \lambda_{w_0}$ 
and $\mu_0 = \mu_{w_0}$.

The minimum value $N_{\rm min}$ that the $N_0$ quantum number can have for an Sp$(3,\Rb)$
representation of a nucleus (with its centre-of-mass contribution removed) is, for example, $118\frac12$ for $^{40}$Ca and $811\frac12$  for $^{166}$Er.
It can be estimated \cite{RosensteelR81} as a function of the nuclear
mass number $A$, from the expression of the mean-square radius of the nuclear ground state given, in harmonic-oscillator units after removal of its centre-of-mass contribution, by
\beq \langle r^2\rangle = \frac{{N}_{\rm min}}{A-1} \frac{\hbar}{M\omega} .\eeq
From the assumed $A^{2/3}$-dependence of $\langle r^2\rangle$
and  the usual $A^{-1/3}$ dependence of $\omega$.  This gives
 \beq {N}_{\rm min} \approx 0.9 A^{4/3} ,\eeq
and exhibits its rapid increase with the mass number $A$. 
Thus, it is expected that good approximations can be obtained for a wide range of nuclei by neglecting terms of order $1/{N}_0$.
This gives the so-called macroscopic limits.

\subsection{\boldmath The U(3)-boson model as a contraction limit of the symplectic model}
\label{sect:U3-boson_limit}
The U(3)-boson algebra  \cite{RosensteelR81,RoweR82}
is a contraction  of the  Sp$(3,\Rb)$ algebra that is realised for states of relatively few bosons in  representations for which  $N_0 \to \infty$.
It was introduced to derive a much simplified version of the symplectic model that is appropriate for  its application to heavy nuclei.
Most fortuitously, as detailed in Section \ref{sect:Sp3R_VCS},
it transpires that the analytical expressions for the matrix elements of the U(3)-boson algebra provide precisely what is required for the accurate calculation of the Sp$(3,\Rb)$ matrix elements.
Applications of the U(3)-boson model were made  by 
Rochford, Casta\~nos and Draayer \cite{RochfordR89,CastanosD89}. 

The U(3)-boson  contraction of the symplectic model algebra was obtained \cite{RosensteelR81} 
by observing that, with $\varepsilon^2 ={3}/{(2N_0)}$, the commutation relations
of the giant-resonance raising and lowering operators are given by
\beqa
[ \varepsilon \hat {\mathcal{B}}_0,\varepsilon \hat {\mathcal{A}}_0]
&=& \frac{2\varepsilon^2}{3}  \hat {\mathcal{C}}_0 
= \frac{1}{N_0} \hat {\mathcal{C}}_0 , \\
{[} \varepsilon \hat {\mathcal{B}}_{2,0}, 
\varepsilon \hat {\mathcal{A}}_{2,0} {]}
&=&  \frac{1}{N_0} \hat {\mathcal{C}}_0 + 
\frac{1}{2N_0}  \hat {\mathcal{Q}}_{2,0} .
\eeqa
When acting on a state of angular momentum $L$ of a U(3) irrep 
$\{N(\lambda\mu)\}$, $\hat {\mathcal{C}}_0$ 
takes the value $N$ and, if $L < L_{\rm max}$ for the SU(3) irrep 
$(\lambda_0 \mu_0)$, $\hat {\mathcal{Q}}_{2,0}$ takes a  value  smaller than $2\lambda+\mu$.
Thus, for states for which $(N-N_0)/N_0 \ll 1$ and 
$ 2\lambda+\mu \ll 2N_0$, the giant-resonance raising and lowering operators contract to boson operators, i.e., 
\bal &\hat {\mathcal{A}}^{(0)} \to \sqrt{\frac{2N_0}{3}} \, s^\dag , 
\quad
\hat {\mathcal{B}}^{(0)} \to \sqrt{\frac{2N_0}{3}} \, s ,\\
&\hat {\mathcal{A}}_{2,\nu} \to \sqrt{\frac{2N_0}{3}} \, d^\dag_\nu , \quad
\hat {\mathcal{B}}_{2,\nu} \to \sqrt{\frac{2N_0}{3}} \, d_\nu ,
\end{align}
that obey the  commutation relations
\beq [s,s^\dag ] = 1 , \quad [d^\mu, d^\dag_\nu] = \delta_{\mu,\nu} \hat I ,\eeq
where $d^\mu = (-1)^\mu d_{-\mu} = (d^\dag_\mu)^\dag$.
However, the contracted raising and lowering operators continue to transform under U(3) in exactly the same way as given by the  boson expansion 
\beq
\hat{\mathcal{C}}_{ij} \equiv \hat \Cb_{ij} + (a^\dag a)_{ij} ,
\label{eq:C}\eeq
of elements of the U(3) Lie algebra given by equation 
(\ref{eq:7.VCS}a).
Thus, the quadrupole tensor of the symplectic model, given in harmonic-oscillator units by equation (\ref{eq:SpQops}), has contraction
\beq 
\hat Q_{2} =  \hat{\mc{Q}}_2+ \sqrt{3} \, (\hat{\mc{A}}_2 + \hat{\mc{B}}_2)  \to \hat{\mc{Q}}_2 + \sqrt{2N_0} \, (d^\dag + d ) ,
\label{eq:8.a2Q} \eeq
and the Sp$(3,\Rb)$ matrix elements have values in the contraction limit given by
\begin{subequations} \label{eq:CGamma}
\bal
&\langle w_0 n \rho w \trib \hat{\mathcal{C}}
 \trib w_0 n' \rho' w'\rangle
=\pm 2 \delta_{n,n'} \delta_{\rho,\rho'}\delta_{w,w'} \nonumber\\
&\hspace{2.5cm} \times
(\lambda^2+\lambda\mu+\mu^2 + 3\lambda+3\mu)^{\frac12} , \\
&\langle w_0 n \rho w \trib \hat{\mathcal{B}}
    \trib w_0 n' \rho' w'\rangle \to \sqrt{2N_0/3}\, \nonumber\\
&\hspace{3.4cm} \times
  (w_0 n \rho w \trib a \trib w_0 n' \rho' w') , \\
&\langle w_0 n' \rho' w' \trib \hat{\mathcal{A}}
\trib w_0 n \rho w\rangle \to \sqrt{2N_0/3}\,\nonumber\\
&\hspace{3.3cm} \times
(w_0 n' \rho' w' \trib a^\dag \trib w_0 n \rho w).
\end{align}
\end{subequations}

\subsection{\boldmath The large $N_0$ asymptotic limit of an Sp$(3,\Rb)$ representation} \label{sect:N0limit}
An asymptotic limit of an Sp$(3,\Rb)$ representation that is approached much more rapidly than that of its contraction
as $(\lambda_0+\mu_0)/N_0 \to 0$ was given in 
\cite{Rowe84,RoweRC84} and its accuracy for relatively small values of $(\lambda_0+\mu_0)/N_0$ was confirmed by Hecht
\cite{Hecht85}.
It follows directly from the observation, implied by the above contraction, that in
the $N_0 \to \infty$ limit, while $\lambda$ and $\mu$ remain finite, the basis states for an irreducible unitary representations of the U(3)-boson algebra need only to be renormalised to become an orthonormal basis for a  VCS representation of Sp$(3,\Rb)$.
Thus,  improved asymptotic matrix elements, 
 for large values of $N_0$, are obtained directly from the VCS  representation of Sp$(3,\Rb)$ elements  in states that are asymptotically proportional to U(3)-boson states, i.e., for which
\beq  |w_0 n\rho w\nu\rangle \sim  k^w_{n\rho} | w_0 n\rho w\nu) .\eeq
A similarity transformation of  
equations (\ref{eq:Gamma}a)-(\ref{eq:Gamma}c), is then obtained by setting
\beq \frac{k^{w'}_{n'\rho'}}{k^w_{n\rho }} = (\Omega_{n'w'}-\Omega_{nw})^{\frac12}, 
 \quad \text{for\;} \Omega_{n'w'} > \Omega_{nw}, \eeq
and leads to the asymptotic matrix elements 
\begin{subequations}  \label{eq:ALGamma}
\bal
&\langle w_0 n \rho w \trib \hat{\mathcal{C}} \trib w_0 n' \rho' w'\rangle 
=2 \delta_{n,n'} \delta_{\rho,\rho'}\delta_{w,w'} \nonumber\\
&\hspace{2.6cm} \times
(\lambda^2+\lambda\mu+\mu^2 + 3\lambda+3\mu)^{\frac12} , \\
&\langle w_0 n \rho w \trib \hat{\mathcal{B}} \trib w_0 n' \rho' w'\rangle \sim
(\Omega_{n' w'}-\Omega_{n w})^{\frac12} \nonumber\\
&\hspace{3.4cm} \times(w_0 n \rho w \trib a \trib w_0 n' \rho' w')  
,\\
&\langle w_0 n' \rho' w' \trib \hat{\mathcal{A}} \trib w_0 n \rho w\rangle \sim
(\Omega_{n' w'}-\Omega_{n w})^{\frac12} \nonumber\\
&\hspace{3.3cm} \times
(w_0 n' \rho' w' \trib a^\dag \trib w_0 n \rho w).
\end{align}
\end{subequations}

These expressions were shown to be accurate to  fourth order in the small parameter $(\lambda_0 + \mu_0)/2N_0$ \cite{Rowe84}.
Moreover, they are  precise for multiplicity-free representations 
$\langle N_O(\lambda_0 \mu_0)\rangle$ with $\lambda_0=\mu_0=0$
as seen from the expressions for these matrix elements in Section
\ref{sect:MFreps} and, as shown in \cite{RoweRC84}, they agree with the results calculated by Casta\~nos \emph{et al.}\ \cite{CastanosCM84}.

\subsection{Comparisons with precisely computed matrix elements}
The motivation for considering the macroscopic limits of the symplectic model is two-fold;
the first is to assess the extent to which they reproduce the results of the standard Bohr-Mottelson-based collective models and the adjustments  to these models implied by their many-nucleon foundations;
the second is to understand their limitations as approximations to precise calculations.

It is emphasised that an important use of the symplectic model and the associated many-nucleon coupling scheme is to identify the many-nucleon spaces needed for realistic descriptions of observed collective states of nuclei.
Exploratory model calculations, e.g., in Section \ref{sect:CRV},
indicate that, in a spherical-harmonic oscillator shell-model basis, 
 these spaces are not small even when restricted to a single 
Sp$(3,\Rb)$ irrep.
Typically, for a rotational rare-earth nucleus, states from 15 to 20 major harmonic-oscillator shells may be required for calculations of the desired level of convergence.
Thus, it appears that in many situations, improved accuracy is likely to be possible by use of asymptotic matrix elements in larger spaces.
It is also recognised  that the restriction of a symplectic model to a single representation is only a beginning. 
One will eventually want to consider calculations with symplectic symmetry-breaking interactions in spaces of mixed representations.

In this section, we compare matrix elements calculated in 
the asymptotic  \eqref{eq:ALGamma} and  U(3)-boson approximations (\ref{eq:CGamma})
with precisely computed values.
In Tables \ref{tab:SpMeA} and \ref{tab:SpMeC}  the asymptotic  (ASp3R) and the U(3)-boson  (U3BC) approximations are compared, respectively, with precisely-computed reduced matrix elements for  $\mc{A}$  and 
$\mc{\mathbb{C}}$.  
The  matrix elements are calculated for the symplectic irrep 
$\langle 824(80,0)\rangle$, as used by Bahri \cite{BahriR00} for the ground state rotational band of $^{166}$Er, and for the irrep
$\langle 836(78,10)\rangle$ which,
 according to Jarrio et al.\ \cite{JarrioWR91}, is the kind of irrep implied by the observed properties of $^{168}$Er.  
The precise matrix elements of Table \ref{tab:SpMeA} are those of 
equation  \eqref{eq:SpME} and the asymptotic and U(3)-boson matrix elements are, respectively,  those of equations 
(\ref{eq:ALGamma}c) and (\ref{eq:CGamma}c).
The reduced matrix elements of $\mathbb{C}$ are obtained by noting from \eqref{eq:C} that $\mathbb{C}_{ij}$ can be expressed in terms of the U(3) elements $\mathcal{C}_{ij}$ of the Sp$(3.\Rb)$ algebra and the boson creation and annihilation operators. 
Its reduced matrix elements are then given in the U(3)-boson basis by
\begin{multline}
(w_0 n'\rho' w'|||\mathbb{C}|||w_0 n\rho w)_{\rho_0}
=(w_0 n'\rho' w'|||\mc{C}|||w_0 n\rho w)_{\rho_0}\\
-(w_0 n'\rho' w'|||a^\dagger a|||w_0 n\rho w)_{\rho_0} .
\label{eq:Cs}
\end{multline}
The corresponding precise and asymptotic matrix elements of 
$\mathbb{C}$ are then obtained from the expansions of the precise and asymptotic bases in the U(3)-boson basis.

The selected matrix elements shown in Tables 
\ref{tab:SpMeA} and  \ref{tab:SpMeC} include those for which the approximations are the most and least accurate for 
each $N_n$ and $N_n'$.
The tables show, for both irreps, that the asymptotic and precise matrix elements  are essentially the same to a high level of accuracy. 
However,  computation of the asymptotic matrix elements,
in the current implementation, is approximately a factor of  
5 times faster.
The U(3)-boson reduced matrix elements do not provide the same accuracy as the asymptotic approximation and are only marginally easier to compute.
However,   and  more significantly as pursued in Section \ref{sect:CRV}, they are amenable to analytical model calculations.

\begin{table*}
\caption{  \label{tab:SpMeA}
Comparison of selected reduced matrix elements 
$\langle \sigma' w' \trib \hat {\mc{A}} \trib \sigma w\rangle$ calculated  in the U(3)-boson contraction (U3BC) and asymptotic limits (ASp3R)
with accurately computed Sp$(3,\Rb)$ matrix elements (Sp3R).
The  state $ |w_0 n \rho w\alpha)$, to which a state 
$|\sigma w\alpha\rangle$ approaches in the asymptotic limit,
is labelled by the U(3)-boson quantum numbers $n\rho w$ shown.}
$\begin{array}{c}
\\
w_0=824(80, 0)\\
\begin{array}{|c|c|cc|c|c|cc||r|r|r| } 
 \hline
w'&\sigma'&n'&\rho'&w & \sigma &n&\rho
&\;\; \text{Sp3R} \;\;\;\;  &\;\; {\rm ASp3R} \;\;\;\;
&\;\; {\rm U3BC}  \;\;\;\;  \\ 
\hline
826\;\;(78, 2)&1&\;2\;(2, 0)&1&824\;\;(80, 0)&1&\;0\;(0, 0)&1&22.2261 &22.2261 &23.4379\\

\hline
830\;\;(82, 2)&2&\;6\;(2, 2)&1&828\;\;(82, 1)&1&\;4\;(4, 0)&1&-15.0319 &-15.0317 &-14.6415\\
830\;\;(80, 0)&1&\;0\;(0, 0)&1&828\;\;(79, 1)&1&\;4\;(0, 2)&1&23.8723 &23.8721 &23.2927\\
830\;\;(86, 0)&1&\;6\;(6, 0)&1&828\;\;(84, 0)&1&\;4\;(4, 0)&1&44.4972 &44.4972 &40.5956\\
\hline
834\;\;(82, 4)&3&10\;(2, 4)&1&832\;\;(82, 3)&2&\;8\;(4, 2)&1&-21.2952 &-21.2947 &-20.7062\\
834\;\;(79, 1)&2&10\;(0, 2)&1&832\;\;(78, 2)&2&\;8\;(2, 0)&1&-30.7897 &-30.7896 &-30.0684\\
834\;\;(90, 0)&1&10(10, 0)&1&832\;\;(88, 0)&1&\;8\;(8, 0)&1&57.6194 &57.6194 &52.4087\\
\hline
838\;\;(82, 6)&2&14\;(2, 6)&1&836\;\;(82, 5)&2&12\;(4, 4)&1&-26.1261 &-26.1254 &-25.3598\\
838\;\;(78, 2)&1&14\;(2, 0)&1&836\;\;(80, 0)&1&12\;(0, 0)&1&31.5594 &31.5595 &33.1461\\
838\;\;(94, 0)&1&14(14, 0)&1&836\;\;(92, 0)&1&12(12, 0)&1&68.3813 &68.3813 &62.0108\\
\hline
842\;\;(86, 6)&2&18(18, 0)&1&840\;\;(88, 4)&2&16(16, 0)&1&17.8690 &17.8680 &18.9966\\
842\;\;(80, 0)&2&18\;(0, 0)&1&840\;\;(79, 1)&3&16\;(0, 2)&1&41.4913 &41.4907 &40.3442\\
842\;\;(98, 0)&1&18(18, 0)&1&840\;\;(96, 0)&1&16(16, 0)&1&77.7689 &77.7689 &70.3136\\
\hline
846\;\;(88, 7)&2&22(22, 0)&1&844\;\;(90, 5)&3&20(20, 0)&1&18.3142 &18.3127 &19.5096\\
846\;\;(79, 1)&2&22\;(0, 2)&1&844\;\;(78, 2)&5&20\;(2, 0)&1&-41.4521 &-41.4518 &-40.3410\\
846\;(102, 0)&1&22(22, 0)&1&844\;(100, 0)&1&20(20, 0)&1&86.2322 &86.2322 &77.7346\\
\hline
\end{array}\\\\
w_0=836 (78, 10)\\
 \begin{array}{|c|c|cc|c|c|cc||r|r|r| } 
 \hline
w'&\sigma'&\;\;\;n'&\rho'&w &\sigma& n&\rho&\;\; \text{Sp3R} \;\;\;\;  &\;\; {\rm ASp3R} \;\;\;\;&\;\; {\rm U3BC}  \;\;\;\;  \\ 
\hline
838\;\;(78, 8)&1&\;2\;(2, 0)&1&836\;(78, 10)&1&\;0\;(0, 0)&1&22.0907 &22.0907 &23.6079\\
\hline
842\;(84, 10)&1&\;6\;(6, 0)&1&840\;(82, 10)&1&\;4\;(4, 0)&1&44.8999 &44.8999 &40.8901\\
842\;(78, 10)&5&\;6\;(0, 0)&1&840\;(77, 11)&1&\;4\;(0, 2)&1&17.6891 &17.6889 &17.3249\\
842\;\;(82, 8)&2&\;6\;(2, 2)&1&840\;\;(81, 9)&1&\;4\;(4, 0)&1&-15.0895 &-15.0892 &-14.7660\\
\hline
846\;(78, 9)&5&10\;(0, 2)&1&844\;\;(78, 8)&2&\;8\;(2, 0)&1&24.1736 &24.1735 &23.4749\\
846\;\;(84, 6)&3&10\;(2, 4)&1&844\;\;(83, 7)&1&\;8\;(4, 2)&1&-21.3766 &-21.3760 &-20.8822\\
846\;(88, 10)&1&10(10, 0)&1&844\;(86, 10)&1&\;8\;(8, 0)&1&58.1378 &58.1378 &52.7889\\
\hline
850\;\;(86, 4)&2&14\;(2, 6)&1&848\;\;(85, 5)&2&12\;(4, 4)&1&-26.2258 &-26.2249 &-25.5754\\
850\;\;(78, 8)&1&14\;(2, 0)&1&848\;(78, 10)&6&12\;(0, 0)&1&31.3687 &31.3688 &33.3866\\
850\;(92, 10)&1&14(14, 0)&1&848\;(90, 10)&1&12(12, 0)&1&68.9928 &68.9928 &62.4607\\
\hline
854\;\;(90, 4)&2&18(18, 0)&1&852\;\;(90, 6)&2&16(16, 0)&1&18.0219 &18.0206 &19.4181\\
854\;(78, 10)&24&18\;(0, 0)&1&852\;(77, 11)&5&16\;(0, 2)&1&30.7439 &30.7434 &30.0076\\
854\;(96, 10)&1&18(18, 0)&1&852\;(94, 10)&1&16(16, 0)&1&78.4602 &78.4602 &70.8237\\
\hline
858\;\;(93, 3)&2&22(22, 0)&1&856\;\;(93, 5)&3&20(20, 0)&1&18.5354 &18.5337 &20.0127\\
858\;(78, 9)&49&22\;(0, 2)&1&856\;\;(78, 8)&32&20\;(2, 0)&1&32.5420 &32.5418 &31.4949\\
858(100, 10)&1&22(22, 0)&1&856\;(98, 10)&1&20(20, 0)&1&86.9943 &86.9943 &78.2986\\
\hline
\end{array}\\\\
\end{array}$
 \end{table*}

\begin{table*}
\caption{  \label{tab:SpMeC}
Comparison of selected reduced matrix elements 
$\langle \sigma' w' \trib \hat{\mc{\mathbb{C}}} \trib \sigma w\rangle$
calculated in the U(3)-boson contraction (U3BC) and asymptotic limits (ASp3R) with accurately computed Sp$(3,\Rb)$ matrix elements
(Sp3R). }
$
\begin{array}{c}
\\
w_0=824(80,0)\\
\begin{array}{|c|c|cc|c|c|cc|c||r|r|r| } \hline
 w'&\sigma'&n'&\rho'&w&\sigma& n&\rho &\rho_0
&\;\; \text{Sp3R} \;\;\;\;  &\;\; {\rm ASp3R} \;\;\;\;
&\;\; {\rm U3BC}  \;\;\;\;  \\ 
\hline
824\;(80, 0)&1&\;0\;(0, 0)&1&824(80, 0)&1&\;0\;(0, 0)&1&1&66.5332 &66.5332 &66.5332\\
\hline
828\;(82, 1)&1&\;4\;(4, 0)&1&828(80, 2)&1&\;4\;(0, 2)&1&1& 0.0150 & 0.0000 & 0.0000\\
828\;(79, 1)&1&\;4\;(0, 2)&1&828(79, 1)&1&\;4\;(0, 2)&1&2& 1.2539 & 1.2539 & 1.2539\\
828\;(82, 1)&1&\;4\;(4, 0)&1&828(80, 2)&2&\;4\;(4, 0)&1&1& 3.1775 & 3.1774 & 2.9623\\
\hline
832\;(82, 3)&2&\;8\;(4, 2)&1&832(80, 4)&1&\;8\;(0, 4)&1&1& 0.0229 & 0.0000 & 0.0000\\
832\;(78, 2)&2&\;8\;(2, 0)&1&832(78, 2)&2&\;8\;(2, 0)&1&2& 2.0109 & 2.0109 & 2.0109\\
832\;(82, 3)&1&\;8\;(8, 0)&1&832(80, 4)&2&\;8\;(8, 0)&1&1& 5.2947 & 5.2947 & 4.9319\\
\hline
836\;(80, 6)&1&12\;(0, 6)&1&836(78, 7)&2&12\;(4, 4)&1&1&-0.0307 & 0.0000 & 0.0000\\
836\;(80, 0)&1&12\;(0, 0)&1&836(80, 0)&1&12\;(0, 0)&1&1&66.5332 &66.5332 &66.5332\\
836\;(82, 5)&3&12(12, 0)&1&836(80, 6)&4&12(12, 0)&1&1& 7.4013 & 7.4012 & 6.8933\\
\hline
840\;(80, 8)&2&16\;(0, 8)&1&840(78, 9)&3&16\;(4, 6)&1&1&-0.0384 & 0.0000 & 0.0000\\
840\;(79, 1)&3&16\;(0, 2)&1&840(79, 1)&3&16\;(0, 2)&1&2& 1.2539 & 1.2539 & 1.2539\\
840\;(82, 7)&2&16(16, 0)&1&840(80, 8)&3&16(16, 0)&1&1& 9.4943 & 9.4942 & 8.8426\\
\hline
844(80, 10)&4&20(0, 10)&1&844(78, 11)&4&20\;(4, 8)&1&1&-0.0461 & 0.0000 & 0.0000\\
844\;(78, 2)&5&20\;(2, 0)&1&844(78, 2)&5&20\;(2, 0)&1&2& 2.0109 & 2.0109 & 2.0109\\
844\;(84, 8)&5&20(20, 0)&1&844(82, 9)&5&20(20, 0)&1&1&11.4033 &11.4031 &10.6084\\
\hline
\end{array}\\\\
w_0=836, (78, 10)\\
\begin{array}{|c|c|cc|c|c|cc|c||r|r|r| } \hline
 w'&\sigma'&n'&\rho'&w&\sigma& n&\rho &\rho_0
&\;\; \text{Sp3R} \;\;\;\;  &\;\; {\rm ASp3R} \;\;\;\;
&\;\; {\rm U3BC}  \;\;\;\;  \\ 
\hline

836(78, 10)&1&\;0\;(0, 0)&1&836(78, 10)&1&\;0\;(0, 0)&1&2& 0.0000 & 0.0000 & 0.0000\\
\hline

840\;(80, 8)&1&\;4\;(0, 2)&1&840\;(79, 7)&1&\;4\;(4, 0)&1&1& 0.0145 & 0.0000 & 0.0000\\
840(77, 11)&1&\;4\;(0, 2)&1&840(77, 11)&1&\;4\;(0, 2)&1&2& 0.8208 & 0.8208 & 0.8208\\
840\;(80, 8)&2&\;4\;(4, 0)&1&840\;(79, 7)&1&\;4\;(4, 0)&1&1&-2.5000 &-2.5000 &-2.3085\\
\hline
844\;(82, 6)&1&\;8\;(0, 4)&1&844\;(81, 5)&2&\;8\;(4, 2)&1&1& 0.0237 & 0.0000 & 0.0000\\
844\;(78, 8)&2&\;8\;(2, 0)&1&844\;(78, 8)&2&\;8\;(2, 0)&1&2&-1.1783 &-1.1783 &-1.1783\\
844\;(81, 5)&1&\;8\;(8, 0)&1&844\;(82, 6)&2&\;8\;(8, 0)&1&1&-4.2020 &-4.2019 &-4.5445\\
\hline
848\;(85, 5)&2&12\;(4, 4)&1&848\;(84, 4)&1&12\;(0, 6)&1&1&-0.0319 & 0.0000 & 0.0000\\
848(78, 10)&6&12\;(0, 0)&1&848(78, 10)&6&12\;(0, 0)&1&2&-0.0000 & 0.0000 & 0.0000\\
848\;(83, 3)&3&12(12, 0)&1&848\;(84, 4)&4&12(12, 0)&1&1&-6.0828 &-6.0826 &-6.5756\\
\hline
852\;(85, 1)&3&16\;(4, 6)&1&852\;(86, 2)&2&16\;(0, 8)&1&1&-0.0391 & 0.0000 & 0.0000\\
852(77, 11)&24&16\;(0, 2)&1&852(77, 11)&24&16\;(0, 2)&1&2& 0.8208 & 0.8208 & 0.8208\\
852\;(84, 0)&2&16(16, 0)&1&852\;(82, 1)&2&16(16, 0)&1&1& 9.3085 & 9.3085 & 8.6389\\
\hline
856\;(88, 0)&4&20(0, 10)&1&856\;(86, 1)&1&20\;(4, 8)&2&1&-0.0565 & 0.0000 & 0.0000\\
856\;(78, 8)&32&20\;(2, 0)&1&856\;(78, 8)&32&20\;(2, 0)&1&2&-1.1783 &-1.1783 &-1.1783\\
856\;(88, 0)&2&20(20, 0)&1&856\;(86, 1)&3&20(20, 0)&1&1&11.7353 &11.7352 &10.8711\\
\hline
\end{array}
\end{array}
$
 \end{table*}

\section{A simple model application}
The above macroscopic limits of the symplectic model as a U(3)-boson model and of its lowest-grade U(3) states as those of a  triaxial rigid rotor, as reviewed in Section \ref{sect:Rotor_SU3}, show that, 
in a combined macroscopic limit, for which both $L/\lambda_0$ and $(\lambda_0+\mu_0)/N_0$ are small, 
the states of an Sp$(3,\Rb)$ representation $\langle N_0 (\lambda_0\mu_0)\rangle$ of angular momentum $L$, 
are those of a coupled rotor-vibrator (CRV) model.
Thus, to assess the usefulness of these  macroscopic limits of the symplectic model in fitting experimental data in a manner that gives information on their many-nucleon structure, we consider a simple symplectic model Hamiltonian that is simply solvable in its macroscopic CRV limit.

The CRV model was introduced by Le Blanc \emph{et al.}\,
\cite{LeBlancCR84,RoweVC89} as a union of rigid-rotor and (giant-resonance) vibrational models.
It is particularly significant in the present context because of its close correspondence with the Bohr-Mottelson unified model, in which intrinsic valence-shell states, described in the symplectic model as lowest-grade U(3) states, are added to a Bohr irrotational-flow vibrational model corresponding to  giant monopole/quadrupole vibrations;
in the absence of valence-shell particles, e.g., for closed-shell nuclei, the giant-resonance vibrational states would have irrotational-flow mass parameters (cf.\,\cite{RoweWood10} Section 8.2.6).

\subsection{The coupled rotor-vibrator model} 
\label{sect:CRV}
We consider  a symplectic model with a  Hamiltonian of the form
\beq  \hat H = \hat H_{\rm su3}
- \kappa_1 \hat\Qb_2 \cdot (\hat{\mc{A}}_2 +  \hat{\mc{B}}_2)
+ \kappa_2 \hat{\mc{A}}_2 \cdot \hat{\mc{B}}_2 
+ \kappa_3 \hat{\mc{A}}_0 \cdot \hat{\mc{B}}_0 , 
\label{eq:H_CRVo} \eeq
in which $\hat\Qb_2$ is the restriction  of the quadrupole tensor
$\hat{\mc{Q}}_{2}$, defined by equations
(\ref{eq:6.L_k2}) and
(\ref{eq:LijQij})
to the lowest-grade SU(3) irrep $(\lambda_0\mu_0)$  of an Sp$(3,\Rb)$ irrep $\langle N_0(\lambda_0 \mu_0)\rangle$; 
its components commute with the giant-resonance operators.
This Hamiltonian is expressed in terms of symplectic model operators and can be diagonalised within the space of an 
 Sp$(3,\Rb)$ irrep.
It is then of interest to interpret the results that emerge
in terms of the corresponding analytically solvable CRV model.

The parameters  of the Hamiltonian (\ref{eq:H_CRVo}) can be adjusted to fit the properties of nuclear states of interest.
They can 
also be determined by mean-field self-consistency methods. 
For example, 
self-consistent vibrating-potential methods  \cite{SuzukiR77b}
have indicated that the parameters should be adjusted to give the  
one-phonon giant-monopole and giant-quadrupole resonance states  at energies of order $\hbar\omega_0 =2\hbar\omega$ 
and $\hbar\omega_2=\sqrt{2}\,\hbar\omega$, respectively,  where $\omega$ is the frequency of the harmonic-oscillator shell-potential.
Mean-field considerations also imply that $\kappa_1$ should be such as to give an effective charge for low-lying states that is twice the natural charge, 
{i.e., the value that would be appropriate if these states were} 
described in an un-renormalized SU(3) model 
\cite{Rowe67,RoweRabida13,Rowe16}.

From equation (\ref{eq:7.VCS}a), the boson expansion of the U(3) algebra is given by
\beq \hat{\mc{C}}_{ij} = \hat{\Cb}_{ij} + \hat{\mc{C}}^{(B)}_{ij}, \eeq
where $\hat{\mc{C}}^{(B)}_{ij} = (a^\dag a)_{ij}$.
Thus, in the U(3)-boson 
limit, in which the quadrupole moments
\beq \hat Q_{2\,\nu} = \hat{\mc{Q}}_{2,\nu} 
+ \sqrt{3}\, (\hat{\mc{A}}_{2,\nu} +  \hat{\mc{B}}_{2,\nu}) 
\eeq
contract to
\beq \hat Q_{2\,\nu} \to  \hat{\Qb}_{2,\nu} +\hat{\mc{Q}}^{(B)}_{2,\nu}
 + \sqrt{2N_0}\, (d^\dag_\nu + d_\nu) ,
\eeq
where
\beq  \hat{\mc{Q}}^{(B)}_{2,\nu}
= 2\sqrt{2}\, (d^\dag_\nu s + s^\dag d_\nu) 
 - \sqrt{14}\, [d^\dag\otimes d]_{2,\nu} ,
\eeq
the Hamiltonian (\ref{eq:H_CRVo}) contracts to
\beq \hat H_{\rm U3BC} = \hat H_{\rm su3} - \kappa\, \hat\Qb_2 \cdot (d^\dag + d) 
+ \hbar\omega_2\, d^\dag \cdot d + \hbar\omega_0\, s^\dag s ,
\label{eq:H_U3BC}\eeq
with
$\kappa = \sqrt{2N_0/3}\,\kappa_1$, 
$\hbar\omega_2 = {(2N_0/3)}\,\kappa_2$, and
$\hbar\omega_0 = {(2N_0/3)}\,\kappa_3$.

The spectrum of this Hamiltonian and the properties of its eigenstates are easily derived in the rotor limit of SU(3).
By setting $\alpha = \kappa/\hbar\omega_2$ and making the substitution
\beq d^\dag_\nu = D^\dag_\nu  + \alpha \hat {\Qb}_{2,\nu} ,
\quad d_\nu = D_\nu  + \alpha \hat {\Qb}_{2,\nu} ,\eeq
the U(3)-boson Hamiltonian (\ref{eq:H_U3BC}) is re-expressed as
\beq \hat H_{\rm CRV} = \hat H_{\rm su3} 
-\alpha^2 \hbar\omega_2\, \hat\Qb_2 \cdot  \hat {\Qb}_2 
 + \hbar\omega_2 D^\dag \cdot D   + \hbar\omega_0 s^\dag s. 
\eeq
Then, with the value of the second-order SU(3) Casimir invariant for the lowest-grade SU(3) irrep given by
\bal  {\cal I}_2(\lambda_0\mu_0) 
&=   \langle\hat{\Qb}_2\cdot \hat{\Qb}_2 + 3 \hat{\Lb}\cdot \hat\Lb \rangle\nonumber\\
&=  4( \lambda_0^2+\lambda_0 \mu_0 + \mu_0^2 + 3 \lambda_0+3\mu_0 ),
\end{align}
where $\hat\Lb$ denotes the restriction of the angular momentum, 
$\hat L$, to states of the lowest-grade SU(3) irrep,  
$\hat H_{\rm CRV}$  simplifies to 
\begin{multline}
 \hat H_{\rm CRV} = \hat H_{\rm su3}  
+ 3\alpha^2 \hbar\omega_2\, \hat{\Lb}\cdot \hat\Lb
- \alpha^2 \hbar\omega_2 {\cal I}_2(\lambda_0\mu_0)\\
 + \hbar\omega_2 D^\dag \cdot D  +\hbar\omega_0 s^\dag s. 
 \label{eq:H_CRV}
\end{multline}
However now, because the SU(3) quadrupole operators 
$\hat\Qb_\nu$ commute with one another  in  the rotor limit,   the operators
$D^\dag_\nu$ and $D_\nu$ are boson operators with commutation relations
\beq [D^\mu,D^\dag_\nu] = [d^\mu,d^\dag_\nu]=\delta_{\mu,\nu} \hat I .
\eeq
Thus, if one already knows the spectrum of the SU(3)-rotor Hamiltonian 
$\hat H_{\rm su3}$, which we suppose to be given in a 
$\{|KLM\rangle\}$ basis with energies $\{E_{KL}\}$, 
it is immediately inferred that the energy levels of the Hamiltonian 
$\hat H_{\rm CRV}$ are given to within terms of order $\alpha^2$ by
\beq 
 E_{n_0n_2KL} = E_{KL} 
 + n_0 \hbar\omega_0 + n_2 \hbar\omega_2 
 + {\rm O}(\alpha^2). \label{eq:CRV_E}
\eeq 
Thus, in the CRV limit, the structure of the energy
spectrum is little changed by introduction of the $\hat\Qb_2 \cdot (\hat{\mc{A}}_2 +  \hat{\mc{B}}_2)$ term in (\ref{eq:H_CRVo}) that couples the intrinsic SU(3) and giant resonance degrees of freedom.
However, there is a  significant change in the quadrupole moments of the low-lying states and the E2 transition matrix elements between them.

Let  $\hat H(\alpha)$ denote the Hamiltonian
\beq \hat H(\alpha) = \hat H_{\rm su3} 
+ \hbar\omega_2 D^\dag \cdot D 
\eeq
with $D_\nu^\dag = d_\nu^\dag- \alpha  \hat\Qb_\nu$
and $D_\nu = d_\nu- \alpha  \hat\Qb_\nu$, where $\{\hat \Qb_\nu\}$ are commuting quadrupole operators.
The transformation between the $d$- and $D$-boson operators, given  by
\beq e^{\alpha\hat\Qb \cdot (d^\dag-d)} d_\nu^\dag e^{-\alpha\hat\Qb \cdot (d^\dag-d)}
= D_\nu^\dag , \eeq
then implies that
\beq  \hat H(\alpha) = 
e^{\alpha\hat\Qb \cdot (d^\dag-d)} \hat H(0) e^{-\alpha\hat\Qb \cdot (d^\dag-d)}.
\eeq
Thus, if $|\varphi\rangle$ is any eigenstate of $\hat H(0)$, the transformed state
\beq |\varphi(\alpha)\rangle = e^{\alpha\hat\Qb \cdot (d^\dag-d)} |\varphi\rangle
\eeq
will be an eigenstate of $\hat H(\alpha)$ with  the same energy, 
as already observed.
The quadrupole moments between the transformed state are similarly obtained.
Within the space of lowest-grade SU(3) (boson-vacuum) states, 
matrix elements of the quadrupole operator  are, by definition,  given by
\beq 
\langle \varphi_1 | \hat Q_{2,\nu} |\varphi_2\rangle
= \langle \varphi_1 | \hat\Qb_{2,\nu} |\varphi_2\rangle ,
\eeq
and in the corresponding space of $D$-boson vacuum states, 
they are given by
\bal 
\langle \varphi_1(\alpha)|\hat Q_{2,\nu} |\varphi_2(\alpha)\rangle
&= \langle \varphi_1| e^{-\alpha\hat\Qb \cdot (d^\dag-d)}
 \hat Q_{2,\nu}  e^{\alpha\hat\Qb \cdot (d^\dag-d)}  |\varphi_2\rangle \nonumber\\
&=  \langle \varphi_1 | \hat \Qb_{2,\nu} 
 - \sqrt{14}\, \alpha^2[\hat \Qb_2 \otimes\hat \Qb_2]_{2,\nu} 
 \nonumber \\
&\quad+2\alpha \sqrt{2N_0}\, \hat\Qb_{2,\nu}  |\varphi_2\rangle . 
 \end{align}
Thus, if $\alpha$ is assigned a value such that $2\alpha = 1/\sqrt{2N_0}$, we obtain
\beq \langle \varphi_1(\alpha)|\hat Q_{2,\nu} |\varphi_2(\alpha)\rangle
= \langle \varphi_1 | 2 \hat \Qb_{2,\nu}    - \frac{\sqrt{14}}{8N_0} 
[\hat \Qb_2\otimes \hat \Qb_2]_{2,\nu}  |\varphi_2\rangle , \eeq
which, to within a term of order $\alpha^2$, is twice the value of 
$\langle  \varphi_1 |  \hat \Qb_{2,\nu}   |\varphi_2\rangle$
and gives the renormalisation of the effective charge of the 
quadrupole operators, due to coupling to higher shells, 
that is predicted by self-consistency mean-field arguments
\cite{RoweRabida13,Rowe16}.

This result has an insightful physical interpretation.
The $D$-boson vacuum states  satisfy the equation
\beq d_\nu |\varphi (\alpha)\rangle = \alpha \hat\Qb_{2,\nu} |\varphi(\alpha)\rangle,
\eeq
which means that, when the  giant-resonance states are regarded as  harmonic-oscillator states, 
the states $|\varphi (\alpha)\rangle$ are interpreted as coherent states or, in other words, polarized vacuum states.  
In fact, we have found that, with a suitable value of the coupling constant $\alpha$,
the states $|\varphi(\alpha)\rangle$ have quadrupole moments with essentially equal contributions to their values coming from the SU(3) rotor and the polarized giant-resonance vacuum.
In pictorial terms, the interaction between the lowest-grade SU(3) rotor and the giant-resonance degrees of freedom  results in a polarisation of the giant-resonance vacuum which then rotates in unison with the SU(3) rotor in much the same way as a tidal wave on the earth's surface follows the moon's orbit.

\subsection{Comparison of symplectic  and CRV model results}

 To examine the interaction between intrinsic SU(3) and
  giant-resonance degrees of freedom, as introduced by the 
  $\hat\Qb_2 \cdot (\hat{\mc{A}}_2 + \hat{\mc{B}}_2)$ term in the Hamiltonian (\ref{eq:H_CRVo}), we compare symplectic model calculations with and without the presence of this term.  
  We use appropriately-defined values of the parameters 
  $\kappa_2$ and $\kappa_3$ so as to place
the giant monopole and quadrupole resonances at reasonable energies  (as described subsection \ref{sect:CRV}).  
Then, we carry out calculations either with $\kappa_1=0$, that is, with the coupling
  between the SU(3) and giant-resonance degrees of freedom suppressed,
  or with $\kappa_1$ chosen to approximately double the matrix
  elements of quadrupole operators, or quadruple the corresponding
  $B$(E2) reduced transition probabilities, for consistency with
  experiment.  Recall that, in the CRV approximation, changing this
  coupling is expected to leave the energy spectrum essentially
  unchanged.

\begin{figure}
\begin{center}
\includegraphics[width=1.0\hsize]{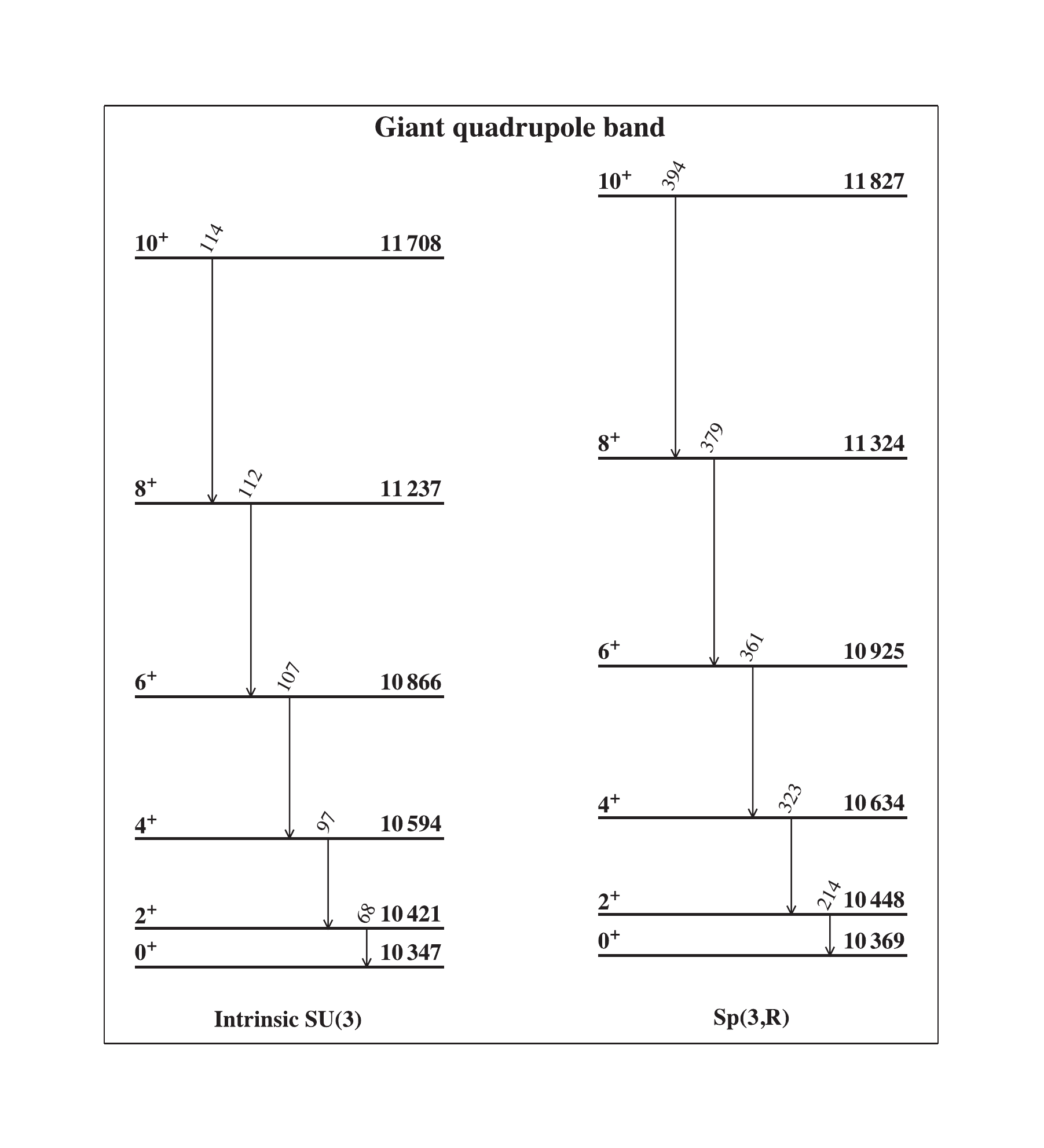} \\
\includegraphics[width=1.0\hsize]{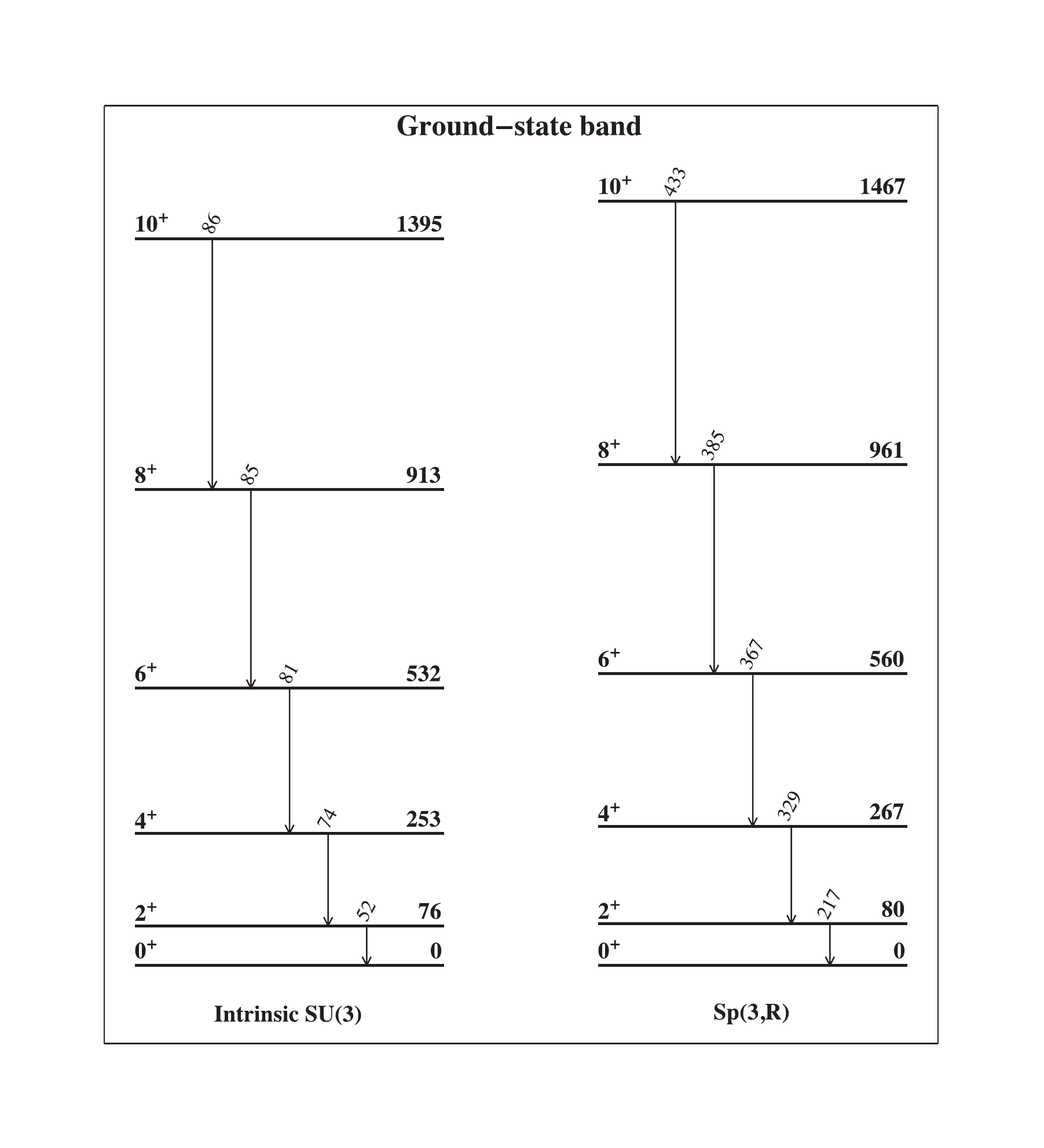}
\end{center}
\caption{Comparison of the energies and $B$(E2)'s calculated obtained from the uncoupled $\kappa_1=0$ (left)  and coupled $\kappa_1= 0.006$ (right) Hamiltonians for the ground state band (bottom) and the giant quadrupole  band (top).  Energies are given in keV and $B$(E2)'s are  in Weisskoff units. \label{fig:compare}
}
\end{figure}

Specifically, we apply the model Hamiltonian \eqref{eq:H_CRVo}
to a description of the  $^{166}$Er ground-state rotational band and
its giant-quadrupole resonance excitations. 
These are treated within  an  
$\langle 824(80\,0)\rangle$ symplectic irrep, resulting in the
energy levels and  $B$(E2) reduced transition probabilities for  the uncoupled  ($\kappa_1=0$) and coupled ($\kappa_1\not=0$) Hamiltonian  shown in Fig.\ \ref{fig:compare}.
In these calculations, the parameters in \eqref{eq:H_CRVo} were set 
as follows.  
For the intrinsic SU(3) Hamiltonian, 
\begin{equation}
\hat{H}_{\rm su3}=H_0-\chi\mathcal{Q}\cdot\mathcal{Q},
\label{eq:intrinsicH}
\end{equation}
$\chi=0.011$ was set to give the excitation energy of the first  $2^+$ state;   
the value  $\hbar\omega=7.36$ MeV was determined by use of the Blomqvist-Molinari formula \cite{BlomqvistM68}; 
values of $\kappa_2=0.019$  and $\kappa_3=0.025$ were chosen to give the  giant quadrupole resonance at $\sqrt{2}\hbar\omega=10.4$MeV and the giant monopole  resonance at $2\hbar\omega=14.7$ MeV;
the parameter $\kappa_1=0.006$ was then adjusted to obtain the experimental value of the $B({\rm E2};2^+\rightarrow 0^+)$  transition for the  ground-state rotational band.   

Because the asymptotic reduced matrix elements are indistinguishably close to the exact reduced matrix elements to the level of accuracy of these calculations, the asymptotic matrix elements were used to construct the Hamiltonian.  

The calculations were carried out in an Sp$(3,\Rb)$ space 
truncated by $
N_n= N_w - N_{w_0}\leq N_{\text{max}}$, with $N_{\text{max}}=20$. 
This truncated space was sufficiently large to provide  convergence of all observables to the accuracy shown in Fig.\ \ref{fig:compare}.

The calculations were carried out for an $N_{max}=20$ truncation.   The energies and $B$(E2)'s of the ground state band and the giant quadrupole resonance band are shown in Fig.\ \ref{fig:compare} for 
both the uncoupled ($\kappa_1=0$) and coupled ($\kappa_1\not=0$) 
Hamiltonian.

The results in Fig.\ \ref{fig:compare} show 
the extent to which the properties of a microscopic model Hamiltonian 
that have been derived in a  large multi-shell calculation can be obtained much more simply from its macroscopic rotor-vibrator limit.
The essential predictions of the CRV limit of the Hamiltonian
(\ref{eq:H_CRVo}) are that the coupling between low-energy U(3) and giant-resonance degrees of freedom has a small effect on the low energy-level spectrum but, 
for an appropriate value of the $\kappa_1$ coupling constant, 
it essentially doubles the effective charge of quadrupole operators.
In accord with these predictions, the figures  show remarkably little change in the energy-level spectra of either the ground state or giant-resonance rotational bands due to a relative strong coupling.
They also show the predicted enhancement of calculated quadrupole
 strengths. 
 Recall that the strength of the coupling was chosen 
 to reproduce  experimental E2 transition rates.  
The enhancement of $\sim1.8$ required to match experiment for the chosen representation is slightly smaller than, but comparable to, 
the factor of $\sim2$  expected on the basis of mean-field self-consistency considerations \cite{RoweRabida13,Rowe16}; 
cf.\  the results of Ref.\ \cite{BahriR00}.
 One must also be aware of the sensitivity of quadrupole moments and E2 transition rates to the tails of wave functions and, hence, to small components of the wave functions lost on truncation of the space and to the fundamental limitations of a very simple model.

\section{Spin-orbit interactions}
 The nuclear spin $S$ and  isospin $T$ are good quantum numbers of the  ${\rm Sp}(3,\Rb) \times {\rm U(2)}_S \times {\rm U(2)}_T$
coupling scheme.
However, the spatial and spin-isospin degrees of freedom  are inextricably intertwined 
by the total antisymmetry of many-nucleon wave functions.
Thus, an Sp$(3,\Rb)$ representation with a given space symmetry can only be combined with spin-isospin states of complementary symmetry. 
It follows that,  if the isospin of an Sp$(3,\Rb)$ representation of an even-even nucleus takes its smallest value $T=(N-Z)/2$, the spin $S$ of that representation is most often uniquely determined by the space symmetry of the  Sp$(3,\Rb)$ representation.

The spherical harmonic-oscillator states of even-even nuclei  that are most deformed and most lowered  in energy by a corresponding deformation of the harmonic-oscillator potential  
are states of maximal space symmetry with wave functions that are Slater-determinants of spin $S=0$ and isospin $T=(N-Z)/2$.
On the basis of shape-consistent mean-field considerations
\cite{Rowe16}, 
it is also expected that the 
${\rm Sp}(3,\Rb) \times {\rm U(2)}_S \times {\rm U(2)}_T$
representations of a given isospin are partially ordered with those of smaller spin  lying lower in energy, and those of $T=(N-Z)/2$ lying lowest.

For representations of non-zero spin, the intertwining of the spin and spatial degrees of freedom  in forming lowest-weight symplectic states is particularly significant because, 
in the macroscopic limit in which a lowest-weight state becomes the intrinsic state of a rotor, 
it generates rotational  states in which the spin and spatial degrees of freedom are strongly coupled as generally presumed in the Bohr-Mottelson-Nilsson model. 

This follows from the observation that an irreducible
${\rm Sp}(3,\Rb) \times {\rm U(2)}_S$ representation has
a macroscopic limit in which it approaches a
${\rm ROT(3)} \times {\rm U(2)}_S$ model
with the same angular momenta and
commuting quadrupole moment operators (also with high-energy giant-resonance vibrational states).
Thus, it is expected that the low-energy rotational states of heavy nuclei with non-zero spin, 
will be close to those  of a strongly coupled rotor model with both SO(3) rotational and SU(2) spin angular momenta
\beq J_k = L_k + S_k .\eeq
By definition, strong-coupling of the intrinsic spin of a rotor  takes place in the intrinsic frame of the rotor as in the Nilsson model.
It is favoured particularly in rotational nuclei in low total angular-momentum states, partly  because of the adiabaticity of the rotational degrees of freedom and, more importantly, because the existence of four spin-isospin states allows the spatial states of nucleon to be maximally aligned \cite{Mottelson62} to form states of large deformation.
This understanding underlies the interpretation of deformation alignment as opposed to rotational alignment and back-bending in nuclei \cite{StephensS72,Stephens75}.
It also counters the concern that, although the spin-orbit interactions do not conserve SU(3) or Sp$(3,\Rb)$ symmetry, their symmetry breaking effects are expected to be suppressed in strongly-deformed nuclei.

\section{Relationships with mean-field models} 
\label{sect:MFmodels}
The microscopic theory of collective dynamics has traditionally been approached in terms of Hartree-Fock (HF) mean-field theory, and its
Hartree-Fock-Bogolyubov (HFB) and time-dependent (TDHF and TDHFB) extensions \cite{Baranger60}.%
\footnote{An elementary introduction to the basic HF, HFB, and TDHF theories is given in \cite{Rowebook70}. 
A recent description of the application of this approach to the low-lying
rotor-vibrator  states of deformed nuclei has been given by
Matsuyanagi \emph{et al.}\ \cite{MatsuyanagiMNHS10}.}
For a deformed nucleus, static HF theory,  returns a minimal energy Slater determinant of single-nucleon states from among a continuous set of such determinants  that is
generally not rotationally invariant.
The rotations and small-amplitude vibrations about such a minimum energy HF equilibrium state are then given by the TDHF extension of HF theory.
A popular model based on this theory is 
Kumar and Baranger's pairing-plus-quadrupole model
\cite{BarangerK65,BarangerK68b,BarangerK68c,%
BarangerK68d,BarangerK68V}.

To appreciate the contribution that mean-field theory can make to the 
understanding of collective states of nuclei, it is useful to first  contrast the HF theory of nuclear states with the theory of many-nucleon quantum mechanics.  
In both cases one starts with a so-called particle-hole vacuum state
which is an anti-symmetric product of a subset of single-nucleon states described as occupied single-particle states.
Such a state is a lowest-weight state for the irrep of the infinite-dimensional Lie algebra of one-body  Hermitian operators, 
in the sense that it is annihilated by all the so-called particle-hole lowering operators.
From this single particle-hole vacuum state, one can use the particle-hole raising operators of the Lie algebra of  one-body operators to generate the whole many-nucleon Hilbert space of the nucleus.
On the other hand, by applying the Lie group of one-body unitary transformations to the particle-hole vacuum state, one generates the continuous HF manifold of all possible Slater determinants  for the nucleus.
The many-nucleon Hilbert space and HF manifold  are very different spaces but have an intimate and complementary relationship.  
Unlike a Hilbert space, the HF manifold is not a vector space which is evident from the fact that the sum of any two Slater determinants cannot be expressed as a single Slater determinant.
In fact, a HF manifold has the properties of a classical phase space whereas the Hilbert space generated by the particle-hole raising operators is its quantisation.
The geometry of a HF manifold as a classical phase and the TDHF dynamics on this space was shown, for example, in  \cite{Rowe82} 
(and references therein) and reviewed in Chapter 8 of 
\cite{RoweWood10}.
However, for present purposes we only need its simple familiar properties.

Some remarkable properties are  seen to emerge if one starts from a lowest-weight state for a symplectic model irrep and restricts the group of one-body unitary transformations to those of the unitary irrep of Sp$(3,\Rb)$ with this lowest-weight state.
It is well known that, for a suitable definition of the 
Sp$(3,\Rb)$  lowest-weight state as  a U(3) highest-weight state, the  Hilbert space of the lowest-grade U(3) representation is spanned by its SO(3) rotations through all angles.
It is similarly known \cite{VassanjiR84}, 
that the whole Hilbert space of the Sp$(3,\Rb)$  irrep with this lowest-weight state, is spanned by the general linear transformations of its lowest-weight state.

By standard mean-field methods, one can find a state on this manifold in which the expectation value of a rotationally invariant Hamiltonian is minimised and which is oriented such that its  monopole/quadrupole moments are aligned to give 
$\langle \hat Q_{ij}\rangle = \delta_{i,j} q_i$.
With respect to the 9-dimensional mean-field dynamics of the time-dependent mean-field equations of motion, the minimum energy state is then an equilibrium state with 3 rotational and 6 vibrational degrees of freedom.
Thus, it can be interpreted  as the intrinsic state of a microscopic version of a unified collective model.
Moreover, in its macroscopic limits, its rotational dynamics is asymptotical that of a triaxial rotor and its vibrational degrees of freedom contract to those of a boson algebra.
However, the quantisation of the microscopic version of this model, which is simply the symplectic model, is given by its irreducible unitary representation.

In this way one regains the CRV model and its relationship to the TDHF and TDHFB methods and the possibility of benefitting from the many insights obtained as, for example, the effects of pairing correlations.
The mean-field perspective on the properties of the symplectic model is further developed in a forthcoming review \cite{Rowe16b}.

\section{Conclusions and challenges for further progress}
Much has been gained from the  complementary perspectives of nuclear structure underlying the Nobel prize winning models of the 1950 years.
The Mayer-Jensen model of independent-particles in a spherical   field led to the nuclear shell model while the Bohr-Mottelson-Rainwater model of particles moving as a fluid in a spheroidal field led to the nuclear collective models.

This review has  focused on the evolution of  many-nucleon models of nuclear collective states and their macroscopic  limits that incorporate both of the early perspectives.
The approach is based  on the Dirac-Weyl
\cite{Dirac30,Weyl50}
formulation of the  quantum mechanics of a system in terms of  unitary representations of its algebra of observables.
The  perspective emerges that, whereas the many-nucleon Hilbert space
is a sum of  shell-model subspaces, it is also expressible as a
sum of collective-model subspaces.
Such collective-model subspaces are  usefully defined as the spaces
for irreducible representations of the direct product group 
${\rm Sp}(3,\Rb) \times {\rm U(2)}_S \times {\rm U(2)}_T$,
where U(2)$_S$ and U(2)$_T$ are, respectively, the nucleon spin and isospin groups.
The decomposition (discussed further in \cite{Rowe16})
corresponds to the use of the subgroup chains
\begin{subequations} \label{eq:SSM-basis}
\bal
&{\rm Sp}(3,\Rb) \supset {\rm U(3)} \supset {\rm SO(3)} ,\\
&{\rm SO(3)}\times {\rm SU(2)}_S \supset {\rm SU(2)}_J,
\end{align}
\end{subequations}
to define a coupling scheme for the many-nucleon Hilbert space.
It has the significant property that there can be no non-zero matrix elements of a nuclear quadrupole moment, or any other element of the
${\rm Sp}(3,\Rb) \times {\rm U(2)}_{S_n} \times {\rm U(2)}_{S_p}$ Lie algebra, 
between states that belong to different collective-model subspaces.
The selection rule associated with this property promises to provide a powerful tool in the experimental identification of the microscopic structure of observed collective states.

To describe the rotational properties  of a deformed nucleus, 
it is appropriate to consider the restriction of a  suitable Hamiltonian to a truncated many-nucleon space of many symplectic-model subspaces.
However, this poses challenges as to the choice of subspaces and  Hamiltonian to be employed.
If it were not for  practical limitations on the sizes of matrices that can be diagonalised, it would be meaningful to work in sufficiently large spherical harmonic-oscillator shell-model bases, that the results for the states of interest would not change to the required level of accuracy if the dimension of the shell-model space were increased.
The physical significance of the results of the calculation could then  be understood subsequently by expanding them in  symplectic-model bases.

In fact, such calculations with realistic interactions have been successfully pursued in multi-shell calculations of  low-energy states of light nuclei by Dytrych and colleagues \cite{DytrychSBDV07a,DytrychSBDV07b,DytrychSBDV08,%
DytrychSDBV08b,DytrychLDMVSCSLC13}.
The calculations were carried out in  
${\rm U(3)} \times {\rm U(2)}_S \times {\rm U(2)}_T$
coupled harmonic-oscillator shell-model  bases and are 
especially valuable because they show the emergence of results consistent with their expression in terms of a small number of
 ${\rm Sp}(3,\Rb) \times {\rm U(2)}_S \times {\rm U(2)}_T$
irreducible representations without \emph{a priori} assumptions that they should do so.
However, already for very light nuclei, such calculations require the use of supercomputers and it is inconceivable that corresponding calculations could be carried out, in the foreseeable future, for heavy nuclei for which the symplectic-model interpretation of collective dynamics was developed.
Thus, a  primary challenge in exploring the symplectic-model-based description of deformed nuclei is  to identify the relevant collective- (i.e., symplectic-) model subspaces that are appropriate for the nuclear states of interest.
This is important because, as discussed in several contributions to a focus issue on nuclear shape coexistence \cite{WoodHeyde}, 
there is strong evidence that many-nucleon states of much larger deformation than one would expect from a spherical shell-model perspective fall into  the low-energy domain, even in light nuclei.

In the  spherical shell model, many-nucleon subspaces are ordered by the sums of the independent-particle energies of $jj$-coupled basis states.
There is then a natural coupling scheme for a number of nucleons in such a subspace given by their pair-coupling \cite{Flowers52,FlowersS64a} symmetries.
However, such a coupling scheme is wildly inappropriate for heavy strongly-deformed nuclei for which the deformation of the mean field has a dominant controlling influence.

Within the framework of the collective model, the Nilsson model
\cite{Nilsson55}  and Mottelson's comparisons of aligned versus pair coupling \cite{Mottelson62} indicate that  deformation-alignment of nucleons  takes place in the intrinsic frame of a rotor.
Such a coupling has a natural realisation in mean-field theory in which states of good angular momentum are obtained by angular-momentum projection from minimal-energy mean-field states.
However, this approach does not, in general, translate easily into a rotationally invariant coupling scheme for the many-nucleon Hilbert space because states projected from different mean-field states do not form complete orthonormal sets of states.

To our knowledge, the only  coupling scheme for the many-nucleon Hilbert space that admits  such deformation-aligned coupling, without restriction to eigenstates of a spherical harmonic oscillator, corresponds to an $LS$-coupled decomposition of  the many-nucleon Hilbert space into a sum of 
${\rm Sp}(3,\Rb) \times {\rm U}(4)
\supset {\rm U(3)}\times  {\rm U(2)}_S \times {\rm U(2)}_T$
invariant subspaces, where U(4) is the Wigner supermultiplet group.
In companions to this review  \cite{Rowe16,Rowe16b}, the ordering of
${\rm Sp}(3,\Rb) \times {\rm U}(4)$ subspaces and the corresponding microscopic descriptions of collective states restricted to these subspaces is considered in terms of 
 mean-field methods. 

In this review, our primary concern has been with the 
interpretation of experimental data.
If the mixing of different collective model subspaces is small, it would appear that, by analysis of  experimental data in terms of the macroscopic limits of low-energy rotational bands as  U(3)-rotor bands with an effective charge,
 it should be possible to identify candidates for the collective-model subspaces involved along the lines pursued by
Jarrio \emph{et al.} \cite{JarrioWR91}.
Such analyses would at least assist in identifying critical data for which more accurate measurements are needed.

Identifying critical data for estimating $K$-band mixing in the description of rotational states is a particular challenge.
It is difficult because the interband E2 transitions between states of good 
$K$ quantum numbers are small in comparison with large in-band transitions for close to axially symmetric nuclei.
This suggests that the way to proceed would be to start with fits of macroscopic limits of the symplectic models, and their mixtures, to  large bodies of experimental data.
The systematics of nuclear shapes of nuclei 
and information gathered from nucleon transfer reactions and their Nilsson model interpretation would  undoubtedly also provide valuable information.

In concluding this summary,  it is emphasised that
an essential difference between the Bohr-Mottelson-Nilsson and the microscopically-based  model presented is that the BMN rotor is axially-symmetric and has low-energy beta- and gamma-vibrational excitations, whereas the macroscopic limit of the symplectic model is a triaxial rotor with only high-energy giant-resonance excitations.
Hopefully the analysis of experimental data will be able to distinguish between the two models.
It must be recognised, however, that with  coherent mixtures of rotational states it may be possible to generate states with some of the properties of beta and gamma vibrational states.
For example, there is a distinct possibility that pairing interaction could effect a partial restoration of the axial symmetry by mixing different symplectic-model representations.
A systematic study of  candidates for one-phonon gamma vibrational states could be  rewarding from this perspective
and that of quasi-dynamical symmetry \cite{RoweRR88,RoweCancun04}.

Major challenges to unravelling the experimental implications for the microscopic structure of the rotational states of nuclei arise both from the mixing of different representations  and the mixing of different rotational $K$ bands within a given representation.
These are huge problems but hopefully they can be tackled from both ends, i.e., by considering theoretical models that produce results consistent with observed data.

An important consideration in the analysis of experimental data
 is that many candidates for
${\rm Sp}(3,\Rb) \times {\rm U(2)}_S \times {\rm U(2)}_T$
representations of the same spin and isospin are  similar, in that their matrix elements are close to being linearly related to one another.
As a result their differences can be accommodated for the most part by choice of an effective charge.
It has also been observed that 
a symmetry breaking interaction has a tendency to mix the representations of a would-be dynamical symmetry in the coherent manner of the previously-mentioned quasi-dynamical symmetry 
\cite{RoweRR88,RoweCancun04}.
When this happens, the states of the mixed representations behave as though they belonged to an average of the mixed representations
and the transitions between such mixed  bands of states  tend to be systematic and relatively easy to interpret.
However, this is a subject that merits further and detailed examination.

It is important to recognise that one can learn as much, perhaps more, from the limitations of a model as from its successes.
Thus, one should always look  for experimental or theoretical observations that disagree and are not naturally explained in terms of a  model.  
In the present context, it promises to be useful to explore the systematics of E2 and other transitions between states that are 
contrary to the expectations of
the simple symplectic model in its asymptotic limits and have the potential to reveal phenomena associated with competing dynamical symmetries.

\begin{acknowledgements}
The perspective of this review has benefitted from numerous interactions and discussions with  John Wood, George Rosensteel, Juliana Carvalho, Joe Repka, 
Ted Hecht, Jerry Draayer, Tomas Dytrych, Kristina Launey, and others.
It was supported in part  by the U.S.\;Department of Energy, Office of Science, under Award Number\;DE-FG02-95ER-40934.  
Computational resources were
provided by the Notre Dame Center for Research Computing.
\end{acknowledgements}



\begin{thebibliography}{100}

\bibitem{Schmidt40}
T.~Schmidt.
\newblock {\em Naturwiss}, 28:565, 1940.

\bibitem{TownesFL49}
C.~H. Townes, H.~M. Foley, and W.~Low.
\newblock Nuclear quadrupole moments and nuclear shell structure.
\newblock {\em Phys. Rev.}, 76:1415--1416, 1949.

\bibitem{Rainwater50}
James Rainwater.
\newblock Nuclear energy level argument for a spheroidal nuclear model.
\newblock {\em Phys. Rev.}, 79:432--434, Aug 1950.

\bibitem{Bohr52}
A.~Bohr.
\newblock The coupling of nuclear surface oscillations to the motion of
  individual nucleons.
\newblock {\em Mat. Fys. Medd. Dan. Vid. Selsk.}, 26(14):1--40, 1952.

\bibitem{Mayer49}
M.~G. Mayer.
\newblock On closed shells in nuclei. ii.
\newblock {\em Phys. Rev.}, 75:1969--1970, 1949.

\bibitem{HaxelJS49}
O.~Haxel, J.~H.~D. Jensen, and H.~E. Suess.
\newblock On the "magic numbers" in nuclear structure.
\newblock {\em Phys. Rev.}, 75:1766--1766, Jun 1949.

\bibitem{BohrM53}
A.~Bohr and B.~R. Mottelson.
\newblock Collective and individual-particle aspects of nuclear structure.
\newblock {\em Mat. Fys. Medd. Dan. Vid. Selsk.}, {27, no. 16}, 1953.

\bibitem{Nilsson55}
S.~G. Nilsson.
\newblock Binding states of individual nucleons in a strongly deformed field.
\newblock {\em Mat. Fys. Medd. Dan. Vid. Selsk.}, {29, no. 16}, 1955.

\bibitem{DavydovF58}
A.~S. Davydov and G.~F. Filippov.
\newblock Rotational states in even atomic nuclei.
\newblock {\em Nucl. Phys.}, 8:237--249, 1958.

\bibitem{GneussG71}
G.~Gneuss and W.~Greiner.
\newblock Collective potential energy surfaces and nuclear structure.
\newblock {\em Nucl. Phys. A}, 171:449--479, 1971.

\bibitem{HessSMG80}
P.~O. Hess, M.~Seiwert, J.~Maruhn, and W.~Greiner.
\newblock General collective model and its application to {$^{238}_{\;\;92}$U}.
\newblock {\em Z. Phys. A}, 296:147--163, 1980.

\bibitem{Stephens75}
F.~S. Stephens.
\newblock Coriolis effects and rotation alignment in nuclei.
\newblock {\em Rev. Mod. Phys.}, 47(1):43, 1975.

\bibitem{RoweWC09}
D.~J. Rowe, T.~A. Welsh, and M.~A. Caprio.
\newblock Bohr model as an algebraic collective model.
\newblock {\em Phys. Rev. C}, 79:054304(1--16), 2009.

\bibitem{BenderHR03}
M.~Bender, P.-H. Heenen, and P.-G. Reinhard.
\newblock Self-consistent mean-field models for nuclear structure.
\newblock {\em Rev. Mod. Phys.}, 75:121--180, 2003.

\bibitem{MatsuyanagiMNHS10}
K.~Matsuyanagi, M.~Matsuo, T.~Nakatsukasa, N.~Hinohara, and K.~Sato.
\newblock Open problems in the microscopic theory of large-amplitude collective
  motion.
\newblock {\em J. Phys. G: Nucl. Part. Phys.}, 37:064018(16), 2010.

\bibitem{Belyaev59}
S.~T. Belyaev.
\newblock Effect of pairing correlations on nuclear properties.
\newblock {\em Mat. Fys. Medd. Dan. Vid. Selsk}, 31(11):1--55, 1959.

\bibitem{KisslingerS63}
L.~S. Kisslinger and R.~A. Sorensen.
\newblock Spherical nuclei with simple residual forces.
\newblock {\em Rev. Mod. Phys.}, 35:853--915, 1963.

\bibitem{BarangerK68V}
K.~Kumar and M.~Baranger.
\newblock Nuclear deformations in the pairing-plus-quadrupole model {(V).}
  {E}nergy levels and electromagnetic moments of the {W, Os} and {Pt} nuclei.
\newblock {\em Nucl. Phys. A}, 122:273--324, 1968.

\bibitem{Mottelson62}
B.~R. Mottelson.
\newblock In G.~Racah, editor, {\em Nuclear Spectroscopy}, {Rendiconti Della
  Scuola Internazionale Di Fisica "Enrico Fermi:" Corso XV}, page~44, Bologna,
  1962. Zanichelli.

\bibitem{BohrMP58}
A.~Bohr, B.~R. Mottelson, and D.~Pines.
\newblock Possible analogy between the excitation spectra of nuclei and those
  of the superconducting metallic state.
\newblock {\em Phys. Rev.}, 110(4):936--938, 1958.

\bibitem{Elliott58ab}
J.~P. Elliott.
\newblock Collective motion in the nuclear shell model.
\newblock {\em Proc. Roy. Soc. (London)}, A245:{128, 562}, {1958}.

\bibitem{Tomonaga55}
S.~Tomonaga.
\newblock Elementary theory of quantum-mechanical collective motion of
  particles.
\newblock {\em Prog. Theor. Phys.}, 13:467--496, 1955.

\bibitem{MiyazimaT56}
T.~Miyazima and T.~Tamura.
\newblock A collective description of the surface oscillation of atomic nuclei.
\newblock {\em Prog. Theor. Phys.}, 15(3):255--272, 1956.

\bibitem{ScheidG68}
W.~Scheid and W.~Greiner.
\newblock {\em Ann. Phys. (N.Y.)}, 48:493--496, 1968.

\bibitem{Cusson68}
R.~Y. Cusson.
\newblock A study of collective motion (i). rigid, liquid and related
  rotations.
\newblock {\em Nucl. Phys. A}, 114:289--308, 1968.

\bibitem{Rowe70}
D.~J. Rowe.
\newblock How do deformed nuclei rotate?
\newblock {\em Nucl. Phys. A}, 152:273--294, 1970.

\bibitem{VillarsC70}
F.~M.~H. Villars and G.~Cooper.
\newblock {\em Ann. Phys. (N.Y.)}, 56:224--258, 1970.

\bibitem{Zickendraht71}
W.~Zickendraht.
\newblock Collective and single-particle coordinates in nuclear physics.
\newblock {\em J. Math. Phys.}, 12:1663--1674, 1971.

\bibitem{DzyublikOSF72}
A.~Y. Dzyublik, V.~I. Ovcharenko, A.~I. Steshenko, and G.~V. Filippov.
\newblock {\em Yad. Fiz.}, 15:869--879, 1972.
\newblock (Sov. J. Nucl. Phys. {\bf 15} 487--492).

\bibitem{GulshaniR76}
P.~Gulshani and D.~J. Rowe.
\newblock Collective motions in nuclei and the spectrum generating algebras
  {$T_5\times$ SO(3), GL(3,R), and CM(3)}.
\newblock {\em Can. J. Phys.}, 54:970--996, 1976.

\bibitem{BuckBC79}
B.~Buck, L.~C. Biedenharn, and R.~Y. Cusson.
\newblock Collective variables for the description of rotational motion in many
  particle systems.
\newblock {\em Nucl. Phys. A}, 317:205--241, 1979.

\bibitem{RoweR79}
D.~J. Rowe and G.~Rosensteel.
\newblock Geometric derivation of the kinetic energy in collective models.
\newblock {\em J. Math. Phys.}, 20:465--468, 1979.

\bibitem{RoweR80}
D.~J. Rowe and G.~Rosensteel.
\newblock On the algebraic formulation of collective models {II}: collective
  and intrinsic submanifolds.
\newblock {\em Ann. Phys. (N.Y.)}, 126:198--233, 1980.

\bibitem{Ui70}
H.~Ui.
\newblock {Quantum mechanical rigid rotator with an arbitrary deformation. I}.
\newblock {\em Prog. Theor. Phys.}, 44:153--171, 1970.

\bibitem{WeaverBC73}
L.~Weaver, L.~C. Biedenharn, and R.~Y. Cusson.
\newblock Rotational bands in nuclei as induced group representation.
\newblock {\em Ann. Phys. (N.Y.)}, 77:250--278, 1973.

\bibitem{RosensteelR77a}
G.~Rosensteel and D.~J. Rowe.
\newblock Nuclear {Sp(3,R)} model.
\newblock {\em Phys. Rev. Lett.}, 38:10--14, 1977.

\bibitem{RosensteelR80}
G.~Rosensteel and D.~J. Rowe.
\newblock On the algebraic formulation of collective models {III}: the
  symplectic shell model of collective motion.
\newblock {\em Ann. Phys. \textup(NY\textup)}, 126:343--370, 1980.

\bibitem{Rowe85}
D.~J. Rowe.
\newblock Microscopic theory of the nuclear collective model.
\newblock {\em Rep. Prog. Phys.}, 48:1419--1480, 1985.

\bibitem{Filippov73}
G.~F. Filippov.
\newblock {\em Fiz. Elem. Chastits At. Yad.}, 4:992--1017, 1973.
\newblock (Sov. J. Part. Nucl. {\bf 4} 405--415).

\bibitem{Filippov78}
G.~F. Filippov.
\newblock {\em Fiz. Elem. Chastits At. Yad.}, 9:1241--1281, 1978.
\newblock (Sov. J. Part. Nucl. {\bf 9} 486--501).

\bibitem{Simonov68}
{Yu}.~A. Simonov.
\newblock {\em Yad. Fiz.}, 7:1210--1220, 1968.
\newblock (Sov. J. Nucl. Phys. {\bf 7} 722--727).

\bibitem{MoshinskyQ71}
M.~Moshinsky and C.~Quesne.
\newblock {Linear canonical transformations and their unitary representations.}
\newblock {\em J. Math. Phys.}, 12:1722--1780, 1971.

\bibitem{Kretzchmar60}
M.~Kretzchmar.
\newblock Gruppentheoretische untersuchungen zum schalenmodell.
\newblock {\em Z. Phys.}, 158:284--303, 1960.

\bibitem{NeudachinS69}
V.~G. Neudachin and {Yu}.~F. Smirnov.
\newblock {\em Nucleon Clusters in Light Nuclei}.
\newblock Nauka, Moscow, 1969.

\bibitem{Vanagas80}
V.~V. Vanagas.
\newblock Microscopic theory of the nucleus in the framework of restricted
  dynamics.
\newblock {\em Fiz. Elem. Chastits At. Yad.}, 11:454--514, 1980.
\newblock (Sov. J. Part. Nucl. {\bf 11} 172--197).

\bibitem{RoweCR12}
D.~J. Rowe, M.~J. Carvalho, and J.~Repka.
\newblock Dual pairing of symmetry groups and dynamical groups in physics.
\newblock {\em Rev. Mod. Phys.}, 84:711--757, 2012.

\bibitem{vonNeumann32}
J.~von Neumann.
\newblock Uber einen satz herrn m. h. stone.
\newblock {\em Annals of Math.}, 33(3):567--573, 1932.

\bibitem{InonuW53}
E.~{\.I}n{\"o}n{\"u} and E.~P. Wigner.
\newblock On the contraction of groups and their representations.
\newblock {\em Proc. Natl. Acad. Sci, (U.S.A.)}, 39:510--524, 1953.

\bibitem{WeaverCB76}
L.~Weaver, R.~Y. Cusson, and L.~C. Biedenharn.
\newblock Nuclear rotational-vibrational collective motion with nonvanishing
  vortex spin.
\newblock {\em Ann. Phys. (N.Y.)}, 102:493--569, 1976.

\bibitem{RosensteelR76}
G.~Rosensteel and D.~J. Rowe.
\newblock The algebraic {CM(3)} model.
\newblock {\em Ann. Phys. \textup(NY\textup)}, 96:1--42, 1976.

\bibitem{FilippovOS72}
G.~F. Filippov, V.~I. Ovcharenko, and A.~I. Steshenko.
\newblock Generalized {Eulerian} angles and the collective motion of
  many-particle system.
\newblock In F.~Calogero and C.~{Ciofi degli Atti}, editors, {\em Proceedings
  of the International Symposium on Present Status and Novel Developments in
  the Nuclear Many-Body Problem: Roma, 1972}, pages 627--668, Bologna, 1973.
  Editrice Compositori.

\bibitem{RoweB00}
D.~J. Rowe and C.~Bahri.
\newblock {C}lebsch-{G}ordan coefficients of {SU(3)} in {SU(2)} and {SO(3)}
  bases.
\newblock {\em J. Math. Phys.}, 41(9):6544--6565, 2000.

\bibitem{DreyfussLDDB13}
A.~Dreyfuss, K.~D. Launey, T.~Dytrych, J.~P. Draayer, and C.~Bahri.
\newblock Hoyle state and rotational features in {C}arbon-12 within a no-core
  shell-model framework.
\newblock {\em Phys. Lett. B}, 727:511--515, 2013.

\bibitem{TobinFLDDDB14}
G.~K. Tobin, M.~C. Ferriss, K.~D. Launey, T.~Dytrych, J.~P. Draayer, A.~C.
  Dreyfuss, and C.~Bahri.
\newblock Symplectic no-core shell-model approach to intermediate-mass nuclei.
\newblock {\em Phys. Rev. C}, 89:034312, Mar 2014.

\bibitem{FilippovO80}
G.~F. Filippov and I.~P. Okhrimenko.
\newblock {\em Sov. J. Nucl. Phys.}, 32:37, 1980.

\bibitem{OkhrimenkoS81}
I.~P. Okhrimenko and A.~I. Steshenko.
\newblock {\em Sov. J. Nucl. Phys.}, 34:488, 1981.

\bibitem{VassanjiR82}
M.~G. Vassanji and D.~J. Rowe.
\newblock The geometric {SO(3) $\times$ D} model.
\newblock {\em Phys. Lett. B}, 115(2):77--80, 1982.

\bibitem{VassanjiR84}
M.~G. Vassanji and D.~J. Rowe.
\newblock The geometric {SO(3) $\times$ D} model; a practical microscopic
  theory of quadrupole collective motion.
\newblock {\em Nucl. Phys. A}, 426:205--221, 1984.

\bibitem{CarvalhoVR87}
M.~J. Carvalho, M.~Vassanji, and D.~J. Rowe.
\newblock Application of the symplectic shell model to the {$L=0$} states of
  {$^4$He}.
\newblock {\em Nucl. Phys. A}, 465:265--273, 1987.

\bibitem{CarvalhoRKB02}
M.~J. Carvalho, D.~J. Rowe, S.~Karram, and C.~Bahri.
\newblock Optimal basis states for a microscopic calculation of intrinsic
  vibrational wave functions of deformed rotational nuclei.
\newblock {\em Nucl. Phys. A}, 703:167--187, 2002.

\bibitem{MayerJ55}
M.~G. Mayer and J.~H.~D. Jensen.
\newblock {\em Elementary Theory of Nuclear Shell Structure}.
\newblock Wiley, New York, 1955.

\bibitem{BassichisR65}
W.~H. Bassichis and G.~Ripka.
\newblock A {H}artree-{F}ock calculation of excited states of {$^{16}$O}.
\newblock {\em Phys. Lett.}, 15(4):320--322, 1965.

\bibitem{BrownG66}
G.~E. Brown and A.~M. Green.
\newblock Even parity states of {$^{16}$O} and {$^{17}$O}.
\newblock {\em Nucl. Phys.}, 75:401--417, 1966.

\bibitem{Suzuki76b}
Y.~Suzuki.
\newblock Structure study of {$T=0$} states in {$^{16}$O} by {$^{12}{\rm
  C}+\alpha$} cluster-coupling model{. II}.
\newblock {\em Prog. Theor. Phys.}, 56(1):111--123, 1976.

\bibitem{RoweTW06}
D.~J. Rowe, G.~Thiamova, and J.~L. Wood.
\newblock Implications of deformation and shape coexistence for the nuclear
  shell model.
\newblock {\em Phys. Rev. Lett.}, 97:202501, 2006.

\bibitem{HeydeW11}
K.~Heyde and J.~L. Wood.
\newblock Shape coexistence in atomic nuclei.
\newblock {\em Rev. Mod. Phys.}, 83:1467--1521, 2011.

\bibitem{JarrioWR91}
M.~Jarrio, J.~L. Wood, and D.~J. Rowe.
\newblock The su(3) structure of rotational states in heavy deformed nuclei.
\newblock {\em Nucl. Phys. A}, 528:409--435, 1991.

\bibitem{Rowe16}
D.~J. Rowe.
\newblock The emergence of deformation and rotational states in the
  many-nucleon quantum theory of nuclei.
\newblock {\em J. of Phys.G: Nucl. Part. Phys.}, {Focus Issue on Shape
  Coexistence in Nuclei} (to be published), 2016.

\bibitem{Rowe16b}
D.~J. Rowe.
\newblock The emergence and use of symmetry in the many-nucleon model of atomic
  nuclei.
\newblock In K.~D. Launey, editor, {\em Emergent phenomena in Atomic Nuclei
  from Large-scale Modeling: a Symmetry-guided Perspective}. World Scientific,
  2016.

\bibitem{FrenchHMW69}
J.~B. French, C.~E. Halbert, J.~B. McGrory, and S.~S.~M. Wong.
\newblock Complex spectroscopy.
\newblock In M.~Baranger and E.~Vogt, editors, {\em Advances in Nuclear
  Physics}, volume~3, pages 193--258. Plenum, New York, 1969.

\bibitem{HechtPang69}
K.~T. Hecht and S.~C. Pang.
\newblock On the wigner supermultiplet scheme.
\newblock {\em J. Math. Phys.}, 10:1571--1616, 1969.

\bibitem{Flowers52}
B.~H. Flowers.
\newblock {Studies in $jj$-coupling. I. Classification of nuclear and atomic
  states.}
\newblock {\em Proc. Roy. Soc. London}, A212:248--263, 1952.

\bibitem{FlowersS64a}
B.~H. Flowers and S.~Szpikowski.
\newblock A generalized quasi-spin formalism.
\newblock {\em Proc. Phys. Soc.}, 84:193--199, 1964.

\bibitem{RoweCancun04}
D.~J. Rowe.
\newblock Embedded representations and quasi-dynamical symmetry.
\newblock In J.~Escher, O.~Casta{\~n}os, J.~Hirsch, S.~Pittel, and
  G.~Stoitcheva, editors, {\em Computational and Group-Theoretical Methods in
  Nuclear Physics}, pages 165--173, Singapore, 2004. World Scientific.
\newblock arXiv:1106.1607 [nucl-th].

\bibitem{Whitehead72}
R.~R. Whitehead.
\newblock A numerical approach to nuclear shell-model calculations.
\newblock {\em Nucl. Phys.}, A182:290--300, 1972.

\bibitem{WhiteheadWCM77}
R.~R. Whitehead, A.~Watt, B.~J. Cole, and I.~Morrison.
\newblock Computational methods for shell-model calculations.
\newblock In M.~Baranger and E.~Vogt, editors, {\em Advances in Nuclear
  Physics}, volume~9, chapter~2, page 123. Plenum, New York, 1977.

\bibitem{SuzukiH86}
Y.~Suzuki and K.~T. Hecht.
\newblock Symplectic and cluster excitations in nuclei: evaluation of
  interaction matrix elements.
\newblock {\em Nucl. Phys. A}, 455:315--343, 1986.

\bibitem{SuzukiH87}
Y.~Suzuki and K.~T. Hecht.
\newblock Spin--orbit and tensor interactions in sp(6,{R})-model calculations.
\newblock {\em Prog. Theor. Phys.}, 77:190--195, 1987.

\bibitem{EscherD98}
J.~Escher and J.~P. Draayer.
\newblock Fermion realization of the nuclear {Sp(6,R)} model.
\newblock {\em J. Math. Phys.}, 29(10):5123--5147, 1998.

\bibitem{AkiyamaD73}
Y.~Akiyama and J.~P. Draayer.
\newblock A user's guide to fortran programs for {Wigner and Racah}
  coefficients of {SU3}.
\newblock {\em Comp. Phys. Comm.}, 5:405--415, 1973.

\bibitem{DraayerA73}
J.~P. Draayer and Y.~Akiyama.
\newblock Wigner and {R}acah coefficients for {SU3}.
\newblock {\em J. Math. Phys.}, 14:1904--1912, 1973.

\bibitem{Millener78}
D.~J. Millener.
\newblock A note on recoupling coefficients for su(3).
\newblock {\em J. Math. Phys.}, 19:1513--1514, 1978.

\bibitem{Braunschweig78}
D.~Braunschweig.
\newblock Reduced su(3) cfp.
\newblock {\em Comput. Phys. Commun.}, 14:109--129, 1978.

\bibitem{Braunschweig78b}
D.~Braunschweig.
\newblock {II}. reduced su(3) matrix elements.
\newblock {\em Comput. Phys. Comm.}, 15:259--273, 1978.

\bibitem{BahriD94}
C.~Bahri and J.~P. Draayer.
\newblock {SU(3)} reduced matrix element package.
\newblock {\em Comput. Phys. Comm.}, 83:59--94, 1994.

\bibitem{DytrychSBDV07a}
T.~Dytrych, K.~D. Sviratcheva, C.~Bahri, J.~P. Draayer, and J~P. Vary.
\newblock Evidence for symplectic symmetry in \textit{ab~Initio} no-core shell
  model results for light nuclei.
\newblock {\em Phys. Rev. Lett.}, 98:162503, Apr 2007.

\bibitem{DytrychSBDV07b}
T.~Dytrych, K.~D. Sviratcheva, C.~Bahri, J.~P. Draayer, and J.~P. Vary.
\newblock Dominant role of symplectic symmetry in \textit{ab initio} no-core
  shell model results for light nuclei.
\newblock {\em Phys. Rev. C}, 76:014315(1--9), Jul 2007.

\bibitem{DytrychSBDV08}
T.~Dytrych, K.~D. Sviratcheva, C.~Bahri, J.~P. Draayer, and J.~P. Vary.
\newblock Highly deformed modes in the \textit{ab initio} symplectic no-core
  model.
\newblock {\em J. of Phys.G: Nucl. Part. Phys.}, 35:095101(11), 2008.

\bibitem{DytrychSDBV08b}
T.~Dytrych, K.~D. Sviratcheva, J.~P. Draayer, C.~Bahri, and J.~P. Vary.
\newblock \textit{Ab initio} symplectic no-core shell model.
\newblock {\em J. of Phys.G: Nucl. Part. Phys.}, 35:123101(47), 2008.

\bibitem{Mackey68}
G.W. Mackey.
\newblock {\em Induced Representations of Groups and Quantum Mechanics}.
\newblock Benjamin, New York, 1968.

\bibitem{Perelomov85}
A.~Perelomov.
\newblock {\em Generalized Coherent States and their Applications}.
\newblock Springer, Berlin, 1986.

\bibitem{RoweR91}
D.~J. Rowe and J.~Repka.
\newblock Vector-coherent-state theory as a theory of induced representations.
\newblock {\em J. Math. Phys.}, 32:2614--2634, 1991.

\bibitem{Klauder63}
J.~R. Klauder.
\newblock Continuous-representation theory.
\newblock {\em J. Math. Phys.}, 4:1055--1074, 1963.

\bibitem{Sudarshan63}
R.~C.~G. Sudarshan.
\newblock Equivalence of semiclassical and quantum mechanical descriptions of
  statistical light beams.
\newblock {\em Phys. Rev. Lett.}, 10(7):277--279, 1963.

\bibitem{Perelomov72}
A.~M. Perelomov.
\newblock Coherent states for arbitrary lie group.
\newblock {\em Commun. Math. Phys.}, 26:222--236, 1972.

\bibitem{Gilmore72}
R.~Gilmore.
\newblock Geometry of symmetrized states.
\newblock {\em Ann. Phys. (N.Y.)}, 74:391--463, 1972.

\bibitem{Onofri75}
E.~Onofri.
\newblock A note on coherent state representations of {Lie} groups.
\newblock {\em J. Math. Phys.}, 16(5):1087--1089, 1975.

\bibitem{KlauderB-S85}
J.~R. Klauder and {{B.-S.} Skagerstam}, editors.
\newblock {\em Coherent states: applications in physics and mathematical
  physics}.
\newblock World Scientific, 1985.

\bibitem{RoweRC84}
D.~J. Rowe, G.~Rosensteel, and R.~Carr.
\newblock {Analytical expressions for the matrix elements of the non-compact
  symplectic algebra.}
\newblock {\em J. Phys. A: Math. Gen.}, 17:L399--L403, 1984.

\bibitem{Rowe84}
D.~J. Rowe.
\newblock Coherent state theory of the non-compact symplectic group.
\newblock {\em J. Math. Phys.}, 25:2662--2671, 1984.

\bibitem{DeenenQ84}
J.~Deenen and C.~Quesne.
\newblock Partially coherent states of the real symplectic group.
\newblock {\em J. Math. Phys.}, 25(8):2354--2366, 1984.

\bibitem{Rowe12}
D.~J. Rowe.
\newblock Vector coherent state representations and their inner products.
\newblock {\em J. Phys A: Math. Theor.}, 45:244003, 2012.

\bibitem{Bargmann61}
V.~Bargmann.
\newblock On a hilbert space of analytic functions and an associated integral
  transform part i.
\newblock {\em Commun. Pure Appl. Math.}, 14:187--214, 1961.

\bibitem{Dobaczewski81}
J.~Dobaczewski.
\newblock A unification theory of boson expansion theories {(I)}. functional
  representations of fermion states.
\newblock {\em Nucl. Phys. A}, 369:213--236, 1981.

\bibitem{Dobaczewski81b}
J.~Dobaczewski.
\newblock A unification theory of boson expansion theories (ii). boson
  expansion as provided by the functional representation method.
\newblock {\em Nucl. Phys. A}, 369:237--257, 1981.

\bibitem{Dobaczewski82}
J.~Dobaczewski.
\newblock A unification theory of boson expansion theories (iii). applications.
\newblock {\em Nucl. Phys.}, 380:1--26, 1982.

\bibitem{Dyson56}
F.~J. Dyson.
\newblock {\em Phys. Rev.}, 102:1217, 1956.

\bibitem{Rowe95}
D.~J. Rowe.
\newblock Resolution of missing label problems; a new perspective on
  {$K$}-matrix theory.
\newblock {\em J. Math. Phys.}, 36:1520--1530, 1995.

\bibitem{Pursey63}
D.~L. Pursey.
\newblock Irreducible representations of the 'unitary symmetry' group.
\newblock {\em Proc. Phys. Soc. A}, 275:284--294, 1963.

\bibitem{BairdB63}
G.E. Baird and L.C. Biedenharn.
\newblock On the representations of the semisimple lie groups. ii.
\newblock {\em J. Math. Phys.}, 4:1449--1466, 1963.

\bibitem{RoweLeBR89}
D.~J. Rowe, R.~{Le Blanc}, and J.~Repka.
\newblock A rotor expansion of the su(3) lie algebra.
\newblock {\em J. Phys. A: Math. Gen.}, 22:L309--L316, 1989.

\bibitem{RoweVC89}
D.~J. Rowe, M.~G. Vassanji, and M.~J. Carvalho.
\newblock The coupled-rotor-vibrator model.
\newblock {\em Nucl. Phys. A}, 504:76--102, 1989.

\bibitem{Rowebook70}
D.~J. Rowe.
\newblock {\em Nuclear Collective Motion: Models and Theory}.
\newblock Methuen, London, 1970.
\newblock (Reprinted by World Scientific 2010).

\bibitem{BohrM75}
A.~Bohr and B.~R. Mottelson.
\newblock {\em Nuclear Structure}, volume~2.
\newblock Benjamin, Reading, Mass, 1975.
\newblock (republished 1998 by World Scientific, Singapore).

\bibitem{Harvey68}
M.~Harvey.
\newblock The nuclear {SU3} model.
\newblock In M.~Baranger and E.~Vogt, editors, {\em Advances in Nuclear
  Physics}, volume~1, page~67. Plenum, New York, 1968.

\bibitem{Hecht65}
K.~T. Hecht.
\newblock {SU(3)} recoupling and fractional parentage in the 2s-1d shell.
\newblock {\em Nucl. Phys.}, 62:1--36, 1965.

\bibitem{Akiyama66}
Y.~Akiyama.
\newblock Tables of the su3 fractional-parentage and clebsch-gordan
  coefficients in the 2s-2d shell.
\newblock {\em Nucl. Data A}, 2:403--428, 1966.

\bibitem{Sebe68}
T.~Sebe.
\newblock {A note on the SU3 couping coefficients in the 2s-1d shell}.
\newblock {\em Nucl. Phys.}, A109:65--80, 1968.

\bibitem{ElliottH63}
J.~P. Elliott and M.~Harvey.
\newblock {Collective motion in the nuclear shell model. III: The calculation
  of spectra.}
\newblock {\em Proc. Roy. Soc. London}, A272:557--577, 1963.

\bibitem{CarvalhoPhD84}
M.~J. Carvalho.
\newblock {\em Algebraic study of rotational features in nuclei}.
\newblock PhD thesis, University of Toronto, Toronto, 1984.

\bibitem{LeBlancCVR86}
R.~{Le Blanc}, J.~Carvalho, M.~G. Vassanji, and D.~J. Rowe.
\newblock An effective shell-model theory of collective states.
\newblock {\em Nucl. Phys. A}, 452:263--276, 1986.

\bibitem{LeschberD86}
Y.~Leschber and J.~P. Draayer.
\newblock Macroscopic limit of the microscopic su(3) $\supset$ so(3) integrity
  basis interaction.
\newblock {\em Phys. Rev. C}, 33:749--751, 1986.

\bibitem{LeschberDR86}
Y.~Leschber, J.~P. Draayer, and G.~Rosensteel.
\newblock Connection between macroscopic and microscopic hamiltonians for
  nuclear rotational motion.
\newblock {\em J. of Phys.G: Nucl. Part. Phys.}, 12:L179--L183, 1986.

\bibitem{CastanosDL88}
O.~Casta{\~n}os, J.~P. Draayer, and Y.~Leschber.
\newblock Shape variables and the shell model.
\newblock {\em Z. Phys. A}, 329:33--43, 1988.

\bibitem{NaqviD90}
H.~A. Naqvi and J.~P. Draayer.
\newblock Shell-model operator for {K-band} splitting.
\newblock {\em Nucl. Phys. A}, 516:351--364, 1990.

\bibitem{LeschberD87}
Y.~Leschber and J.~P. Draayer.
\newblock Algebraic realization of rotor dynamics.
\newblock {\em Phys. Lett. B}, 190:1--6, 1987.

\bibitem{RosensteelR77b}
G.~Rosensteel and D.J. Rowe.
\newblock On the shape of deformed nuclei.
\newblock {\em Ann. Phys. (N.Y.)}, 104:134--144, 1977.

\bibitem{RoweT08}
D.~J. Rowe and G.~Thiamova.
\newblock {Construction of SU(3) irreps in canonical SO(3)-coupled bases}.
\newblock {\em J. Phys A: Math. Theor.}, 41:065206(1--9), 2008.

\bibitem{Rosensteel80}
G.~Rosensteel.
\newblock A recursion formula for {sp(3,R)} matrix elements.
\newblock {\em J. Math. Phys.}, 21:924--927, 1980.

\bibitem{RosensteelR81}
G.~Rosensteel and D.~J. Rowe.
\newblock The u(3)-boson model of nuclear collective motion.
\newblock {\em Phys. Rev. Lett.}, 47:223--226, 1981.

\bibitem{CastanosCM84}
O.~Casta{\~n}os, E.~Chac{\'o}n, and M.~Moshinsky.
\newblock Analytic expressions for the matrix elements of generators of sp(6)
  in an sp(6) $\supset$ u(3) basis.
\newblock {\em J. Math. Phys.}, 25(5):1211--1218, 1984.

\bibitem{Quesne81}
C.~Quesne.
\newblock Matrix elements of operators in symmetric
  {U(6)$\supset$U(3)$\supset$U(2)$\supset$U(1)} and
  {U(6)$\supset$SU(3)$\supset$SO(3)$\supset$SO(2)} basis.
\newblock {\em J. Math. Phys.}, 22(7):1482--1496, 1981.

\bibitem{RosensteelR83}
G.~Rosensteel and D.~J. Rowe.
\newblock An analytical formula for u(3)-boson matrix elements.
\newblock {\em J. Math. Phys.}, 24(10):2461--2463, 1983.

\bibitem{LeBlancR87}
R.~{Le Blanc} and D.~J. Rowe.
\newblock Heisenberg-weyl algebras of symmetric and anti-symmetric bosons.
\newblock {\em J. Phys A: Math. Theor.}, 20:L681--687, 1987.

\bibitem{RoweWB85}
D.~J. Rowe, B.~G. Wybourne, and P.~H. Butler.
\newblock {Unitary representations, branching rules and matrix elements for the
  non-compact symplectic groups.}
\newblock {\em J. Phys. A: Math. Gen.}, 18:939--953, 1985.

\bibitem{DavydovF59}
A.~S. Davydov and G.~F. Filippov.
\newblock On the shape of even atomic nuclei.
\newblock {\em Nucl. Phys.}, 10:654--662, 1959.

\bibitem{RoweR82}
D.~J. Rowe and G.~Rosensteel.
\newblock Rotational bands in the u(3)-boson model.
\newblock {\em Phys. Rev. C}, 25:3236--3238, 1982.

\bibitem{RochfordR89}
P.~Rochford and D.~J. Rowe.
\newblock Exploration of {U(3)}-boson dynamics within the nuclear symplectic
  model.
\newblock {\em Nucl. Phys. A}, 492:253--274, 1989.

\bibitem{CastanosD89}
O.~Casta{\~n}os and J.~P. Draayer.
\newblock Contracted symplectic model with $ds$-shell applications.
\newblock {\em Nucl. Phys. A}, 491:349--372, 1989.

\bibitem{Hecht85}
K.~T. Hecht.
\newblock An approximation formula for the {K-}matrix elements of the
  symplectic algebra {Sp(6,R)}.
\newblock {\em J. Phys. A: Math. Gen.}, 18:L1003--1008, 1985.

\bibitem{BahriR00}
C.~Bahri and D.~J. Rowe.
\newblock {SU(3)} quasi-dynamical symmetry as an organizational mechanism for
  generating nuclear rotational states.
\newblock {\em Nucl. Phys. A}, 662:125--147, 2000.

\bibitem{LeBlancCR84}
R.~{Le Blanc}, J.~Carvalho, and D.~J. Rowe.
\newblock A coupled rotor-vibrator model as the macroscopic limit of the
  microscopic symplectic model.
\newblock {\em Phys. Lett. B}, 140:155--158, 1984.

\bibitem{RoweWood10}
D.~J. Rowe and J.~L. Wood.
\newblock {\em Fundamentals of Nuclear Models: Foundational Models.}
\newblock World Scientific, Singapore, 2010.

\bibitem{SuzukiR77b}
T.~Suzuki and D.~J. Rowe.
\newblock The splitting of giant multipole states of deformed nuclei.
\newblock {\em Nucl. Phys. A}, 289:461--474, 1977.

\bibitem{Rowe67}
D.~J. Rowe.
\newblock Schematic interactions for nuclear random-phase approximation
  calculations.
\newblock {\em Phys. Rev.}, 162(4):866--871, 1967.

\bibitem{RoweRabida13}
D.~J. Rowe.
\newblock The fundamental role of symmetry in nuclear models.
\newblock {\em AIP Conf. Proc.}, 1541:104--136, 2013.
\newblock (arXiv:1304.6115 [nucl-th]).

\bibitem{BlomqvistM68}
J.~Blomqvist and A.~Molinari.
\newblock Collective {$0^-$} vibrations in even spherical nuclei with tensor
  forces.
\newblock {\em Nucl. Phys. A}, 106:545--569, 1968.

\bibitem{StephensS72}
F.~S. Stephens and R.S. Simon.
\newblock Coriolis effects in the yrast states.
\newblock {\em Nucl. Phys. A}, 183:257--284, 1972.

\bibitem{Baranger60}
Michel Baranger.
\newblock Extension of the shell model for heavy spherical nuclei.
\newblock {\em Phys. Rev.}, 120:957--968, Nov 1960.

\bibitem{BarangerK65}
M.~Baranger and K.~Kumar.
\newblock Nuclear deformations in the pairing-plus-quadrupole model {(I).}
\newblock {\em Nucl. Phys. A}, 62:113, 1965.

\bibitem{BarangerK68b}
M.~Baranger and K.~Kumar.
\newblock Nuclear deformations in the pairing-plus-quadrupole model {(II).}
  {D}iscussion of valididty of the model.
\newblock {\em Nucl. Phys. A}, 110:490--528, 1968.

\bibitem{BarangerK68c}
K.~Kumar and M.~Baranger.
\newblock Nuclear deformations in the pairing-plus-quadrupole model {(III).}
  {S}tatic nuclear shapes in the rare-earth region.
\newblock {\em Nucl. Phys. A}, 110:529--554, 1968.

\bibitem{BarangerK68d}
M.~Baranger and K.~Kumar.
\newblock Nuclear deformations in the pairing-plus-quadrupole model {(IV).}
  {T}heory of collective motion.
\newblock {\em Nucl. Phys. A}, 122:241--272, 1968.

\bibitem{Rowe82}
D.~J. Rowe.
\newblock Constrained quantum mechanics and a coordinate independent theory of
  the collective path.
\newblock {\em Nucl. Phys. A}, 391:307--326, 1982.

\bibitem{Dirac30}
P.~A.~M. Dirac.
\newblock {\em The Principles of Quantum Mechanics}.
\newblock Oxford University Press, 4 edition, 1967.
\newblock (first published in 1930).

\bibitem{Weyl50}
H.~Weyl.
\newblock {\em The Theory of Groups and Quantum Mechanics}.
\newblock Dover, originally published in 1931 by Methuen, 2 edition, 1950.
\newblock translated by H. P. Robertson from the 1928 German edition,
  Gruppentheorie und Quantenmechanik.

\bibitem{DytrychLDMVSCSLC13}
T.~Dytrych, K.~D. Launey, J.~P. Draayer, P.~Maris, J.~P. Vary, E.~Saule,
  U.~Catalyurek, M.~Sosonkina, D.~Langr, and M.~A. Caprio.
\newblock Collective modes in light nuclei from first principles.
\newblock {\em Phys. Rev. Lett.}, 111:252501(1--5), 2013.

\bibitem{WoodHeyde}
{``Focus Issue on Shape Coexistence in Nuclei"}.
\newblock {\em J. of Phys.G: Nucl. Part. Phys.}, {to be published}.

\bibitem{RoweRR88}
D.~J. Rowe, P.~Rochford, and J.~Repka.
\newblock Dynamic structure and embedded representation in physics: the group
  theory of the adiabatic approximation.
\newblock {\em J. Math. Phys.}, 29:572--577, 1988.

\end{thebibliography}
\end{document}